\documentclass[14pt]{article}
\pdfoutput=1
\usepackage{amsmath,amssymb,amsfonts,amsthm,bm,bbm,cancel,wasysym}
\usepackage{epsfig,graphics,graphicx,epstopdf,caption,subcaption}
\graphicspath{{Charts/}}
\usepackage{array,booktabs,colortbl,colordvi,multirow}
\usepackage{colordvi,color,xcolor}
\usepackage{hyperref}
\usepackage{rotating}
\usepackage{comment}
\usepackage{feynmf}

\usepackage{verbatim}
\usepackage{cite}
\usepackage{setspace}
\usepackage{url}
\usepackage[percent]{overpic}
\usepackage{slashed}
\usepackage{authblk}
\usepackage{xspace}
\usepackage{fullpage}

\def\ie{{\it i.e.}}
\def\eg{{\it e.g.}}

\newskip\zatskip \zatskip=0pt plus0pt minus0pt
\def\matth{\mathsurround=0pt}
\def\lsim{\mathrel{\mathpalette\atversim<}}
\def\gsim{\mathrel{\mathpalette\atversim>}}
\def\atversim#1#2{\lower0.7ex\vbox{\baselineskip\zatskip\lineskip\zatskip
  \lineskiplimit 0pt\ialign{$\matth#1\hfil##\hfil$\crcr#2\crcr\sim\crcr}}}

\begin{document}


\begin{flushright}
SLAC-PUB-17609\\
\today
\end{flushright}
\vspace*{5mm}

\renewcommand{\thefootnote}{\fnsymbol{footnote}}
\setcounter{footnote}{1}

\begin{center}

{\Large {\bf Forbidden Scalar Dark Matter and Dark Higgses}}\\

\vspace*{0.75cm}

{\bf George N. Wojcik$^{1,2}$ and Thomas G. Rizzo$^1$}~\footnote{gwojcik@wisc.edu,  rizzo@slac.stanford.edu}

\vspace{0.5cm}

{$^1$SLAC National Accelerator Laboratory, Menlo Park, CA, 94025 USA}\\
{$^2$Department of Physics, University of Wisconsin-Madison, Madison, WI 53703 USA}

\end{center}
\vspace{.5cm}

\begin{abstract}
 
\noindent 

As experimental searches for WIMP dark matter continue to yield null results, models beyond the WIMP paradigm have proliferated in order to elude ever improving observational constraints, among them that of sub-GeV dark matter mediated by a massive vector portal (a dark photon) associated with a new dark $U(1)$ gauge symmetry. It has been previously noted that for a significant range of the parameter space of this class of models, the annihilation of dark matter particles into a pair of dark photons can dominate the freeze-out process even when this process is kinematically forbidden for dark matter at rest-- this is known as the ''forbidden dark matter'' (FDM) regime. Prior studies of this regime, however, assume that any ``dark Higgs'' associated with breaking the dark $U(1)$ and imparting mass to the dark photon is decoupled from the dark matter and as such plays no role in the freeze-out process. In this paper, we explore the effects of a dark Higgs on sub-GeV dark matter phenomenology in this FDM regime by considering the simplest possible construction in which there exist non-trivial dark matter-dark Higgs couplings: a model with a single complex scalar DM candidate coupled directly to the dark Higgs field. We find that for a wide range of parameter space, the dark Higgs can alter the resulting relic abundance by many orders of magnitude, and that this effect can remain significant even for a small dark matter-dark Higgs coupling constant. Considering measurements from direct detection and measurements of the CMB, we further find that points in this model's parameter space which recreate the appropriate dark matter relic abundance suffer only mild constraints from other sources at present, but may become accessible in near-future direct detection experiments.

\end{abstract}

\renewcommand{\thefootnote}{\arabic{footnote}}
\setcounter{footnote}{0}
\thispagestyle{empty}
\vfill
\newpage
\setcounter{page}{1}



\section{Introduction}\label{Section:Intro}

In spite of ample evidence for its existence from astrophysical and cosmological data, the precise identity of dark matter (DM) remains an ongoing mystery in physics. As the parameter space for traditional DM candidates, such as Weakly Interacting Massive Particles (WIMPs) \cite{Arcadi:2017kky,Roszkowski:2017nbc} and axions \cite{Kawasaki:2013ae,Graham:2015ouw,Irastorza:2018dyq} become more and more experimentally constrained without appearing, other models to describe the nature of DM have proliferated \cite{Battaglieri:2017aum,Alexander:2016aln} that evade some or all of the constraints on the traditional candidates. Generally, the observed relic abundance of DM from \emph{Planck} \cite{Planck:2018vyg} suggests that, just as in the traditional models, many of these other models of DM require that the dark sector of particles interacts with the SM through some mechanism other than gravity. A wide swath of these models, then, can be classified simply by the mechanism through which this interaction occurs. In this paper, we focus on the so-called ``vector portal/kinetic mixing'' scenario \cite{Holdom:1985ag,Holdom:1986eq,Pospelov:2007mp,Izaguirre:2015yja,Essig:2013lka,Curtin:2014cca}, in which this interaction proceeds through a new $U(1)$ ``dark'' force. Dark matter particles are charged under this dark $U(1)$ and uncharged under the Standard Model (SM) gauge group, while SM particles are uncharged under the dark $U(1)$ (and, of course, charged under the SM gauge group). Interaction between the dark sector and the SM then proceeds via kinetic mixing between the SM hypercharge field and the dark $U(1)$; this occurs due to a term of the form
\begin{align}
    \mathcal{L}_{\textrm{KM}} \sim \frac{\epsilon}{2 c_W}F^D_{\mu \nu} F^Y_{\mu \nu},
\end{align}
where $F^D$ and $F^Y$ denote the dark $U(1)$ and SM hypercharge field strength tensors, respectively, while $c_W$ is the cosine of the Weinberg angle. If absent at the tree level, this kinetic mixing can occur at the one- or two-loop level due to so-called ``portal matter'' at some high scale \cite{Holdom:1985ag,Holdom:1986eq,Gherghetta:2019coi,Rizzo:2018vlb,Rueter:2020qhf,Kim:2019oyh}, in which case the kinetic mixing parameter $\epsilon$ is naturally of $O(10^{-(3-4)})$. The simplest realization of this construction introduces only two new particles -- a DM field $\phi$ and a dark photon $A_D$ -- and is fully defined by a handful of parameters: The dark coupling constant $g_D$, the kinetic mixing parameter $\epsilon$, and the masses of the DM and the dark photon (the gauge boson associated with the dark $U(1)$ force). It has been found that in this simple setup, the observed relic abundance of DM can be recreated for a significant range of parameters assuming $g_D \sim O(1)$, $\epsilon \sim O(10^{-(3-4)})$ and that the DM and dark photon masses both lie in the range $\sim 100 \; \textrm{MeV}$ and $\sim 1 \; \textrm{GeV}$; in this regime, DM will annihilate into SM final states through the dark photon, generating the relic abundance via the familiar freeze-out mechanism \cite{Saikawa:2020swg,Gondolo:1990dk}. Often, the simple construction described above is augmented by an additional complex scalar $S$ which acquires a vacuum expectation value in order to break the dark $U(1)$ gauge symmetry and imbue the dark photon $A_D$ with a mass, leaving an additional physical particle in the form of a real scalar ``dark Higgs'' $h_D$.\footnote{Because the dark $U(1)$ is Abelian, an alternative to introducing a dark Higgs would be to simply imbue the dark photon with a mass from the Stueckelberg mechanism. However, we shall see that the dark Higgs allows for a substantially more complex and interesting phenomenology in the construction we consider in this paper.} Often, the effect of the dark Higgs is not explicitly included in discussions of the DM phenomenology of these models, because unless its mass is smaller than that of the dark photon, in which case it can be long-lived and potentially phenomenologically relevant, the dark Higgs often has little bearing on the thermal history of the universe or modern detection prospects for DM \cite{Darme:2017glc}.

The most well-explored version of the kinetic mixing/vector portal scenario assumes that the dominant DM annihilation process at freeze-out is the annihilation of a DM pair into a pair of SM particles via an $s$-channel exchange of a dark photon, the direct parallel of the usual annihilation process for WIMP DM. However, the parameter space of the model does permit other experimentally viable regimes which display substantially different phenomenology: Notably, when the dark photon mass lies between 1 and 2 times the DM mass, the annihilation of a pair of DM particles into a pair of on-shell dark photons, both of which then decay into SM particles, can become significant. Rather than being suppressed by the small kinetic mixing parameter $\epsilon$, as in the case of the WIMP-like annihilation process, the thermal average of this ``forbidden DM'' (FDM) cross section suffers an exponential Boltzmann suppression because the DM is less massive than that of the dark photon, and hence the process can only occur for DM with an energy above the kinematic threshold.

The FDM scenario, where this kinematically forbidden process dominates freeze-out, was first discussed in \cite{Griest:1990kh} and is further explored at the weak scale in \cite{Delgado:2016umt}. It is discussed in the context of sub-GeV vector portal/kinetic mixing DM in \eg \cite{DAgnolo:2015ujb,Cline:2017tka,Fitzpatrick:2020vba}. The effect of the dark Higgs in this FDM regime, however, has been left largely unexplored.\footnote{See, however, \cite{Hara:2021lrj}, which explores forbidden sub-GeV DM with a scalar portal, albeit outside of the kinetic mixing paradigm.} Given the fact that a broken dark $U(1)$ strongly motivates the existence of such a dark Higgs, it is not unreasonable to consider if there exist constructions in which the dark Higgs plays a significant role in model phenomenology, and further to consider how finely-tuned these constructions are. To that end, in this paper we present the simplest construction of a sub-GeV vector portal/kinetic mixing model in which the dark Higgs directly couples to the DM: The SM augmented by a dark $U(1)$ group, a complex scalar DM candidate, and a second complex scalar that achieves a vev in order to break the dark $U(1)$ (containing the dark Higgs).\footnote{There do exist more complicated constructions in this framework such that the DM has a significant coupling to the dark Higgs. For example, with fermionic DM, one can realize significant Yukawa-like dark Higgs-DM couplings by selecting the dark $U(1)$ charges of the new particles appropriately. However, this selection requires either two chiral DM fermions of different dark $U(1)$ charge or that the dark Higgs vev imparts a Majorana mass term for the Dirac fermion, \ie, the psuedo-Dirac setup. The former case requires multiple additional chiral fields to avoid gauge anomalies, while the latter will split the Dirac fermion DM into two Weyl fermions with non-degenerate masses. In either case, the constructions are substantially more complicated than the complex scalar DM scenario discussed here.} Even in this simple construction, the dark Higgs provides for the addition of rich phenomenology to the FDM paradigm.
We shall find that for a significant range of dark Higgs masses, the dark Higgs effects are potentially enormous, even altering the predicted relic density for these constructions by as much as three orders of magnitude. Additionally, we find that these effects are remarkably resilient against changes in the coupling between the dark Higgs and the DM: Even very small couplings of the dark Higgs to the DM can result in potentially very large effects on the DM relic abundance.

Our paper is laid out as follows. In Section \ref{Section:ModelSetup}, we introduce the vector portal/kinetic mixing DM setup that we  employ, including the dark Higgs, and list all the free parameters in our model. In Section \ref{Section:RelicDensity}, we outline the methodology we employ for computing the DM relic abundance in this system, including outlining which annihilation processes we have determined to dominate freeze-out. In Section \ref{Section:Analysis}, we give numerical results for DM relic abundance calculations, quantifying the effect of various model parameters at different benchmark points and producing benchmark points in parameter space that recreate the observed relic abundance. In Section \ref{Section:Constraints}, we discuss the dominant experimental constraints that arise on this model, from direct detection and the cosmic microwave background. Finally, in Section \ref{Section:Conclusion}, we summarize our findings and discuss avenues for future inquiry.

\section{Model Setup}\label{Section:ModelSetup}

Our setup is a straightforward realization of a nearly minimal model of vector portal kinetic mixing dark matter (DM), in the manner of, \eg \cite{DAgnolo:2015ujb,Darme:2017glc}. The gauge group of the SM is extended by a new Abelian dark gauge symmetry $U(1)_D$, under which all SM particles are neutral. The dark sector itself consists of the $U(1)_D$ gauge boson $A_D$ (\ie, the ``dark photon''), a stable complex scalar DM particle $\phi$, and a second complex dark scalar $S$, which acquires a vacuum expectation value (vev) and breaks $U(1)_D$. Both $\phi$ and $S$ are SM singlets. The dark sector is coupled to the SM via the small kinetic mixing between $U(1)_D$ and the SM hypercharge -- to an excellent approximation the main effect of this kinetic mixing is to grant SM particles a coupling term to the dark photon equal to $\epsilon \sim 10^{-(3-4)}$ times their usual coupling to the SM photon. The DM particle $\phi$, can then achieve its relic abundance from conventional thermal freeze-out via annihilation into SM fermions, dark photons, or dark Higgs particles $h_D$ that will emerge as the remaining physical part of the scalar field $S$ after spontaneous symmetry breaking. As discussed in Section \ref{Section:Intro}, we are particularly interested in the so-called ``forbidden DM'' (FDM) regime of parameter space, in which the ratio of the dark photon mass to the DM mass, $r_\phi \equiv m_{A_D}/m_\phi$, is between 1 and 2, in which case the annihilation process $\phi^* \phi \rightarrow A_D A_D$ can play a central role in producing the correct relic abundance. Furthermore, we limit our consideration of the mass of the dark Higgs $h_D$ to the case in which $m_{h_D} > m_{A_D}$, so that even if any other decays are kinematically disallowed, $h_D$ can still decay promptly through $h_D \rightarrow A_D^* A_D \rightarrow \overline{f} f A_D$, where $f$ denotes some SM fermion. Allowing $h_D$ to be less massive than this would leave kinematically allowed only highly suppressed decay processes, such as the one-loop decay $h_D \rightarrow \overline{f} f$, leading to an extremely long-lived $h_D$ subject to cosmological and possibly BBN constraints \cite{Darme:2017glc}.

Our choice of a scalar DM candidate has important implications for our model building: Unlike the equivalent process for Dirac fermion DM, the annihilation cross section for the process $\phi \phi^* \rightarrow f \overline{f}$, where $f$ denotes an SM fermion, is $p$-wave rather than $s$-wave, and so is suppressed by a velocity-squared factor. This velocity suppression in turn allows a scalar DM model such as this one to trivially evade CMB constraints \cite{Planck:2018vyg} from DM annihilations to SM fermions at the epoch of recombination. Furthermore, the velocity suppression experienced by this cross section is enough to limit the strength of the annihilation cross section to SM fermions relative to that of the kinematically forbidden transition $\phi^* \phi \rightarrow A_D A_D$ even at temperatures near freeze-out, (depending, \eg, upon the size of $g_D$). As a result, unlike the analyses done of a similar model with Dirac fermion DM in \cite{DAgnolo:2015ujb,Fitzpatrick:2020vba}, we find that the annihilation cross section of the standard WIMP-like annihilation process $\phi^* \phi \rightarrow \overline{f} f$ does \emph{not} necessarily dominate over that of the kinematically forbidden process $\phi^* \phi \rightarrow A_D A_D$ when we allow the kinetic mixing term $\epsilon$ to be as large as $O(10^{-(3-4)})$, for reasonable selections of $g_D$ between $0.1$ and $1$. 

The action of the dark sector can then written as 
\begin{align}\label{eq:darkAction}
    S_{\textrm{dark}} = \int d^4 x \bigg\{ |D_\mu \phi|^2 + |D_\mu S|^2 -\frac{1}{4} (A_D)^{\mu \nu} (A_D)_{\mu \nu} - V(S, \phi) \bigg\},
\end{align}
where $D_\mu$ denotes the usual covariant derivative, $(A_D)_{\mu\nu}$ denotes the usual field strength tensor, and $V(S, \phi)$ is the potential associated with the two scalars. When choosing the form of $V(S, \phi)$, we need to consider the characteristics which we want the scalars $S$ and $\phi$ to have. First, we note that we are exploring the parameter space around which the dark photon $A_D$ acquires a mass on the order of $m_{A_D} \sim 0.1-1 \; \textrm{GeV}$ via the Abelian Higgs mechanism. Therefore, one or both of the $U(1)_D-$charged scalars, $S$ and $\phi$, have to acquire vev's of roughly the magnitude of $m_{A_D}$. In order to ensure the stability of our DM candidate $\phi$, we require that it does \emph{not} acquire a vev, and so the $U(1)_D$ symmetry is entirely broken by $S$. In order to achieve this arrangement of vev's, we assume that $V(S,\phi)$ takes the form
\begin{align}\label{eq:VSPhi}
    V(S, \phi) = m_{\phi,0}^2 |\phi |^2 - \mu_S^2 |S|^2 + \lambda_\phi |\phi |^4 + \eta |\phi |^2 |S |^2 + \lambda_S |S |^4,
\end{align}
where $m_{\phi,0}$, $\mu_S$, $\lambda_\phi$, $\eta$, and $\lambda_S$ are all real parameters. This can easily be seen to be the most general scalar potential that we can write given two assumptions: First, that cross terms such as $\phi^\dagger S$ or $\phi^\dagger S^3$ are forbidden, and second, that any terms mixing the dark scalars with the SM Higgs are small enough that they are negligible. The first assumption can easily be justified by either assigning appropriate differing $U(1)_D$ charges for $\phi$ and $S$ or requiring the action to be symmetric under the $Z_2$ transformation $\phi \rightarrow -\phi$. Meanwhile, the second assumption is motivated by harsh observational constraints: Any mixing term between the SM Higgs and the dark scalars is well-known to be very small to avoid an excessive invisible branching fraction for the SM Higgs since $B_{inv} \lsim 0.1$ \cite{ATLAS:2019cid,Rizzo:2018vlb}. 

As long as the bounded-from-below conditions $\lambda_S>0$, $\lambda_\phi>0$, and $\eta > -2 \sqrt{\lambda_S \lambda_\phi}$ are met and $m_{\phi,0}^2 +\eta \mu_S^2/(2 \lambda_S) >0$, the potential of Eq.(\ref{eq:VSPhi}) will achieve a minimum when
\begin{align}
    \langle \phi \rangle = 0, \;\;\; \langle S \rangle= \frac{v_S}{\sqrt{2}} \equiv \sqrt{\frac{\mu_S^2}{2 \lambda_S}}.
\end{align}
After spontaneous symmetry breaking, the dark photon, $A_D$, will acquire a mass $m_{A_D} = g_D |Q_S| v_S$, where $g_D$ is the coupling constant for $U(1)_D$ and $Q_S$ is the $U(1)_D$ charge of the $S$ field, in units of the $U(1)_D$ charge of the DM field $\phi$, which without loss of generality we set equal to unity. The scalar field $S$ will have only one physical degree of freedom remaining, a real scalar which we'll call $h_D$, the ``dark Higgs''. Meanwhile, the DM $\phi$ will acquire an additional mass term from the $\eta |\phi|^2 |S|^2$ term in the potential. The full mass of $\phi$, which we'll denote by $m_\phi$, is then given by
\begin{align}
    m_\phi^2= m_{\phi,0}^2+\frac{\eta}{2}v_S^2.
\end{align}
When $\eta>0$, the mass term $\eta v_S^2/2$ can be intuitively written in terms of a new parameter $b$, defined with
\begin{align}
    b m_\phi^2 = \frac{\eta}{2}v_S^2.
\end{align}
When $\eta>0$, $b$ can range from 0 to 1, with $b=0$ indicating that none of the $\phi$ particle's mass comes from the vev of $S$, while $b=1$ indicates that \emph{all} of the tree-level mass comes from this vev. Roughly, $b$ can be thought of as the fraction of the DM mass squared that that comes from the vev of $S$. In the case of negative $\eta$ (or equivalently, negative $b$), the intuitive understanding of $b$ must differ somewhat. In principle, $b$ can achieve larger negative magnitudes than 1, as long as $m_{\phi,0}$ is large enough to ensure that the total mass squared of the DM is still positive. In practice, however, satisfying bounded-from-below conditions on the potential and assuming from naturalness that the original coefficients $\lambda_S$, $\lambda_\phi$, and $\eta$ are of $O(1)$ limits the reasonable magnitude of negative $b$ to roughly $O(1)$, dependent on the specific selections of parameters. We shall find that for the physical processes which depend on $b$, the effects of allowing negative $b$ are straightforward to qualitatively determine from our results for positive $b$: Apart from a more complicated upper bound on the magnitude of the $b$ parameter, the only effect on the rates of physical processes will be the sign of some interference terms in cross section calculations. For the sake of simplicity, therefore, we will restrict our quantitative analysis here to positive $b$ (equivalently positive $\eta$), and discuss the qualitative effects of allowing negative $b$ where they become relevant.

In the unitary gauge, the dark sector action can be written as
\begin{align}\label{eq:BrokenAction}
    S_{\textrm{dark}} =\int d^4 x \bigg\{ &|\partial_\mu \phi |^2-m_\phi^2 |\phi|^2-\frac{1}{4}(A_D)^{\mu \nu} (A_D)_{\mu \nu} + \frac{m_{A_D}^2}{2}(A_D)^\mu(A_D)_\mu + \frac{1}{2}(\partial_\mu h_D)^2-\frac{m_{h_D}^2}{2} h_D^2 \nonumber\\
    &- i g_D (A_D)^\mu (\phi^* \partial_\mu \phi - \phi \partial_\mu \phi^*)+g_D^2 |\phi |^2 (A_D)^\mu(A_D)_\mu + \lambda_\phi |\phi |^4\\
    &+ \frac{g_D^2 Q_S^2}{2}h_D^2 (A_D)^\mu (A_D)_\mu + g_D |Q_S| m_{A_D} h_D (A_D)^\mu (A_D)_\mu \nonumber\\
    &+\frac{g_D^2 Q_S^2}{8} \frac{m_{h_D}^2}{m_{A_D}^2} h^4+\frac{g_D Q_S m_{h_D}^2}{2 m_{A_D}} h_D^3- \frac{b g_D^2 Q_S^2}{2}\frac{m_\phi^2}{m_{A_D}^2} h_D^2 |\phi|^2-b g_D |Q_S| m_\phi \frac{m_\phi}{m_{A_D}} h_D |\phi |^2 \bigg\}, \nonumber
\end{align}
where for convenience we have written the action solely in terms of the gauge coupling as well as the DM mass $m_\phi$, the dark photon mass $m_{A_D}$, and the dark Higgs mass $m_{h_D}$, taking $Q_\phi=1$ as previously noted. Up to kinetic mixing, which we briefly discussed earlier in this Section, this expression contains the complete action for the dark sector of the model.

\section{Relic Density Calculation}\label{Section:RelicDensity}

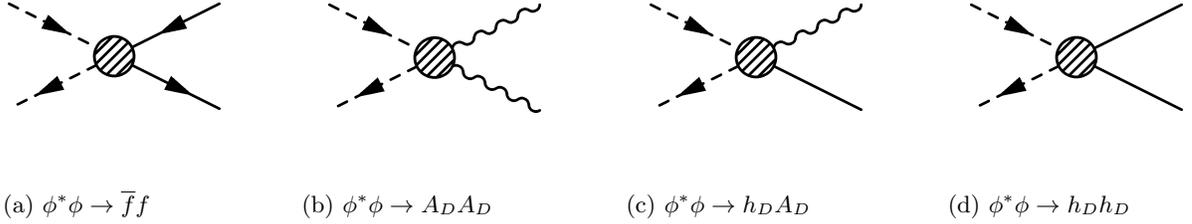
\begin{figure}
  \centering
  \begin{subfigure}[b]{0.2\textwidth}
    \centering
    \fmfframe(10,25)(10,25){
    \begin{fmfgraph*}(100,40)
        \fmfleft{i1,i2}
        \fmfright{o1,o2}
        \fmf{scalar}{i2,v1,i1}
        \fmfblob{0.15w}{v1}
        \fmf{fermion}{o2,v1,o1}
        \fmflabel{$\phi$}{i2}
        \fmflabel{$\phi^*$}{i1}
        \fmflabel{$\overline{f}$}{o2}
        \fmflabel{$f$}{o1}
    \end{fmfgraph*}
    }
    \caption{$\phi^* \phi \rightarrow \overline{f} f$}
  \end{subfigure}
  \hfill
  \begin{subfigure}[b]{0.2\textwidth}
    \centering
    \fmfframe(10,25)(10,25){
    \begin{fmfgraph*}(100,40)
        \fmfleft{i1,i2}
        \fmfright{o1,o2}
        \fmf{scalar}{i2,v1,i1}
        \fmfblob{0.15w}{v1}
        \fmf{boson}{o2,v1,o1}
        \fmflabel{$\phi$}{i2}
        \fmflabel{$\phi^*$}{i1}
        \fmflabel{$A_D$}{o2}
        \fmflabel{$A_D$}{o1}
    \end{fmfgraph*}
    }
    \caption{$\phi^* \phi \rightarrow A_D A_D$}
  \end{subfigure}
  \hfill
  \begin{subfigure}[b]{0.2\textwidth}
    \centering
    \fmfframe(10,25)(10,25){
    \begin{fmfgraph*}(100,40)
        \fmfleft{i1,i2}
        \fmfright{o1,o2}
        \fmf{scalar}{i2,v1,i1}
        \fmfblob{0.15w}{v1}
        \fmf{boson}{o2,v1}
        \fmf{plain}{o1,v1}
        \fmflabel{$\phi$}{i2}
        \fmflabel{$\phi^*$}{i1}
        \fmflabel{$A_D$}{o2}
        \fmflabel{$h_D$}{o1}
    \end{fmfgraph*}
    }
    \caption{$\phi^* \phi \rightarrow h_D A_D$}
  \end{subfigure}
  \hfill
  \begin{subfigure}[b]{0.2\textwidth}
    \centering
    \fmfframe(10,25)(10,25){
    \begin{fmfgraph*}(100,40)
        \fmfleft{i1,i2}
        \fmfright{o1,o2}
        \fmf{scalar}{i2,v1,i1}
        \fmfblob{0.15w}{v1}
        \fmf{plain}{o2,v1,o1}
        \fmflabel{$\phi$}{i2}
        \fmflabel{$\phi^*$}{i1}
        \fmflabel{$h_D$}{o2}
        \fmflabel{$h_D$}{o1}
    \end{fmfgraph*}
    }
    \caption{$\phi^* \phi \rightarrow h_D h_D$}
  \end{subfigure}
    \caption{The four processes which dominantly contribute to the thermally averaged annihilation cross section $\sum_X \langle \sigma v \rangle_{\phi^* \phi \rightarrow X}$ in the Boltzmann equation written in Eq.(\ref{eq:OriginalBoltzmann}). Note that $\phi^* \phi \rightarrow h_D A_D$ and $\phi^* \phi \rightarrow h_D h_D$ are only included in our calculations when the dominant $h_D$ decay is $h_D \rightarrow A_D^* A_D \rightarrow \overline{f} f A_D$, as discussed in the text.}
    \label{fig1}
\end{figure}

Having set up our model, we now must compute the relic abundance of the dark matter (DM) candidate $\phi$, the only stable SM singlet in our model, achieves. In the region of parameter space we are considering, $\phi$ achieves its relic abundance via a conventional freeze-out mechanism \cite{Boehm:2003hm}. In this case, the number density of the DM $n_\phi$ (noting that this refers to the combined number density of both $\phi$ and its antiparticle $\phi^*$, each of which should have the same number density equal to $n_\phi/2$) is given by the solution to the Boltzmann equation,
\begin{align}\label{eq:OriginalBoltzmann}
    \dot{n_\phi}+3 H n_\phi = -\frac{1}{2}(n_\phi^2-n_{\phi,0}^2) \sum_{X} \langle \sigma v \rangle_{\phi^* \phi \rightarrow X},
\end{align}
where $X$ denotes any final state such that the $\phi^*$ and $\phi$ in the initial state are no longer present, $n_{\phi,0}$ represents the equilibrium number density of the DM  and $\langle \sigma v \rangle$ denotes a thermal averaging of the annihilation cross section, discussed later in this section. The factor of $1/2$ in front of the collision term here correctly accounts for both the fact that $n_\phi$ describes the combined number density of $\phi$ particles and antiparticles, and each annihilation process eliminates two total DM particles: both a $\phi$ and a $\phi^*$. To determine what selections of model parameters will reproduce the observed relic abundance of DM, then, we only need to compute all the $\langle \sigma v \rangle_{\phi^* \phi \rightarrow X}$ and solve Eq.(\ref{eq:OriginalBoltzmann}) numerically. In our case, there are four major processes that can contribute to $\langle \sigma v \rangle_{\phi^* \phi \rightarrow X}$: The WIMP-like $s$-channel annihilation $\phi^* \phi \rightarrow \overline{f} f$ (where $f$ denotes a SM fermion), the classic ``forbidden DM'' annihilation process $\phi^* \phi \rightarrow A_D A_D$, and the processes with dark Higgses in the final state: $\phi^* \phi \rightarrow h_D A_D$ and $\phi^* \phi \rightarrow h_D h_D$. For convenience, we have depicted each process graphically in Figure \ref{fig1}. Note that while we have selected a mass range for $\phi$ and $A_D$ such that the process $\phi^* \phi \rightarrow A_D A_D$ will always reduce the number of DM particles in the system by 2 (specifically, we've chosen $m_{A_D}$ and $m_\phi$ such that $A_D$ will always decay into SM particles), the processes $\phi^* \phi \rightarrow h_D A_D$ and $\phi^* \phi \rightarrow h_D h_D$ will only contribute to the DM number density in the manner depicted in Eq.(\ref{eq:OriginalBoltzmann}) when $m_{h_D}< 2 m_\phi$. Otherwise, the dominant decay for $h_D$ will be either $h_D \rightarrow \phi^* \phi$ or, if kinematically accessible, $h_D \rightarrow A_D A_D$. If the dominant decay of $h_D$ is $h_D \rightarrow \phi^* \phi$, then the process $\phi^* \phi \rightarrow h_D A_D$ will actually \emph{preserve} the number of DM particles, while $\phi^* \phi \rightarrow h_D h_D$ will \emph{increase} it. In practice, however, processes with final-state dark Higgses have a negligible contribution to the collision term in Eq.(\ref{eq:OriginalBoltzmann}) whenever $h_D$ is massive enough to decay into dark sector particles, due to the severe kinematic suppression of the cross section of these processes; as $h_D$ becomes heavier relative to $\phi$, cross sections which produce $h_D$ in the final state from $\phi-\phi^*$ collisions become more and more exponentially suppressed. For the purposes of our calculation, we limit ourselves to including the processes $\phi^* \phi \rightarrow h_D A_D$ and $\phi^* \phi \rightarrow h_D h_D$ only when $m_{h_D} < 2 m_\phi$, so that $h_D$ will dominantly decay via $h_D \rightarrow A_D^* A_D \rightarrow \overline{f} f A_D$, where $f$ again denotes an SM fermion. When $m_{h_D}> 2 m_\phi$, a reader may also be concerned with the effect on freeze-out of the $2 \rightarrow 1$ process $\phi^* \phi \rightarrow h_D$: If $m_{h_D}< 2 m_{A_D}$, this process even undergoes less Boltzmann suppression than the $2 \rightarrow 2$ process $\phi^* \phi \rightarrow A_D A_D$. However, because $h_D$ in this regime can decay into DM pairs, the number-changing effect of the process $\phi^* \phi \rightarrow h_D$ can be simply included in the processes $\phi^* \phi \rightarrow A_D A_D$ and $\phi^* \phi \rightarrow A_D A_D^* \rightarrow A_D \overline{f} f$ (the latter of which we omit from our calculations because it suffers both $\alpha_D = g_D^2/(4 \pi)$ and three-body phase space suppression relative to the WIMP-like process $\phi^* \phi \rightarrow \overline{f} f$).

\begin{fmffile}{diagram1}
\end{fmffile}

In writing Eq.(\ref{eq:OriginalBoltzmann}), we have made several tacit assumptions about the dominant processes governing the number density of the DM; it is useful for us to explicitly state and justify them here. First, we have assumed that the dominant processes governing the DM are straightforward $2 \rightarrow 2$ $\phi-\phi^*$ annihilations which leave no DM in the final state. It is easy to see that, if we restrict our attention to $2 \rightarrow 2$ annihilations, the only interactions in the action of Eq.(\ref{eq:BrokenAction}) which reduce the number of $\phi$'s are particle-antiparticle annihilations. Our choice to limit our attentions to $2 \rightarrow 2$ processes, however, requires some more scrutiny. In particular, it has been found \cite{Cline:2017tka} that in models where kinematically forbidden processes like $\phi^* \phi \rightarrow A_D A_D$ dominate the DM annihilation cross section to SM fermions, it is feasible that $3\rightarrow 2$ processes such as $\phi \phi \phi^* \rightarrow A_D \phi$ may play a significant role in the thermal freeze-out process. Specifically, in the region of parameter space in which $r_\phi = m_{A_D}/m_\phi \lsim 1.5$, it was found that the cross section of the most significant $3 \rightarrow 2$ process is exponentially suppressed relative to the $2 \rightarrow 2$ kinematically forbidden process, $\phi^* \phi \rightarrow A_D A_D$, but as $r_\phi$ gets larger, the $3 \rightarrow 2$ process will come to dominate. Generally in our numerical study we shall focus on regions of parameter space in which the kinematically forbidden $\phi^* \phi \rightarrow A_D A_D$ dominates over all other $2 \rightarrow 2$ processes, and therefore we shall find that points in the parameter space that are of interest to us are usually comfortably in the $r_\phi \lsim 1.5$ range. However, we also find that there are regions of our parameter space in which our relic abundance calculation with only $2 \rightarrow 2$ processes still has the forbidden $\phi^* \phi \rightarrow A_D A_D$ dominate, but has $r_\phi \gsim 1.5$. In this case, our choice to exclude the $3 \rightarrow 2$ processes merits some further discussion.

Semi-quantitatively, we can estimate without explicit computation that in our setup the $3\rightarrow 2$ processes will likely be dominated by the WIMP-like annihilation cross section $\phi^* \phi \rightarrow \overline{f} f$ in regions of the parameter space in which $r_\phi \gsim 1.5$, in contrast to the analogous case with smaller $\epsilon$ considered in \cite{DAgnolo:2015ujb,Cline:2017tka,Fitzpatrick:2020vba}. Ultimately, this is due to the fact that we have assumed a larger value of the kinetic mixing parameter, letting $\epsilon \sim 10^{-(3-4)}$ rather than $< 10^{-6}$. Accommodating Boltzmann factors, the contribution of the process $\phi^* \phi \phi \rightarrow A_D \phi$ to the collision term of the Boltzmann equation will be suppressed by a factor of $\sim \exp[-m_\phi/T] (\alpha_D)^2/(\alpha_{\textrm{em}} \epsilon^2)$ relative to the contribution of the process $\phi^* \phi \rightarrow \overline{f} f$, where $\alpha_D=g_D^2/(4 \pi)$, $\alpha_{\textrm{em}} \approx 1/137$ is the electromagnetic fine structure constant, and $T$ is the universe's temperature. For the regions of parameter space we will consider in this work, where $\epsilon \sim 10^{-(3-4)}$, $\alpha_D \sim \alpha_{\textrm{em}} \sim 10^{-2}$, and freeze-out tends to occur at $m_\phi/T \gsim 20$, we find that this suppression factor roughly suggests that the $3 \rightarrow 2$ process should be subordinate to the process $\phi^* \phi \rightarrow \overline{f} f$. While the preceding argument might motivate our ad hoc omission of $3\rightarrow 2$ processes in our calculation, it is hardly rigorous. However, once we have obtained numerical results considering only the $2 \rightarrow 2$ processes, we check several points in parameter space in which the $3 \rightarrow 2$ process is most likely to be dominant (namely, when $g_D$ is as large as possible, and $r_\phi = m_{A_D}/m_\phi$ is as close to 2 as possible) and find numerically that for model points that reproduce the appropriate relic abundance, the contribution of $3 \rightarrow 2$ processes to the collision term in the Boltzmann equation near freeze-out are several orders of magnitude below the contributions of the $2 \rightarrow 2$ processes. The only region of parameter space at which the $3 \rightarrow 2$ process might play a dominant role in our analysis occurs when $m_{h_D}/m_\phi = r_h r_\phi \approx 2$, in which case the amplitude $\phi^* \phi \phi \rightarrow A_D \phi$ enjoys a resonant enhancement, while the $g_D$ is kept as large as we consider it and $m_{A_D}$ is kept as small as we consider it. However, we shall see in Section \ref{Section:Constraints} that for points that recreate the relic abundance with $r_\phi \gsim 1.5$ (namely, $r_\phi$ large enough to overcome the exponential suppression of the $3\rightarrow 2$ cross sections relative to those of the $2\rightarrow 2$ processes), these these regions of parameter space are excluded by constraints arising from measurements of the cosmic microwave background. As such, while the explicit contribution of $3 \rightarrow 2$ processes to the relic abundance of the DM is omitted in our calculations, we can safely estimate that its effect is negligible.

The final assumption we have made when writing Eq.(\ref{eq:OriginalBoltzmann}) is that throughout freeze-out, the only particle out of thermal equilibrium will be the DM candidate $\phi$. In general, this assumption is not unreasonable, especially because the other dark sector particles (the dark photon $A_D$ and the dark Higgs $h_D$) both have large decay rates. In the case of $A_D$, the decay rate proportional to $\epsilon^2 \alpha_{\textrm{em}}$ is in general large enough to keep $A_D$ easily in thermal equilibrium throughout freeze-out when we allow ``large'' $\epsilon \sim 10^{-(3-4)}$, as seen in \cite{Cline:2017tka}. A more interesting scenario emerges in the case of $h_D$, in particular when $m_{h_D}<2 m_\phi$, so that the dominant decay process of $h_D$ is the $\epsilon$-suppressed $h_D \rightarrow A_D^* A_D \rightarrow \overline{f} f A_D$. In this case, the narrow decay width of $h_D$ may cause it to acquire a non-equilibrium distribution before or during the freeze-out process of $\phi$, much like the dark photon does for much smaller $\epsilon$ values in \cite{DAgnolo:2015ujb,Cline:2017tka,Fitzpatrick:2020vba}. If the effect of non-equilibrium $h_D$ number densities were significant, then the single Boltzmann equation of Eq.(\ref{eq:OriginalBoltzmann}) would need to be extended to a coupled set of equations to be solved for both the number density of the DM $\phi$ and that of the dark Higgs $h_D$. Instead, however, we find numerically that for the smallest value of $m_{h_D}$ (and therefore the smallest decay width) that we consider, $(m_{h_D}-m_{A_D})/m_{A_D} \sim 10^{-2}$, the effect of including a non-equilibrium $h_D$ number density has at most a percent level effect on the final computed value of the relic density of $\phi$. We can better explain the minuteness of this effect in light of the results of our numerical studies: In Section \ref{Section:Analysis}, we shall observe that the DM annihilation processes which are affected by a non-equilibrium distribution of dark Higgses, namely $\phi^* \phi \rightarrow h_D h_D$ and $\phi^* \phi \rightarrow h_D A_D$, have at most an $O(10\%)$ effect on relic abundance when $h_D$ is assumed to be in equilibrium with the SM bath. Therefore, in order to have a discernible effect on the final relic density, the number density of $h_D$ must dramatically depart from its equilibrium value before the DM $\phi$ freezes out. Any departure of the dark Higgs from its equilibrium density turns out not to be significant enough to make the annihilation processes with final-state dark Higgses dominate over the process $\phi^* \phi \rightarrow A_D A_D$.


Having now justified our assumptions, we can move on to the work of computing the DM relic abundance that emerges from this model. It is well-known that the Boltzmann equation in Eq.(\ref{eq:OriginalBoltzmann}) is much simpler to work with when one solves for $Y_\phi \equiv n_\phi/\tilde s$, where $\tilde s$ is the entropy per comoving volume of the universe, instead of $n_\phi$; we can rewrite Eq.(\ref{eq:OriginalBoltzmann}) as
\begin{align}\label{eq:YBoltzmann}
    \frac{d Y_\phi}{d x} = - \frac{1}{2} \bigg( \frac{45}{\pi}G \bigg)^{-\frac{1}{2}} \frac{g_*^{1/2} m_\phi}{x^2} (Y_\phi^2 - Y_{\phi,0}^2) \sum_X \langle \sigma v \rangle_{\phi^* \phi \rightarrow X}, \; \; x \equiv m_\phi/T,
\end{align}
where $G$ is the gravitational constant and $g_*^{1/2}$ is a relativistic degrees of freedom parameter which we extract from \cite{Hindmarsh:2005ix} (the use of other sources for the values of this parameter alter our results for relic densities at the percent level at most). Our remaining task is then to compute the thermally averaged annihilation cross sections for each of the four processes we consider here. To do so, we follow \cite{Gondolo:1990dk}, which gives a convenient single-integral formula for the thermally averaged annihilation cross section:
\begin{align}
    \langle \sigma v \rangle_{\phi^* \phi \rightarrow A B} = \frac{2 x}{K_2^2 (x)}\int_{\varepsilon_{\textrm{min}}}^\infty d \varepsilon \; \varepsilon^{1/2}(1+2 \varepsilon)K_1 (2 x \sqrt{1+\varepsilon}) ~\sigma v_{\textrm{lab}}, \; \; \varepsilon \equiv \frac{s}{4 m_\phi^2}-1,
\end{align}
where $s$ denotes the usual Mandelstam variable (not to be confused with the co-moving entropy density) and $A$ and $B$ denote any two final-state particles. $\sigma v_{\textrm{lab}}$ denotes the Lorentz-invariant cross section $\sigma$ of a given process multiplied by the velocity of one DM particle in the other's rest frame, $v_{\textrm{lab}}$. We note that this expression only applies when one can approximate $\phi$ as following a Maxwell-Boltzmann distribution in equilibrium rather than a Bose-Einstein distribution. As noted in \cite{Gondolo:1990dk}, this approximation is very good when $x \gsim 3-4$, which shall hold for our analysis. The lower bound on the integration variable, $\varepsilon_{\textrm{min}}$, differs for each process based on kinematics -- each process can only take place provided $s$ is large enough to produce the final state particles. Specifically, for the annihilation of a $\phi$ and a $\phi^*$ into two final-state particles $A$ and $B$, we have
\begin{align}\label{eq:GondoloThermalAverage}
    \varepsilon_{\textrm{min}} = \textrm{max}\bigg( 0, \frac{(m_A + m_B)^2}{4 m_\phi^2} - 1\bigg).
\end{align}
For $\phi^* \phi \rightarrow \overline{f} f$, $\varepsilon_{\textrm{min}}$ is 0 (at least for kinematically accessible annihilations to SM particles, which will dominate over those which are kinematically forbidden), while the other processes we consider here will have positive $\varepsilon_{\textrm{min}}$, the precise values of which depend on the relative masses of the $\phi$, $A_D$, and $h_D$ particles. The positive values of $\varepsilon_{\textrm{min}}$ in turn provide for the exponential suppression of the kinematically forbidden processes (that is, $\phi^* \phi \rightarrow A_D A_D$, $\phi^* \phi \rightarrow h_D A_D$, and $\phi^* \phi \rightarrow h_D h_D$). This is most easily seen in the non-relativistic limit, in which case Eq.(\ref{eq:GondoloThermalAverage}) becomes
\begin{align}\label{eq:GondoloNR}
    \langle \sigma v \rangle_{\textrm{NR}} \approx 2 \sqrt{\frac{x^3}{\pi}} \int_{\varepsilon_{\textrm{min}}}^\infty d \varepsilon \; \varepsilon^{1/2} e^{-x \varepsilon} \sigma v_{\textrm{lab}}.
\end{align}
Since the higher-$\varepsilon$ collisions are suppressed by exponential Boltzmann factors, the cutoff at $\varepsilon_{\textrm{min}}$ will exponentially suppress, but not eliminate, the kinematically forbidden annihiliation cross sections, in agreement with \cite{Griest:1990kh,DAgnolo:2015ujb,Cline:2017tka,Fitzpatrick:2020vba}. The exact degree of the exponential suppression is of course difficult to surmise from the integral form of Eq.(\ref{eq:GondoloNR}), but can be easily approximated using the principle of detailed balance: In terms of the thermal average for the reverse process (which is of course kinematically allowed) at equilibrium, $A B \rightarrow \phi^* \phi$, the thermal average for the forbidden process $\phi^* \phi \rightarrow A B$ must follow
\begin{align}
    \langle \sigma v \rangle_{\phi^* \phi \rightarrow A B} = \frac{n_{A,0} n_{B,0}}{n_{\phi,0}^2} \langle \sigma v \rangle_{A B \rightarrow \phi^* \phi}, 
\end{align}
where $n_{A,0}$, $n_{B,0}$, and $n_{\phi,0}$ are the number densities of particles $A$, $B$, and $\phi$ at thermal equilibrium, respectively. In the non-relativistic approximation, this indicates that the Boltzmann suppression of the forbidden cross section should be (up to non-exponential terms) $\sim \exp [-(m_{A}+m_{B}-2 m_\phi)/T]$, the well-known result in, \eg, \cite{DAgnolo:2015ujb,Cline:2017tka,Fitzpatrick:2020vba}. While the non-relativistic approximations discussed here are useful for building an intuition for the exponential suppression of the kinematically forbidden annihilation processes, for the remainder of this work and for all numerical calculations we shall use the relativistic expression of Eq.(\ref{eq:GondoloThermalAverage}).

To find the thermally averaged cross sections, then, all we require are the $\sigma v_{\textrm{lab}}$ expressions for each of the four processes included in our calculation. To start, we have the WIMP-like annihilation cross section into SM fermions, given as a function of $m_{A_D}$ and the kinematic variable $\varepsilon$ as
\begin{align}\label{eq:WIMPxsection}
    (\sigma v_{\textrm{lab}})_{\overline{f} f} = \frac{\alpha_{\textrm{em}} g_D^2 \epsilon^2 s (s-4 m_\phi^2)}{3(s-2 m_\phi^2)((s- m_{A_D}^2)+ m_{A_D}^2 \Gamma_{A_D}^2)} \bigg( 1+ O(m_f^2/m_{A_D}^2) \bigg).
\end{align}
Here, $\Gamma_{A_D}$ is the decay width of the dark photon, while $s$ is the familiar Mandelstam variable.
In Eq.(\ref{eq:WIMPxsection}), we have explicitly worked in the limit where the final state SM fermions are far less massive than the dark photon (or the DM); for the mass ranges we are considering, this works to excellent approximation when the SM fermion $f$ is assumed to be an electron, since $m_e \sim 500 \; \textrm{keV} \ll m_{A_D} \sim 0.1-1 \; \textrm{GeV}$. For simplicity, we have taken the liberty of assuming that this annihilation process will \emph{only} leave electron final states in our numerical work, even in the cases in which other final states, such as muons or hadronic final states, might be kinematically accessible. In practice this should have a negligible effect on our numerical results, since in this study we have explicitly focused on regions of parameter space in which the kinematically forbidden processes dominate over this WIMP-like process at freeze-out: We shall find that the only regions of parameter space that we study in which the WIMP-like cross section is of comparable magnitude to the kinematically forbidden ones will occur for DM masses that are light enough that only the $e^+ e^-$ final state is kinematically accessible for these WIMP-like annihilation processes.

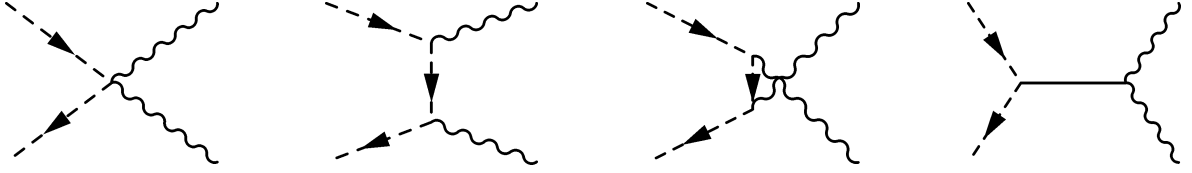
\begin{figure}
  \begin{subfigure}[b]{0.2\textwidth}
    \centering
    \fmfframe(10,25)(10,25){
    \begin{fmfgraph*}(100,60)
        \fmfleft{i1,i2}
        \fmfright{o1,o2}
        \fmf{scalar}{i2,v1,i1}
        \fmf{boson}{o2,v1,o1}
        \fmflabel{$\phi$}{i2}
        \fmflabel{$\phi^*$}{i1}
        \fmflabel{$A_D$}{o2}
        \fmflabel{$A_D$}{o1}
    \end{fmfgraph*}
    }
  \end{subfigure}
  \hfill
  \begin{subfigure}[b]{0.2\textwidth}
    \centering
    \fmfframe(10,25)(10,25){
    \begin{fmfgraph*}(100,60)
        \fmfleft{i1,i2}
        \fmfright{o1,o2}
        \fmf{scalar}{i2,v2,v1,i1}
        \fmf{boson}{o1,v1}
        \fmf{boson}{o2,v2}
        \fmflabel{$\phi$}{i2}
        \fmflabel{$\phi^*$}{i1}
        \fmflabel{$A_D$}{o2}
        \fmflabel{$A_D$}{o1}
    \end{fmfgraph*}
    }
  \end{subfigure}
  \hfill
  \begin{subfigure}[b]{0.2\textwidth}
    \centering
    \fmfframe(10,25)(10,25){
    \begin{fmfgraph*}(100,60)
        \fmfleft{i1,i2}
        \fmfright{o1,o2}
        \fmf{scalar}{i2,v2,v1,i1}
        \fmf{phantom}{o1,v1}
        \fmf{phantom}{o2,v2}
        \fmf{phantom}{v1,i1}
        \fmf{phantom}{i2,v2}
        \fmf{boson}{o2,v1}
        \fmf{boson}{o1,v2}
        \fmflabel{$\phi$}{i2}
        \fmflabel{$\phi^*$}{i1}
        \fmflabel{$A_D$}{o1}
        \fmflabel{$A_D$}{o2}
    \end{fmfgraph*}
    }
  \end{subfigure}
  \hfill
  \begin{subfigure}[b]{0.2\textwidth}
    \centering
    \fmfframe(10,25)(10,25){
    \begin{fmfgraph*}(100,60)
        \fmfleft{i1,i2}
        \fmfright{o1,o2}
        \fmf{scalar}{i2,v1,i1}
        \fmf{plain,label=$h_D$}{v1,v2}
        \fmf{boson}{o2,v2,o1}
        \fmflabel{$\phi$}{i2}
        \fmflabel{$\phi^*$}{i1}
        \fmflabel{$A_D$}{o2}
        \fmflabel{$A_D$}{o1}
    \end{fmfgraph*}
    }
  \end{subfigure}
    \caption{The Feynman diagrams which contribute to the $\phi^* \phi \rightarrow A_D A_D$ amplitude at tree level. Note the presence of the fourth diagram, which emerges from the $s$-channel exchange of the dark Higgs scalar $h_D$.}
    \label{fig2}
\end{figure}

Next, we consider the cross section for the process $\phi^* \phi \rightarrow A_D A_D$. For the convenience of the reader, we have depicted the Feynman diagrams which contribute to the process at tree level in Figure \ref{fig2}. We arrive at
\begin{align}\label{eq:ForbiddenDarkPhotonxsection}
    (\sigma v_{\textrm{lab}})_{A_D A_D} &= \frac{1}{2}\frac{\sqrt{1-\frac{4m_{A_D}^2}{s}}}{32 \pi (s-2 m_\phi^2)} \int_{-1}^{1} d \cos \theta_{cm} \; |\mathcal{M}|^2_{A_D A_D},
\end{align}
where $\theta_{cm}$ is the azimuthal center-of-mass scattering angle for the process, and $|\mathcal{M}|^2_{A_D A_D}$ is the squared amplitude. The factor of $1/2$ in front of Eq.(\ref{eq:ForbiddenDarkPhotonxsection}) of course accounts for the two identical particles in the final state. For clarity, we have written the expressions for this process (and the remaining DM annihilation processes that we discuss) as integrals over the center-of-mass angle, in which form many of the important characteristics of the amplitudes become apparent and most easily legible. For our numerical work, we of course evaluate these integrals symbolically in order to produce the total cross section. For the process $\phi^* \phi \rightarrow A_D A_D$, we find that the squared matrix element $|\mathcal{M}|^2_{A_D A_D}$ is given by
\begin{align}\label{eq:ForbiddenDarkPhotonMatrixElement}
    |\mathcal{M}|^2_{A_D A_D} = 4 g_D^4 &\bigg\{ 2+ \frac{2(m_{A_D}^2-2 m_\phi^2)s}{(t-m_\phi^2)(u-m_\phi^2)} + \frac{(4 m_\phi^2 -m_{A_D}^2)^2(s-2 m_{A_D}^2)^2}{4(u-m_\phi^2)^2(t-m_\phi^2)^2}\nonumber\\
    &+\frac{b m_\phi^2 Q_S^2 (s-m_{h_D}^2)}{(s-m_{h_D}^2)^2+m_{h_D}^2 \Gamma_{h_D}^2}\bigg(8-\frac{(s+2 m_{A_D}^2-8 m_\phi^2)(s-2 m_{A_D}^2)}{(u-m_\phi^2)(t-m_\phi^2)} \bigg) \\
    &+\frac{4 b^2 m_\phi^4 Q_S^4}{(s-m_{h_D}^2)^2+ m_{h_D}^2 \Gamma_{h_D}^2} \bigg( 2+ \frac{(s-2 m_{A_D}^2)^2}{4 m_{A_D}^4}\bigg) \bigg\}, \nonumber
\end{align}
where $\Gamma_{h_D}$ is the decay width of the dark Higgs $h_D$, and $s$, $t$, and $u$ are the usual Mandelstam variables. Notably, the cross section in Eqs.(\ref{eq:ForbiddenDarkPhotonxsection}) and (\ref{eq:ForbiddenDarkPhotonMatrixElement}) contains significant terms that stem from the $s$-channel exchange of a dark Higgs (the fourth diagram in Figure \ref{fig2}), namely those terms in the expression which are proportional to powers of the parameter $b$ which governs the interaction strength between the DM and the dark Higgs in the second and third line of Eq.(\ref{eq:ForbiddenDarkPhotonMatrixElement}). These terms can be subject to a significant resonant enhancement when $s \approx m_{h_D}^2$. In our thermal average, they will therefore be most pronounced when $m_{h_D} \approx 2 m_{A_D}$. As we shall see in Section \ref{Section:Analysis}, these terms can have a substantial numerical effect on the overall cross section when the relative masses of the dark photon and the dark Higgs are near this resonance peak.

\begin{fmffile}{diagram2}
\end{fmffile}

\begin{figure}
  \begin{subfigure}[b]{0.2\textwidth}
    \centering
    \fmfframe(10,25)(10,25){
    \begin{fmfgraph*}(100,60)
        \fmfleft{i1,i2}
        \fmfright{o1,o2}
        \fmf{scalar}{i2,v1,i1}
        \fmf{boson,label=$A_D$}{v1,v2}
        \fmf{plain}{o2,v2}
        \fmf{boson}{o1,v2}
        \fmflabel{$\phi$}{i2}
        \fmflabel{$\phi^*$}{i1}
        \fmflabel{$h_D$}{o2}
        \fmflabel{$A_D$}{o1}
    \end{fmfgraph*}
    }
  \end{subfigure}
  \hfill
  \begin{subfigure}[b]{0.2\textwidth}
    \centering
    \fmfframe(10,25)(10,25){
    \begin{fmfgraph*}(100,60)
        \fmfleft{i1,i2}
        \fmfright{o1,o2}
        \fmf{scalar}{i2,v2,v1,i1}
        \fmf{boson}{o1,v1}
        \fmf{plain}{o2,v2}
        \fmflabel{$\phi$}{i2}
        \fmflabel{$\phi^*$}{i1}
        \fmflabel{$h_D$}{o2}
        \fmflabel{$A_D$}{o1}
    \end{fmfgraph*}
    }
  \end{subfigure}
  \hfill
  \begin{subfigure}[b]{0.2\textwidth}
    \centering
    \fmfframe(10,25)(10,25){
    \begin{fmfgraph*}(100,60)
        \fmfleft{i1,i2}
        \fmfright{o1,o2}
        \fmf{scalar}{i2,v2,v1,i1}
        \fmf{phantom}{o1,v1}
        \fmf{phantom}{o2,v2}
        \fmf{phantom}{v1,i1}
        \fmf{phantom}{i2,v2}
        \fmf{plain}{o2,v1}
        \fmf{boson}{o1,v2}
        \fmflabel{$\phi$}{i2}
        \fmflabel{$\phi^*$}{i1}
        \fmflabel{$A_D$}{o1}
        \fmflabel{$h_D$}{o2}
    \end{fmfgraph*}
    }
  \end{subfigure}
    \caption{The Feynman diagrams which contribute to the $\phi^* \phi \rightarrow h_D A_D$ amplitude at tree level.}
    \label{fig3}
\end{figure}
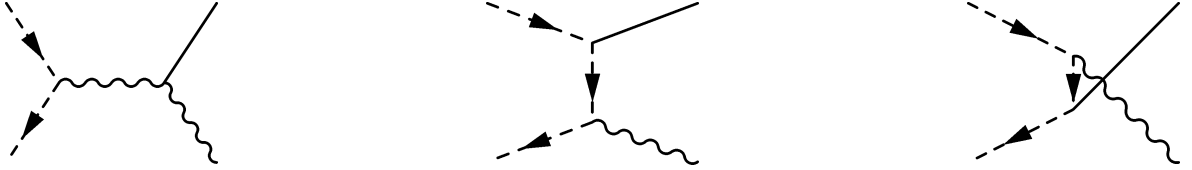

Next, we consider the cross section for the process $\phi^* \phi \rightarrow h_D A_D$. For convenience, we again include the Feynman diagrams which contribute to this process at tree level, this time in Figure \ref{fig3}. We find the lab-frame cross section for this process is given by
\begin{align}\label{eq:ForbiddenMixedxsection}
    (\sigma v_{\textrm{lab}})_{h_D A_D} = \frac{\sqrt{1- \frac{(m_{A_D}+m_{h_D})^2}{s}} \sqrt{1 - \frac{(m_{A_D}-m_{h_D})^2}{s}}}{32 \pi (s-2 m_\phi^2)} \int_{-1}^{1} d \cos \theta_{cm} \; |\mathcal{M}|^2_{A_D h_D},
\end{align}
where the squared amplitude $|\mathcal{M}|^2_{A_D h_D}$ is given by
\begin{align}\label{eq:ForbiddenMixedMatrixElement}
    |\mathcal{M}|^2_{A_D h_D} = 4 Q_S^2 g_D^4 m_{A_D}^2 &\bigg\{ \frac{s-4 m_\phi^2+ \frac{1}{4 m_{A_D}^2}(t-u)^2}{(s-m_{A_D}^2)^2} \nonumber\\
    &- \frac{2 b m_\phi^2}{m_{A_D}^2(s-m_{A_D}^2)} \bigg( \frac{(s-4 m_\phi^2)(s-m_{A_D}^2-m_{h_D}^2)-\frac{1}{2}(t-u)^2}{(t-m_\phi^2)(u-m_\phi^2)}\bigg)\\
    &+\frac{b^2 m_\phi^4}{m_{A_D}^4 (t-m_\phi^2)(u-m_\phi^2)} \bigg( 4(s-m_{A_D}^2) -\frac{(4m_\phi^2-m_{A_D}^2)(s-m_{A_D}^2-m_{h_D}^2)}{(t-m_\phi^2) (u-m_\phi^2)} \bigg) \bigg\}. \nonumber
\end{align}
Unlike the $\phi^* \phi \rightarrow A_D A_D$ process, the cross section in Eqs.(\ref{eq:ForbiddenMixedxsection}) and(\ref{eq:ForbiddenMixedMatrixElement}) does not feature any terms which can enjoy a resonant enhancement.

\begin{fmffile}{diagram3}
\end{fmffile}

\begin{figure}
  \begin{subfigure}[b]{0.2\textwidth}
    \centering
    \fmfframe(10,25)(10,25){
    \begin{fmfgraph*}(100,60)
        \fmfleft{i1,i2}
        \fmfright{o1,o2}
        \fmf{scalar}{i2,v1,i1}
        \fmf{plain}{o2,v1,o1}
        \fmflabel{$\phi$}{i2}
        \fmflabel{$\phi^*$}{i1}
        \fmflabel{$h_D$}{o2}
        \fmflabel{$h_D$}{o1}
    \end{fmfgraph*}
    }
  \end{subfigure}
  \hfill
  \begin{subfigure}[b]{0.2\textwidth}
    \centering
    \fmfframe(10,25)(10,25){
    \begin{fmfgraph*}(100,60)
        \fmfleft{i1,i2}
        \fmfright{o1,o2}
        \fmf{scalar}{i2,v2,v1,i1}
        \fmf{plain}{o1,v1}
        \fmf{plain}{o2,v2}
        \fmflabel{$\phi$}{i2}
        \fmflabel{$\phi^*$}{i1}
        \fmflabel{$h_D$}{o2}
        \fmflabel{$h_D$}{o1}
    \end{fmfgraph*}
    }
  \end{subfigure}
  \hfill
  \begin{subfigure}[b]{0.2\textwidth}
    \centering
    \fmfframe(10,25)(10,25){
    \begin{fmfgraph*}(100,60)
        \fmfleft{i1,i2}
        \fmfright{o1,o2}
        \fmf{scalar}{i2,v2,v1,i1}
        \fmf{phantom}{o1,v1}
        \fmf{phantom}{o2,v2}
        \fmf{phantom}{v1,i1}
        \fmf{phantom}{i2,v2}
        \fmf{plain}{o2,v1}
        \fmf{plain}{o1,v2}
        \fmflabel{$\phi$}{i2}
        \fmflabel{$\phi^*$}{i1}
        \fmflabel{$h_D$}{o1}
        \fmflabel{$h_D$}{o2}
    \end{fmfgraph*}
    }
  \end{subfigure}
  \hfill
  \begin{subfigure}[b]{0.2\textwidth}
    \centering
    \fmfframe(10,25)(10,25){
    \begin{fmfgraph*}(100,60)
        \fmfleft{i1,i2}
        \fmfright{o1,o2}
        \fmf{scalar}{i2,v1,i1}
        \fmf{plain,label=$h_D$}{v1,v2}
        \fmf{plain}{o2,v2,o1}
        \fmflabel{$\phi$}{i2}
        \fmflabel{$\phi^*$}{i1}
        \fmflabel{$h_D$}{o2}
        \fmflabel{$h_D$}{o1}
    \end{fmfgraph*}
    }
  \end{subfigure}
    \caption{The Feynman diagrams which contribute to the $\phi^* \phi \rightarrow h_D h_D$ amplitude at tree level.}
    \label{fig4}
\end{figure}
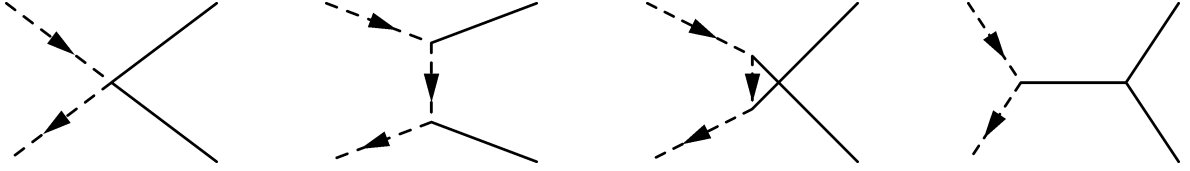

Finally, we consider the cross section for the process $\phi^* \phi \rightarrow h_D h_D$, where the leading Feynman diagrams contributing to this process are pictured in Figure \ref{fig4}. We arrive at the cross section
\begin{align}\label{eq:ForbiddenHiggsxsection}
    (\sigma v_{\textrm{lab}})_{h_D h_D} &= \frac{\sqrt{1-\frac{4m_{h_D}^2}{s}}}{64 \pi (s-2 m_\phi^2)} \int_{-1}^{1} d \cos \theta_{cm} \; |\mathcal{M}|^2_{h_D h_D},
\end{align}
where
\begin{align}
    |\mathcal{M}|^2_{h_D h_D} = \frac{4 b^2 m_\phi^4 Q_S^4 g_D^4}{m_{A_D}^4} &\bigg\{ 1- \frac{6 m_{h_D}^2}{(s-m_{h_D}^2)}+\frac{9 m_{h_D}^4}{(s-m_{h_D}^2)^2} \\
    &- \frac{4 b m_\phi^2 (s-2 m_{h_D}^2) (s-4 m_{h_D}^2)}{(t-m_\phi^2)(u-m_\phi^2)(s - m_{h_D}^2)} + \frac{4 b^2 m_\phi^4 (s- 2 m_{h_D}^2)^2}{(t-m_\phi^2)^2 (u-m_\phi^2)^2} \bigg\}. \nonumber
\end{align}

\begin{fmffile}{diagram4}
\end{fmffile} 

Armed with these expressions, the thermally averaged cross sections for each process can be straightforwardly computed using Eq.(\ref{eq:GondoloThermalAverage}), where each integral over $\varepsilon$ can be done numerically. With the thermally averaged cross sections, the differential equation of Eq.(\ref{eq:YBoltzmann}) can be also then performed numerically, allowing us to compute relic abundance in the usual manner.

\section{Analysis}\label{Section:Analysis}
Using the procedure outlined in Secion \ref{Section:RelicDensity}, we can perform a numerical study of the parameter space of this model to understand how the presence of the dark scalar $h_D$ affects the relic density of the dark matter (DM). From the annihilation cross sections Section \ref{Section:RelicDensity}, we see that the relic density of the DM in our setup depends on the following parameters: The mass of the dark photon $m_{A_D}$, the DM-dark Higgs coupling parameter $b$, the dark gauge coupling $g_D$, and the mass ratios $r_\phi \equiv m_{A_D}/m_\phi$ (the ratio of the dark photon mass to that of the DM) and $r_h \equiv m_{h_D}/m_{A_D}$ (the ratio of the dark Higgs mass to that of the dark photon). There are two additional parameters, $Q_S$ (the $U(1)_D$ charge of the scalar $S$ that contains the dark Higgs) and $\epsilon$ (the kinetic mixing parameter) that we keep fixed in our analysis below, because their qualitative effect on our results is extremely limited. Inspection of the annihilation cross sections in Section \ref{Section:RelicDensity} shows that changing $Q_S$ by an $O(1)$ factor will have much the same effect as varying $b$, especially in the dominant $\phi^* \phi \rightarrow A_D A_D$ cross section given in Eqs.(\ref{eq:ForbiddenDarkPhotonxsection}) and (\ref{eq:ForbiddenDarkPhotonMatrixElement}), in which each factor of $b$ is always accompanied by a factor of $Q_S^2$. Similarly, the kinetic mixing factor $\epsilon$ will only have a significant effect on the subdominant $\phi^* \phi \rightarrow \overline{f} f$ cross section, provided that the kinetic mixing is still large enough to ensure that the dark photon and dark Higgs particles remain in thermal equilibrium throughout DM freeze-out, which we have argued is true for the $\epsilon \sim 10^{-(3-4)}$ case we consider here. For definiteness, we specify $Q_S = 1$ (that is, the scalar $S$ has the same $U(1)_D$ charge as the DM field $\phi$) and $\epsilon = 3 \times 10^{-4}$, so that for a $g_D \sim 0.3$ (that is, close to the electroweak coupling constant $g$), the product $g_D \epsilon \sim 10^{-4}$, which will be roughly consistent with constraints from DM direct detection \cite{Essig:2015cda,Aprile:2019xxb,Agnes:2018oej,Essig:2012yx,DarkSide:2018ppu}. Furthermore, this selection for the value of $\epsilon$ satisfies current upper limits from searches for visible dark photon decays from collider and fixed-target experiments in the mass range of $m_{A_D} \sim 100-200 \; \textrm{MeV}$ \cite{Fabbrichesi:2020wbt,Merkel:2014avp,NA482:2015wmo,LHCb:2019vmc}, the range in which the majority of our numerical work shall take place, while remaining within an inconsequential (at least insofar as the relic abundance calculation is concerned) $O(1)$ factor of the dominant LHCb upper limit \cite{LHCb:2019vmc} for the cases we consider with larger $m_{A_D}$.

\begin{figure}[htbp]
\centerline{\includegraphics[width=3.5in]{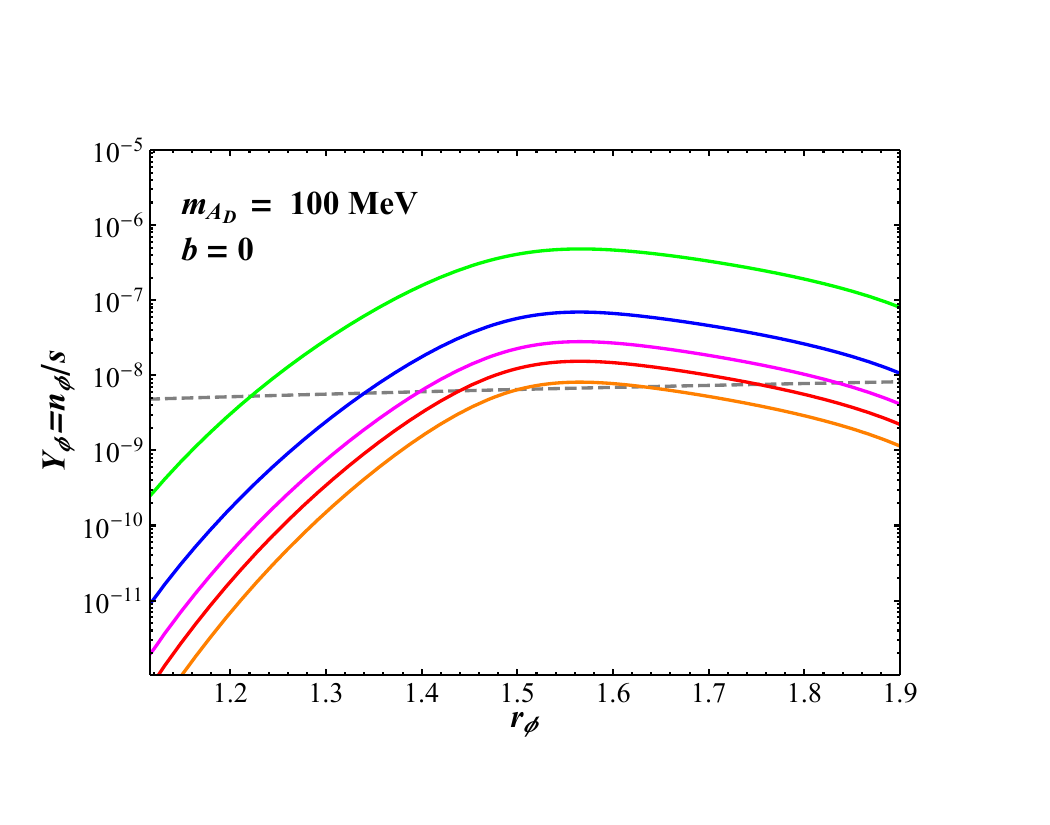}
\hspace{-0.75cm}
\includegraphics[width=3.5in]{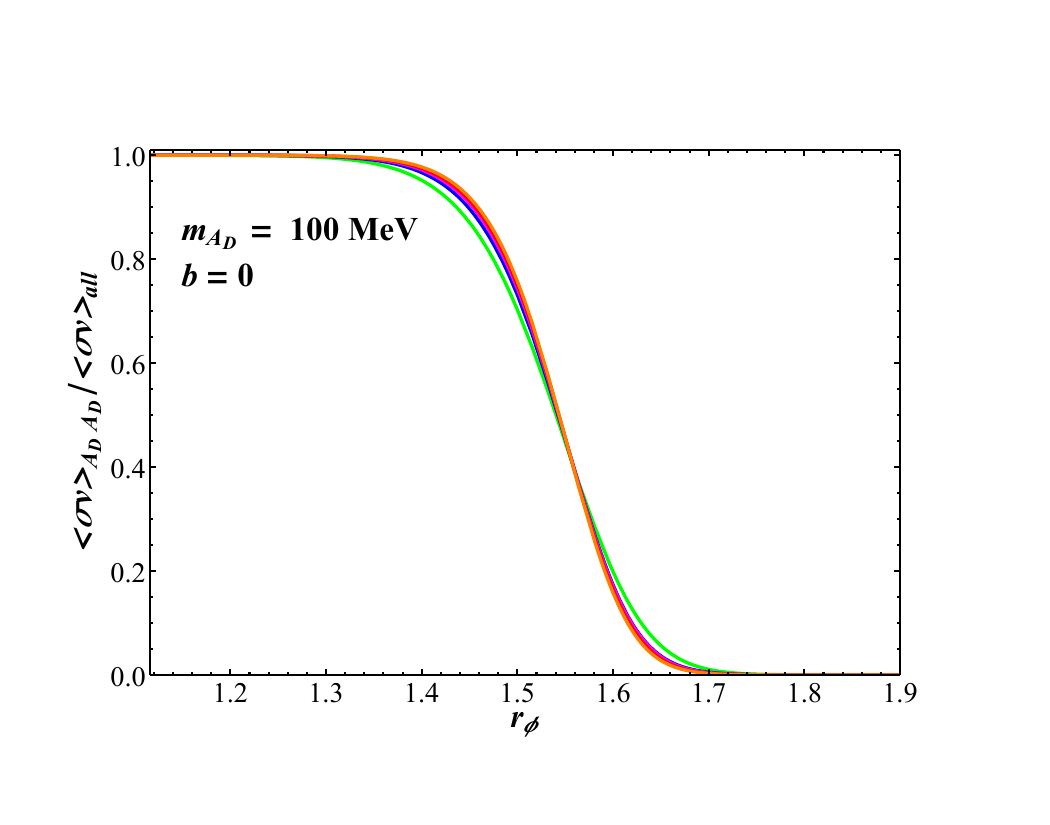}}
\vspace*{-0.75cm}
\centerline{\includegraphics[width=3.5in]{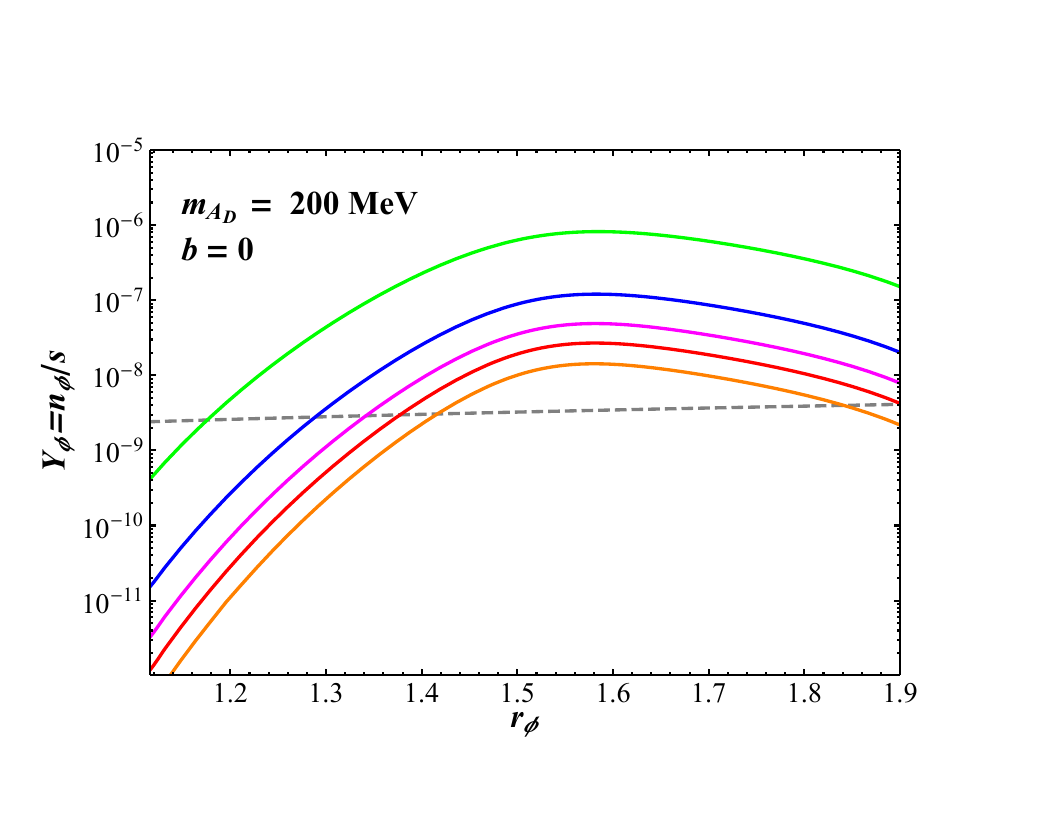}
\hspace{-0.75cm}
\includegraphics[width=3.5in]{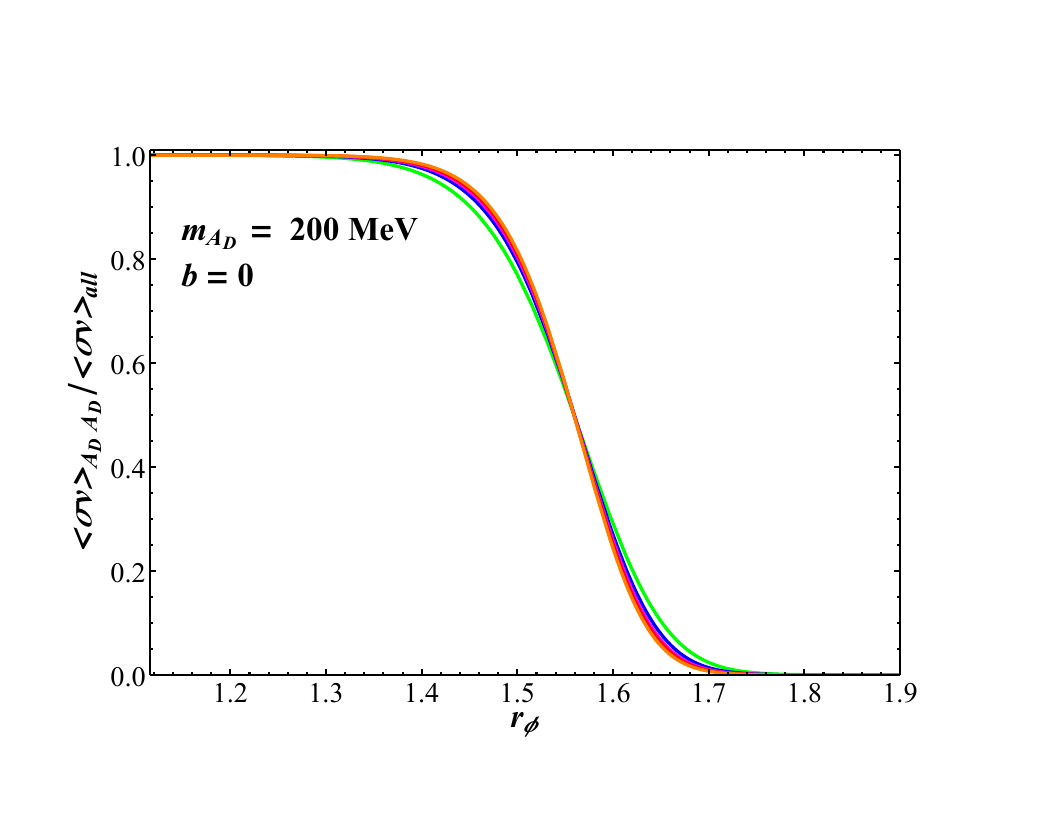}}
\vspace*{-0.75cm}
\centerline{\includegraphics[width=3.5in]{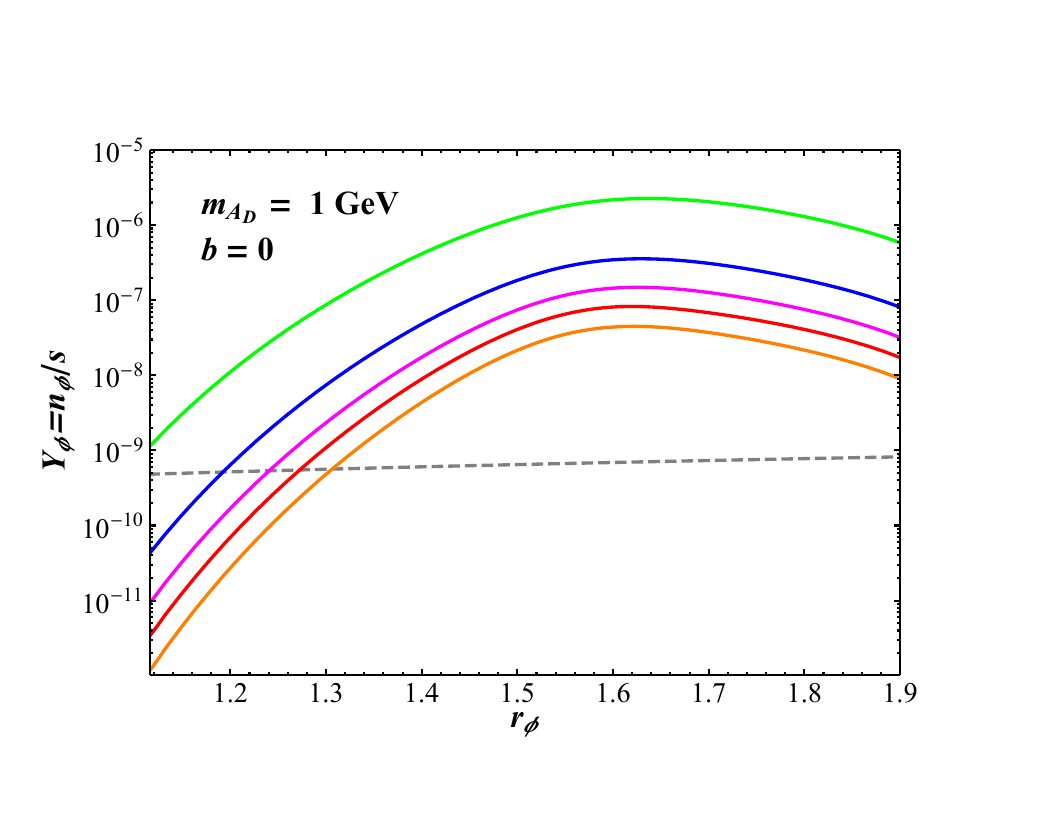}
\hspace{-0.75cm}
\includegraphics[width=3.5in]{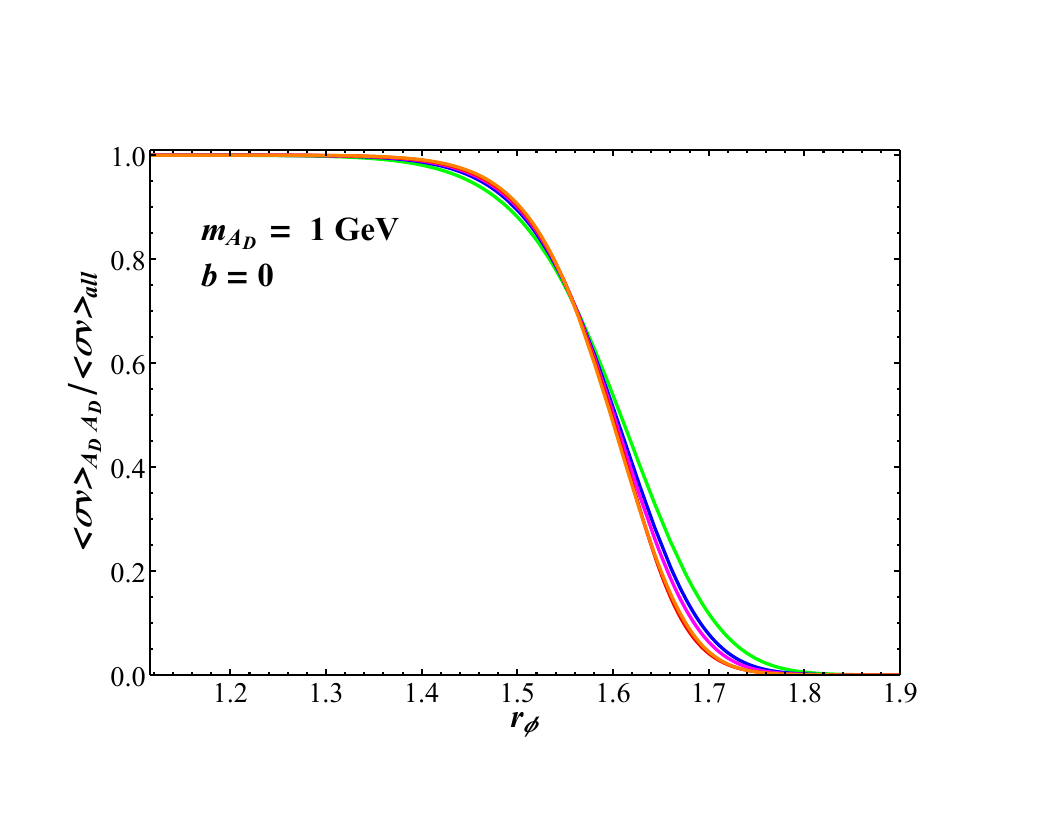}}
\caption{(Left) The DM yield $Y_\phi$ as a function of $r_\phi$ in the absence of any dark Higgs contributions to the DM annihilation cross section ($b=0$) for $m_{A_D} = 100 \; \textrm{MeV}$ (Top), $m_{A_D} = 200 \; \textrm{MeV}$ (Middle), and $m_{A_D} = 1 \; \textrm{GeV}$ (Bottom). Each chart depicts the results for three different selections of $g_D$: $g_D=0.1$ (Green), $g_D=0.3$ (Blue), $g_D = 0.5$ (Magenta), $g_D=0.7$ (Red), and $g_D=1$ (Orange). The $Y_\phi$ value to recreate the observed relic abundance, $Y_\phi = (4.32 \times 10^{-10})\times (\textrm{GeV}/m_\phi)$ \cite{Edsjo:1997bg}, is displayed on each chart as a dashed gray line. (Right) Same as the right, except the fraction of the thermally averaged annihilation cross section that stems from the kinematically forbidden process $\phi^* \phi \rightarrow A_D A_D$ at freeze-out is charted.}
\label{fig5}
\end{figure}

Our task now remains to examine the behavior of the relic density as the parameters $m_{A_D}$, $b$, $g_D$, $r_\phi$, and $r_h$ are modified. Before considering the main topic of our analysis, the effect of the dark Higgs $h_D$ on the relic abundance, it is useful to get a feel for the behavior of the system \emph{without} dark Higgs effects included -- in practice, this simply amounts to setting the parameter $b$ to 0. In this case, the model closely resembles that of \cite{DAgnolo:2015ujb}, albeit with scalar DM in lieu of fermions. In particular, in agreement with \cite{DAgnolo:2015ujb}, we find that the forbidden cross section $\phi^* \phi \rightarrow A_D A_D$ will dominate freeze-out even for $O(1)$ mass splittings between the DM and the dark photon (that is, $r_\phi-1$ that can be a significant fraction of 1), in contrast to the requirement that the mass splitting be minute (that is, $r_\phi-1 \ll 1$) in the case of heavier forbidden DM \cite{Griest:1990kh}. To get a sense of the behavior of the relic abundance in the absence of $h_D$, in Figure \ref{fig5} we depict the final DM yield $Y_\phi$ for several different selections of $g_D$ and $m_{A_D}$ as a funcion of the parameter $r_\phi$. To better clarify the relative contributions of the different annihilation processes to freeze-out, Figure \ref{fig5} also depicts the ratio of the thermally averaged cross sections $\langle \sigma v \rangle_{A_D A_D}/\langle \sigma v \rangle_{all}$ at the freeze-out temperature,\footnote{We follow the definition of a specific temperature for freeze-out used in micrOMEGAs \cite{Belanger:2006is}, namely, we define the freeze-out temperature as the temperature at which the number density per unit comoving entropy $Y_\phi$ is equal to 2.5 times its equilibrium value.} that is, the relative contribution of the thermally averaged annihilation cross section of the kinematically forbidden DM process to the total DM annihilation cross section at freeze-out.

From Figure \ref{fig5}, we can note that, consistent with intuition, DM freeze-out process is divided into two regimes, roughly defined by $r_\phi \lesssim 1.6$ and $r_\phi \gsim 1.6$ regardless of the choices of $m_{A_D}$ and $g_D$ in the range we consider. When $r_\phi \lesssim 1.6$, the dominant process governing freeze-out is $\phi^* \phi \rightarrow A_D A_D$, the kinematically forbidden transition. Since this cross section falls exponentially as $r_\phi$ increases, the DM yield rapidly increases with higher $r_\phi$. In the $r_\phi \gsim 1.6$ regime, the exponential suppression of the kinetically forbidden process results in the WIMP-like annihilation $\phi^* \phi \rightarrow \overline{f} f$ dominating the annihilation cross section at freeze-out. In this regime, the relic abundance \emph{decreases} with increasing $r_\phi$, as the $s$-channel dark photon exchange enjoys a resonant enhancement as $r_\phi$ approaches 2.

\begin{figure}[htbp]
\centerline{\includegraphics[width=3.5in]{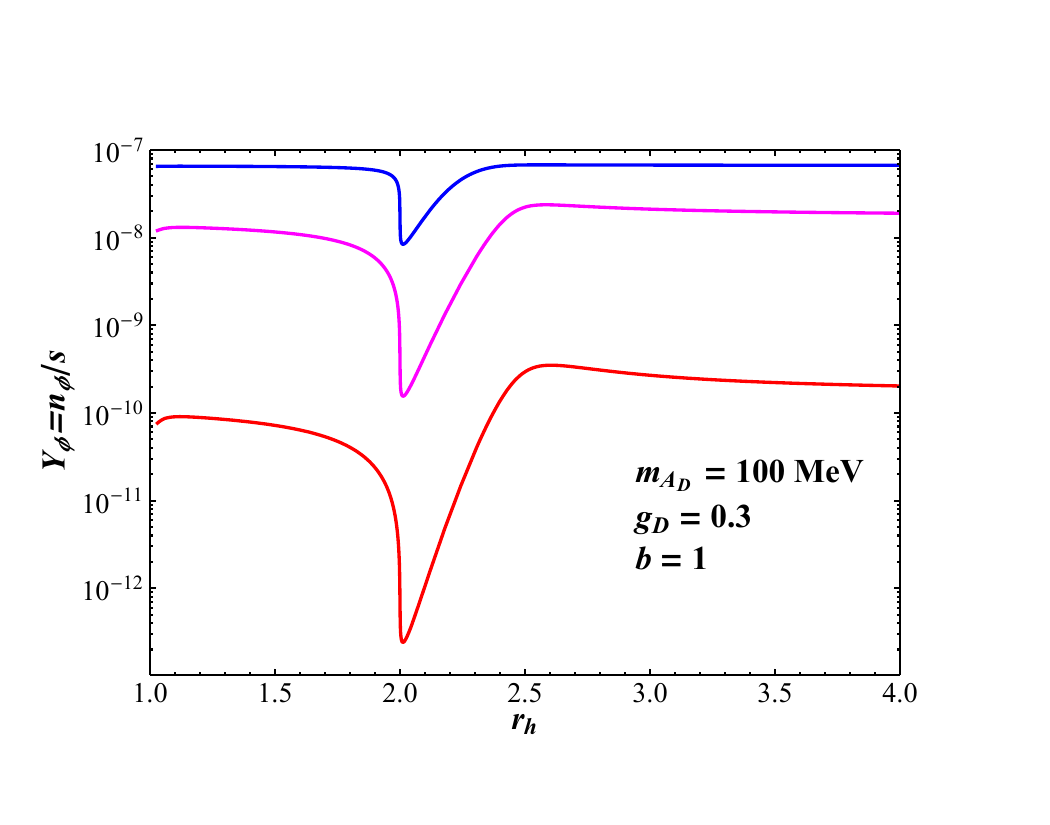}
\hspace{-0.75cm}
\includegraphics[width=3.5in]{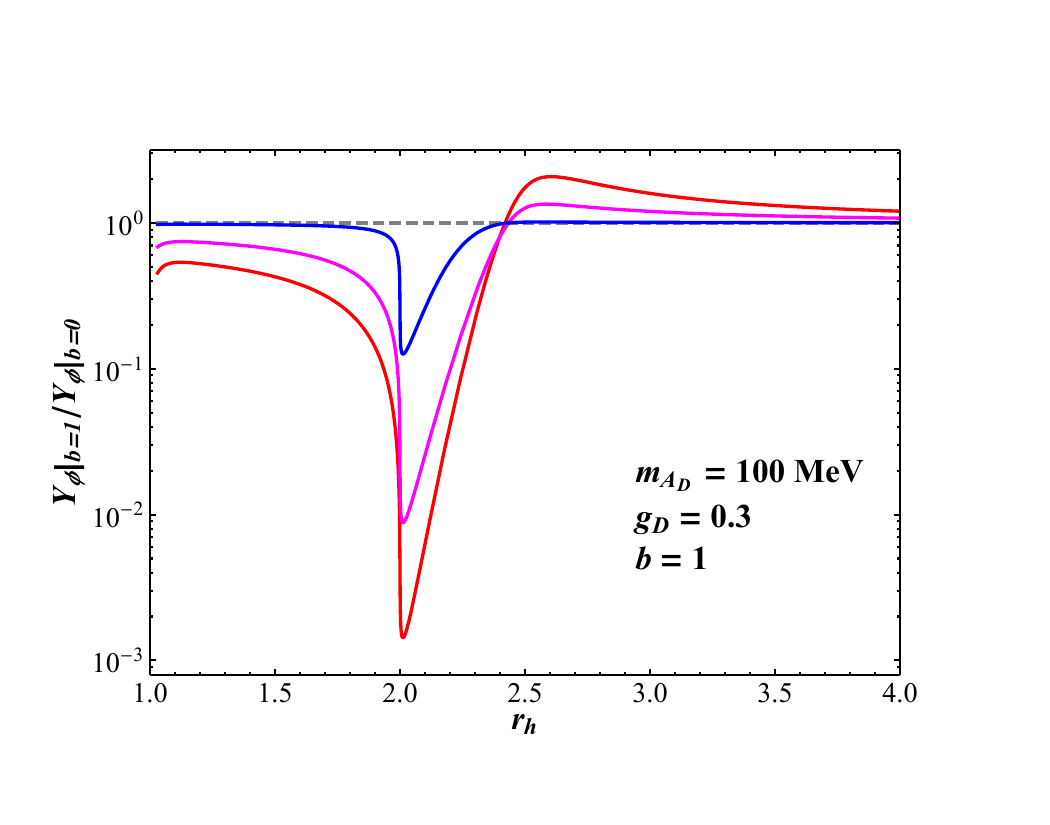}}
\caption{(Left) The DM yield as a function of $r_h=m_{h_D}/m_{A_D}$, assuming a dark photon mass $m_{A_D}=100 \; \textrm{MeV}$, a dark gauge coupling of $g_D = 0.3$, and a dark Higgs-DM coupling parameter $b=1$, as discussed in the text. Selections of $r_\phi=m_{A_D}/m_\phi$ are $r_\phi=1.2$ (Red), $r_\phi=1.4$ (Magenta), and $r_\phi=1.6$ (Blue). (Right) The same as the left, but now each yield is scaled by the value for equivalent parameter choices \emph{without} dark Higgs contributions being included (that is, with $b=0$).}
\label{fig6}
\end{figure}

Having established a baseline understanding of the model parameter space when the dark Higgs is omitted from the model, we can now examine how reintroducing the dark Higgs alters our phenomenology. The dark Higgs $h_D$ has two major effects on freeze-out: First, the $s$-channel exchange of a dark Higgs results in significant additional terms in the cross section $\langle \sigma v \rangle_{A_D A_D}$, the annihilation of DM particles into a pair of dark photons (as seen in the fourth diagram of Figure \ref{fig2}). Second, for choices of $m_{h_D}$ such that the dark Higgs decays dominantly into SM particles (namely, when $m_{h_D}< 2 m_\phi$), the additional kinematically forbidden annihilation processes $\phi^* \phi \rightarrow A_D h_D$ and $\phi^* \phi \rightarrow h_D h_D$ can contribute to freeze-out. Both of these effects impact kinematically forbidden processes, so we can expect that they will only be apparent in the region of parameter space in which these processes dominate freeze-out; from the case without the dark Higgs in Figure \ref{fig5}, this would suggest that $r_\phi$ must be sufficiently far from 2. To get a quantitative sense of this behavior, we depict the DM yield as a function of the ratio $r_h$ (which we remind the reader is the ratio $m_{h_D}/m_{A_D}$) for differing values of $r_\phi$ in Figure \ref{fig6}. Here, in order to maximize the effect of the dark Higgs, we have set the coupling parameter $b$ equal to 1, that is, we have assumed that $100\%$ of the DM particle's mass comes from the vev of the dark Higgs field.

From Figure \ref{fig6}, we can already identify some salient features of the dark Higgs's effects on the relic abundance computation. First, we see that the largest effect clearly stems from the resonance peak of the $s$-channel dark Higgs exchange in $\langle \sigma v \rangle_{A_D A_D}$ at $r_h=2$. Depending heavily on the selection of $r_\phi$, this effect can alter the relic abundance computation by several orders of magnitude. However, as $r_\phi$ increases and the Boltzmann suppression of $\langle \sigma v \rangle_{A_D A_D}$ becomes more severe, the effect of the resonance (and indeed all dark Higgs effects) are seriously diminished: For the parameter selections used in Figure \ref{fig6}, we see that when $r_\phi=1.2$, resonant $s$-channel dark Higgs exchanges can reduce the DM yield by a factor of $10^3$, while for $r_\phi=1.4$ this reduction factor becomes only $10^2$, and at $r_\phi = 1.6$ it becomes merely a factor of 10. We also note that the resonance peak itself is quite broad for $r_h > 2$. For our sample points, for example, we see that $r_h$ can be as large as $\sim 2.25$ and still effect an order-of-magnitude decrease in the resulting yield.

In spite of the overall primacy of the resonance effect in the results of Figure \ref{fig6}, we also can observe some other results of interest before moving on. First, on either side of the sharp resonance peak (so, for $r_h \lesssim 1.5$ and $r_h \gtrsim 2.5$), the dark Higgs contributions can still provide significant $O(1)$ corrections to the projected DM yield; for the $r_\phi=1.2$ case, as much as a $100\%$ correction can be achieved in these regions. Intriguingly, the dark Higgs contribution appears to \emph{negatively} interfere with the DM annihilation cross section for $r_h \gtrsim 2.4$. This negative interference can be intuitively understood by considering the cross section for the process $\phi^* \phi \rightarrow A_D A_D$, given in Eqs.(\ref{eq:ForbiddenDarkPhotonxsection}) and (\ref{eq:ForbiddenDarkPhotonMatrixElement}). Away from resonance, the dominant contribution of this cross section to the thermally averaged annihilation cross section will be near the kinematic threshold for the process, namely $s = 4 m_{A_D}^2$. At this threshold, the term corresponding to interference between the $s$-channel dark Higgs exchange amplitude and the other diagrams in the process (given as the second line in Eq.(\ref{eq:ForbiddenDarkPhotonMatrixElement}) is, up to positive multiplicative factors, proportional to $b (4-r_h^2)$. Given that we have chosen positive $b$ (see Section \ref{Section:ModelSetup} for a discussion of this choice), it is clear that the $\phi^* \phi \rightarrow A_D A_D$ annihilation cross section will suffer some negative interference from the dark Higgs exchange amplitude when $r_h > 2$, consistent with what we observe in Figure \ref{fig6}. If we were to instead consider the scenario in which $b < 0$, we instead would observe \emph{positive} interference from this term for $r_h >2$, that is, the relic abundance would be decreased in this regime relative to the $b=0$ case, while the negative interference would be observed for $r_h <2$. Because this discrepancy represents the sole numerically significant effect of allowing $b<0$, while as mentioned in Section \ref{Section:ModelSetup} moving into this regime creates substantial additional difficulties regarding the stability of the scalar potential, we omit a quantitative discussion of the $b<0$ regime here, contenting ourselves with these qualitative observations.

The final effect of the dark Higgs scalars evinced in Figure \ref{fig6} is extremely slight: Namely, the relic abundance decreases slightly (by an $O(10\%)$ factor) when $r_h$ gets extremely close to 1. This decrease is the result of the contributions of the subdominant annihilation processes $\phi^* \phi \rightarrow A_D h_D$ and $\phi^* \phi \rightarrow h_D h_D$ to freeze-out: For most of our parameter space, these contributions are severely curtailed by their exponential Boltzmann suppression relative to the process $\phi^* \phi \rightarrow A_D A_D$. In principle, therefore, when $r_h = 1$ the cross section of these otherwise-suppressed annihilation processes should be roughly comparable to those of the process $\phi^* \phi \rightarrow A_D A_D$. Since we require $r_h$ to be slightly greater than 1, in order to kinematically permit the decay $h_D \rightarrow A_D A_D^* \rightarrow A_D \overline{f} f$, where the lightest SM fermion $f$ in this process is the electron, the lowest $r_h$ we actually consider in Figure \ref{fig6} actually corresponds to a percent level splitting betweeen $m_{h_D}$ and $m_{A_D}$, that is, $r_h-1 \sim O(10^{-2})$. A finer mass splitting than this would both raise fine-tuning concerns and, for $m_{A_D} = 100 \; \textrm{MeV}$, kinematically disallow the $h_D \rightarrow A_D \overline{f} f$ decay channel. We see that within this finely tuned range of $r_h$, there is an $O(10\%)$ reduction in the relic abundance from annihilations with dark Higgses in the final state, however this contribution vanishes rapidly as $r_h$ increases. We shall discuss the effect of these interactions on the relic abundance more quantitatively later on in this work.

\begin{figure}[htbp]
\centerline{\includegraphics[width=5.0in]{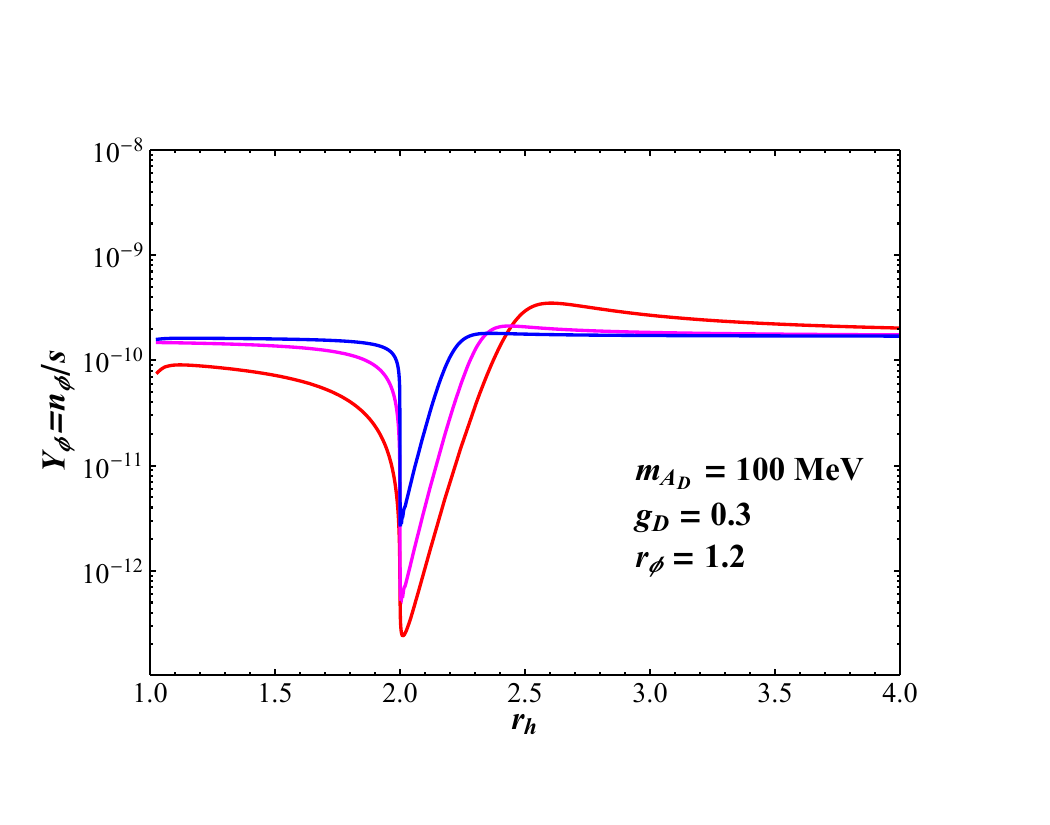}}
\caption{The DM yield $Y_\phi$ for $b=1$ (Red), $b=0.2$ (Magenta), and $b=0.05$ (Blue) as a function of $r_h$, scaled by the yield when the effects of the dark Higgs are omitted ($b=0$). Here, we have chosen $r_\phi=1.2$, $g_D=0.3$, and $m_{A_D}= 100 \; \textrm{MeV}$.}
\label{fig7}
\end{figure}

The resonance peak in Figure \ref{fig6} is quite pronounced, however, we do note that the selection that the parameter $b=1$ in this figure certainly optimizes the peak's contribution: Referencing the squared amplitude in Eq.(\ref{eq:ForbiddenDarkPhotonMatrixElement}), we see that the resonantly enhanced term in the amplitude is proportional to $b^2$. Naively, we might anticipate that a modest reduction in $b$, say, to $b=0.1$, might reduce the resonant cross section at freeze-out by a factor of $10^{-2}$, which would in turn increase the DM yield $Y_\phi$ by a similar factor-- that is, $Y_\phi \propto b^{-2}$. Since $b$ itself can range between 0 and 1 (although assuming an $O(1)$ coupling constant between the DM particle $\phi$ and the scalar $S$ which contains the dark Higgs, we might assume that very small $b$, say $\lesssim O(10^{-2})$, likely requires some fine tuning), this suggests that the effect of the resonance peak is only significant over a narrow range of $b$ values. However, the naive expectation that $Y_\phi \propto b^{-2}$ when the resonant $h_D$ exchange dominates the cross section $\langle \sigma v \rangle_{A_D A_D}$ actually fails here. In Figure \ref{fig7}, we depict the DM yield for a given sample point in model parameter space as a function of $r_h$, for various selections of $b$.

Notably, even when $b=0.05$, 20 times smaller than our $b=1$ benchmark, the resonant peak \emph{still} reduces the DM yield by 2 orders of magnitude. Given that the $b=1$ line in Figure \ref{fig7} only displays a yield reduction of 3 orders of magnitude at the resonance peak, we can see that the effect of resonant Higgs exchange is surprisingly robust against modifications to the $b$ parameter.
In fact, the robustness of the resonance peak against changes in $b$ stems from the exponential dependence of the dominant annihilation cross section $\langle \sigma v \rangle_{A_D A_D}$ on temperature: Intuitively, as $b$ (or any parameter which multiplicatively scales $\langle \sigma v \rangle_{A_D A_D}$) increases, the temperature parameter $x=m_\phi/T$ at which freeze-out occurs increases, which reduces the cross section $\langle \sigma v \rangle_{A_D A_D}$ at freeze-out, compensating for the change in $b$. We can explore this phenomenon for a general rescaling of an exponentially suppressed cross section more quantitatively in the instantaneous-freeze-out approximation of, \eg, \cite{Gondolo:1990dk}. To start, we assume that for some system, the dominant annihilation cross section is given by
\begin{align}\label{eq:YGenericCrossSection}
    \langle \sigma v \rangle = \frac{\alpha f(x) e^{-\beta x}}{m_\phi^2},
\end{align}
where $f(x)$ is some $O(1)$ function, $m_\phi$ is the DM mass, and $\alpha$ and $\beta$ are dimensionless (positive) constants. Following the treatment of \cite{Gondolo:1990dk}, we can recast the freeze-out condition as occurring instantaneously when the DM yield $Y_\phi(x_f)=(1+\delta)Y_{\phi,0}(x_f)$, where $\delta$ is an $O(1)$ number,$Y_{\phi, 0}(x)$ is the DM yield in thermal equilibrium, and $x_f$ is the point at which freeze-out occurs. For $x$ before freeze-out, $d(Y_\phi-Y_{\phi,0})/dx \approx 0$, while for $x$ after freeze-out the $Y_{\phi,0}$ term in the Boltzmann equation Eq.(\ref{eq:YBoltzmann}) becomes numerically insignificant and can be ignored, allowing for the equation to be solved in closed form. In this approximation, freeze-out occurs at the point $x_f$ for which
\begin{align}\label{eq:YGenericFOCondition}
    \langle \sigma v \rangle \approx \sqrt{\frac{45}{\pi}G}\frac{\rho(x_f)}{m_\phi}\frac{1}{Y_{\phi,0}(x_f)} \approx \sqrt{\frac{45}{\pi} G} \frac{\overline{\rho}(x_f)}{m_\phi}e^{x_f},
\end{align}
where $\rho(x)$ is an $O(1)$ dimensionless function of $x$, and in the second expression we have used the non-relativistic approximation for $Y_{\phi,0}$, $Y_{\phi,0} \propto e^{-x}$, while absorbing the non-exponential prefactors in $Y_\phi$ into the new function $\overline{\rho}(x)$. We can insert Eq.(\ref{eq:YGenericCrossSection}) into Eq.(\ref{eq:YGenericFOCondition}) to get a transcendental equation for the freeze-out temperature parameter $x_f$,
\begin{align}\label{eq:FOConditionGeneric}
    e^{(1+\beta)x_f} \approx \bigg(\frac{45}{\pi} G\bigg)^{-1/2}\frac{\alpha f(x_f)}{m_\phi \overline{\rho}(x_f)}.
\end{align}
The DM yield in the present day, $Y_\infty$, can then be approximately found by integrating the Boltzmann equation from freeze-out to the present-day temperature (effectively $x = \infty$) with the assumption that $Y_\phi(x_f) \gg Y_\infty$, which gives us
\begin{align}\label{eq:YInfinityGeneric}
    Y_\infty \approx 2\sqrt{\frac{45}{\pi} G} \bigg/ \bigg( \int_{x_f}^{\infty} \frac{g_*^{1/2} m_\phi \langle \sigma v \rangle_{A_D A_D}}{x^2}\bigg) = 2 \sqrt{\frac{45}{\pi} G} \bigg/ \bigg( \frac{\alpha e^{- \beta x} F(x)}{m_\phi} \bigg\rvert_{x=x_f}^{x=\infty} \bigg),
\end{align}
where in the second equality here we have absorbed all non-exponential dependence of the integral on $x$ into a single function, $F(x)$. Since $F(x)$ is merely polynomial in $x$, the specific value of $F(x)$ will not matter for our purposes. Assuming that the $x=x_f$ limit of the integral will numerically dominate over the $x=\infty$ limit, which given the exponential factor of $e^{-\beta x}$ is reasonable, we can see from Eq.(\ref{eq:YInfinityGeneric}) that
\begin{align}\label{eq:YProptoRelation}
    Y_\infty \propto \frac{e^{\beta x_f}}{\alpha}.
\end{align}
However, from Eq.(\ref{eq:FOConditionGeneric}), we see that the freeze-out temperature $x_f$ follows the relation,
\begin{align}\label{eq:xfProptoRelation}
    e^{\beta x_f} \propto \alpha^{\frac{\beta}{1+\beta}}. 
\end{align}
Inserting Eq.(\ref{eq:xfProptoRelation}) into Eq.(\ref{eq:YProptoRelation}) in turn yields
\begin{align}\label{eq:ExponentialScalingBehavior}
    \langle \sigma v \rangle \propto \alpha e^{- \beta x} \implies Y_\infty \propto \alpha^{-1/(1+\beta)},
\end{align}
that is, for an annihilation cross section which scales linearly with a parameter $\alpha$ and exponentially as $e^{-\beta x}$ with the temperature parameter, the relic abundance should approximately scale as $\alpha^{-1/(1+\beta)}$ instead of the naive expectation, $\alpha^{-1}$. Since $\beta$ is positive, we see that this scaling behavior is always less pronounced than the naive expectation. While this discussion relies heavily on approximation, and is clearly not numerically rigorous, it does provide a useful semi-quantitative framework for understanding the blunted scaling behavior of the relic abundance with various parameters. 

We can apply the arguments above straightforwardly to the relic abundance scaling behavior with the parameter $b$ near resonance. When $r_h \geq 2$, we can use the narrow width approximation to write the thermally averaged cross section $\langle \sigma v \rangle_{A_D A_D}$ as
\begin{align}
    \langle \sigma v \rangle_{A_D A_D}|_{r_h \geq 2} \approx \frac{x K_{1}(r_h r_\phi x)}{K_{2}^2(x)} \frac{b^2 g_D^4}{m_{h_D} \Gamma_h} f(r_\phi, r_h),
\end{align}
where $f(r_\phi, r_h)$ is an $O(1)$ function only of $r_h$ and $r_\phi$. In the limit of large $x$ (roughly, the non-relativisitic limit), the Bessel functions can be approximated by asymptotic forms, yielding
\begin{align}\label{eq:ApproxCrossSection}
    \langle \sigma v \rangle_{A_D A_D}|_{r_h \geq 2} \approx \frac{x^{3/2} e^{-(r_h r_\phi -2) x}(1+ O(x^{-1}))}{\sqrt{r_h r_\phi}}\frac{b^2 g_D^4}{m_{h_D} \Gamma_h} f(r_\phi, r_h).
\end{align}
Following the relation in Eq.(\ref{eq:ExponentialScalingBehavior}), we can see that
\begin{align}
    Y_\phi |_{r_h \geq 2} \propto \bigg( \frac{b^2 g_D^2}{\Gamma_h} \bigg)^{-1/(r_h r_\phi-1)}.
\end{align}
Our final task to determine the scaling of $Y_\phi$ with $b$ in this regime then becomes finding the scaling behavior of $\Gamma_h$, the decay width of the dark Higgs. The two dominant decay channels for the dark Higgs are simply $h_D \rightarrow \phi^* \phi$ and $h_D \rightarrow A_D A_D$. Since the $h_D \rightarrow \phi^* \phi$ partial width scales as $b^2$ while the $h_D \rightarrow A_D A_D$ partial width is independent of $b$, we can suggestively write
\begin{align}
    \Gamma_h = m_h g_D^2 ( b^2 p +q), \;\; p \equiv \frac{\sqrt{1-4 r_h^{-2}}}{128 \pi} (r_h^2+12 r_h^{-2}-4), \;\; q \equiv \frac{\sqrt{1-4 r_h^{-2} r_\phi^{-2}}Q_S^2}{4 \pi r_\phi^4 r_h^2}.
\end{align}
where the $p$ term stems from the partial decay width for the channel $h_D \rightarrow \phi^* \phi$, while the $q$ term emerges from the partial width for $h_D \rightarrow A_D A_D$. Inserting this result into Eq.(\ref{eq:ExponentialScalingBehavior}), we have
\begin{align}\label{eq:YResScalingBehaviorGamma}
    Y_\phi|_{r_h \geq 2} \propto \bigg[ \bigg( \frac{45}{\pi} G\bigg)^{-1/2} \frac{b^2 g_D^2}{m_{h_D} ( p b^2 + q)}\bigg]^{-1/(r_h r_\phi -1)}.
\end{align}
Notably, we see from Eq.(\ref{eq:YResScalingBehaviorGamma}) that the explicit dependence of $Y_\phi$ on $b$ will depend on the relative values of $p$ and $q$. If $p \gg q$ (that is, the decay channel $h_D \rightarrow \phi^* \phi$ is the dominant channel unless $b \ll 1$), then the $b$ dependence in the numerator and denominator of Eq.(\ref{eq:YResScalingBehaviorGamma}) cancel, and $Y_\phi$ actually remains approximately constant in $b$. As $q$ becomes larger, the $b$ dependence of the denominator in Eq.(\ref{eq:YResScalingBehaviorGamma}) becomes less pronounced; the strongest possible dependence of $Y_\phi$ on $b$ will occur when $q \gg p$, at which point $Y_\phi$ will scale as
\begin{align}
    Y_\phi |_{r_h \geq 2} \propto b^{-2/(r_h r_\phi -1)}.
\end{align}
Because $r_h > 2$ and $r_\phi > 1$, we see that the scaling behavior of $Y_\phi$ with $b$ is, as the generic case indicated \emph{always} more mild than our naive estimate, $Y_\phi \propto b^{-2}$. An analogous derivation can be performed in the region in which $r_h < 2$: In this case, because the exact resonance peak $s = m_h^2$ is no longer attainable, the cross section $\langle \sigma v \rangle_{A_D A_D}$ scales as $\exp [-(2 r_\phi -2) x]$, rather than $\exp [-(r_h r_\phi -2) x]$. When $r_h < 2$, then, we derive that the scaling behavior of $Y_\phi$ is at most
\begin{align}
    Y_\phi|_{r_h < 2} \propto b^{-2/(2 r_\phi -1)},
\end{align}
which, because $r_\phi > 1$, also results in a less extreme scaling with $b$ than the naive estimate.

We have seen thus far that the presence of the contributions of the dark Higgs can have a substantial impact on the relic abundance, particularly near the resonance peak at $r_h = 2$. It would be of more phenomenological interest, however, for us to restrict our further examination of the parameter space of the model to points which recreate the observed relic density, which we note occurs when the final yield of the DM achieves the value $Y_\phi \approx (4.32 \times 10^{-10}) (\textrm{GeV}/m_\phi)$. Given a set of parameters $g_D$, $m_{A_D}$, $b$, and $r_h$, it is straightforward to identify a value of $r_\phi$ that recreates the observed DM relic density. Referencing the behavior of the DM yield as a function of $r_\phi$ seen in Figure \ref{fig5}, we note that there can generally be up to two possible $r_\phi$ values which produce the correct yield: A smaller $r_\phi$ for which the kinematically forbidden processes dominate freeze-out, and a larger $r_\phi$ for which the resonantly-enhanced WIMP-like process $\phi^* \phi \rightarrow \overline{f} f$ dominates. As we have discussed before, the effect of the dark Higgs on the WIMP-like annihilation cross section is negligible, so when identifying points of the model parameter space that recreate the observed relic abundance, we shall invariably select the \emph{lowest} $r_\phi$ value that does so, in the event of ambiguity. This should in general restrict us to the more interesting region of parameter space, in which the kinematically forbidden processes control freeze-out.

\begin{figure}[htbp]
\centerline{\includegraphics[width=3.5in]{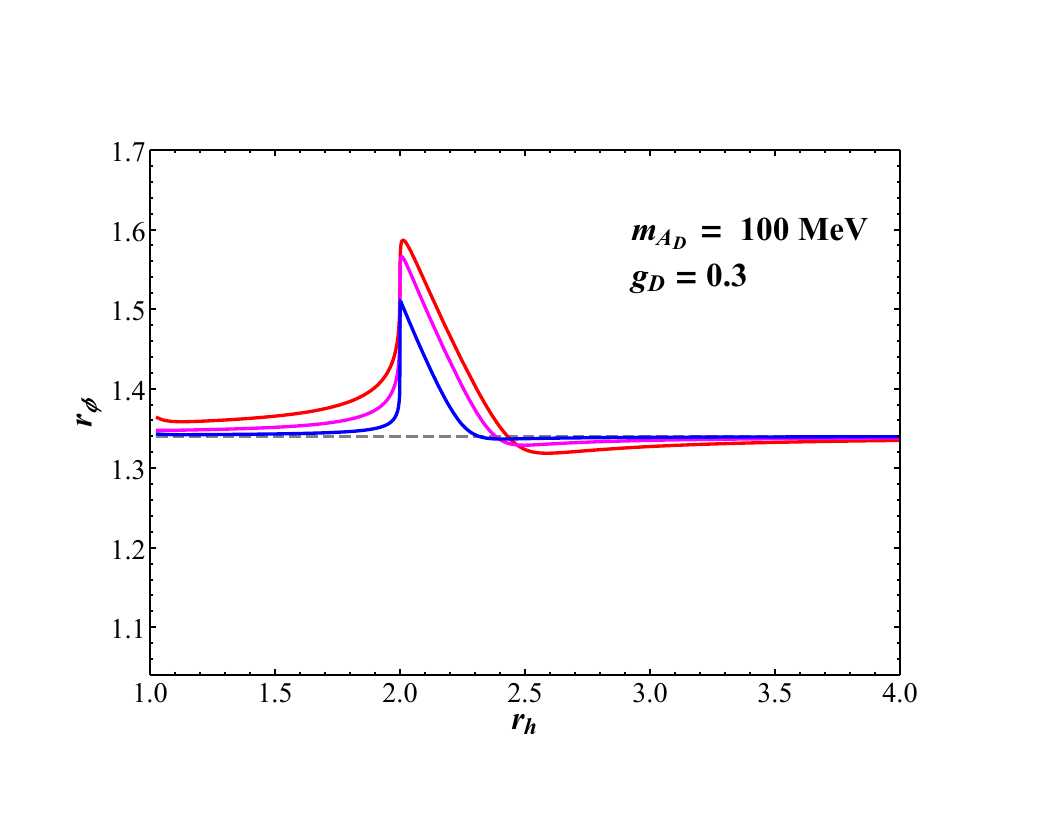}
\hspace{-0.75cm}
\includegraphics[width=3.5in]{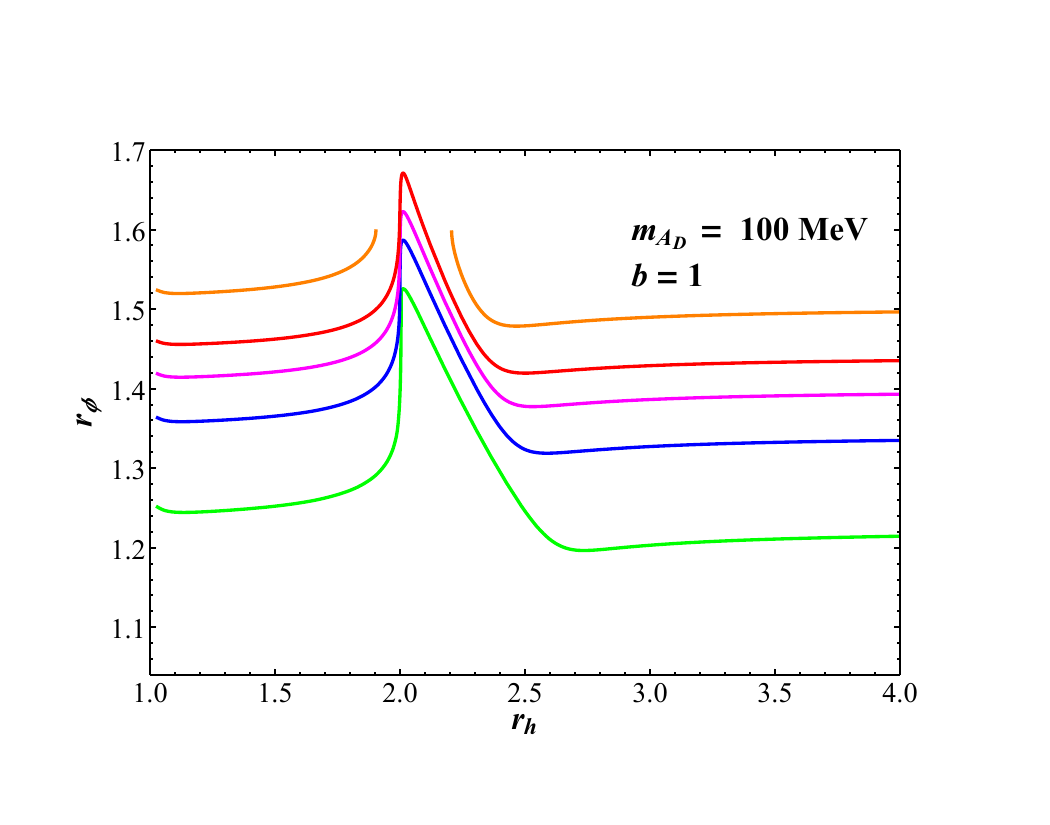}}
\vspace*{-0.75cm}
\centerline{\includegraphics[width=3.5in]{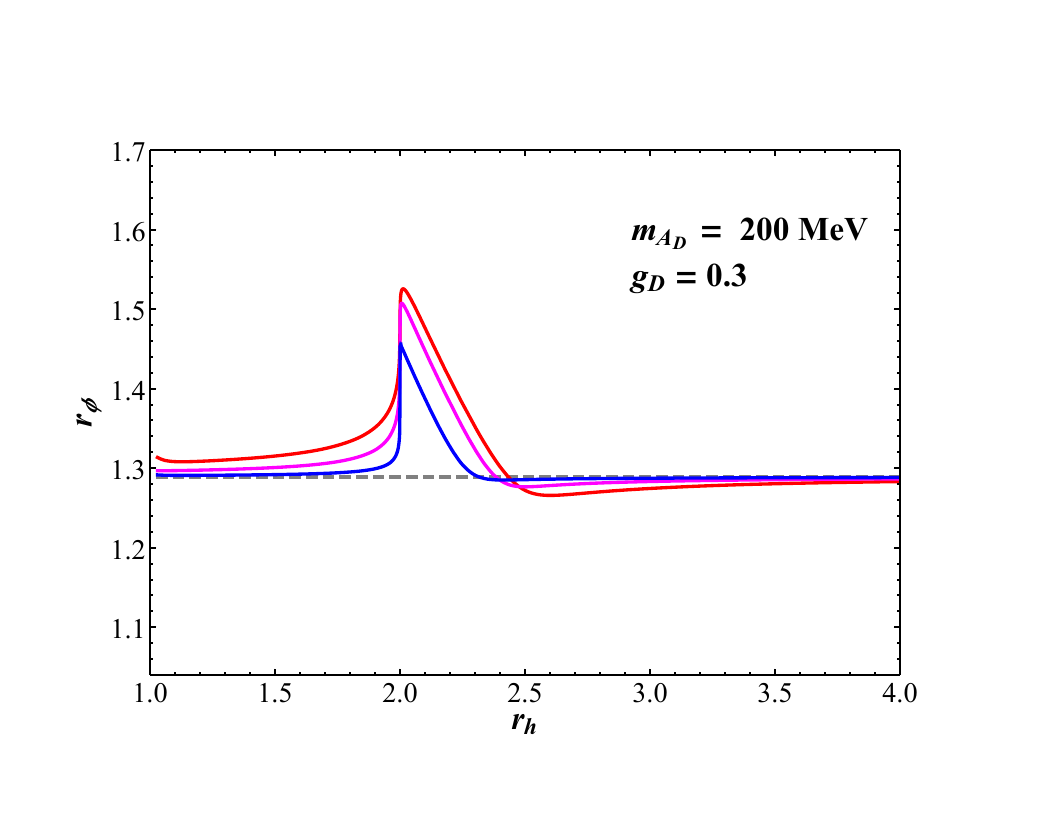}
\hspace{-0.75cm}
\includegraphics[width=3.5in]{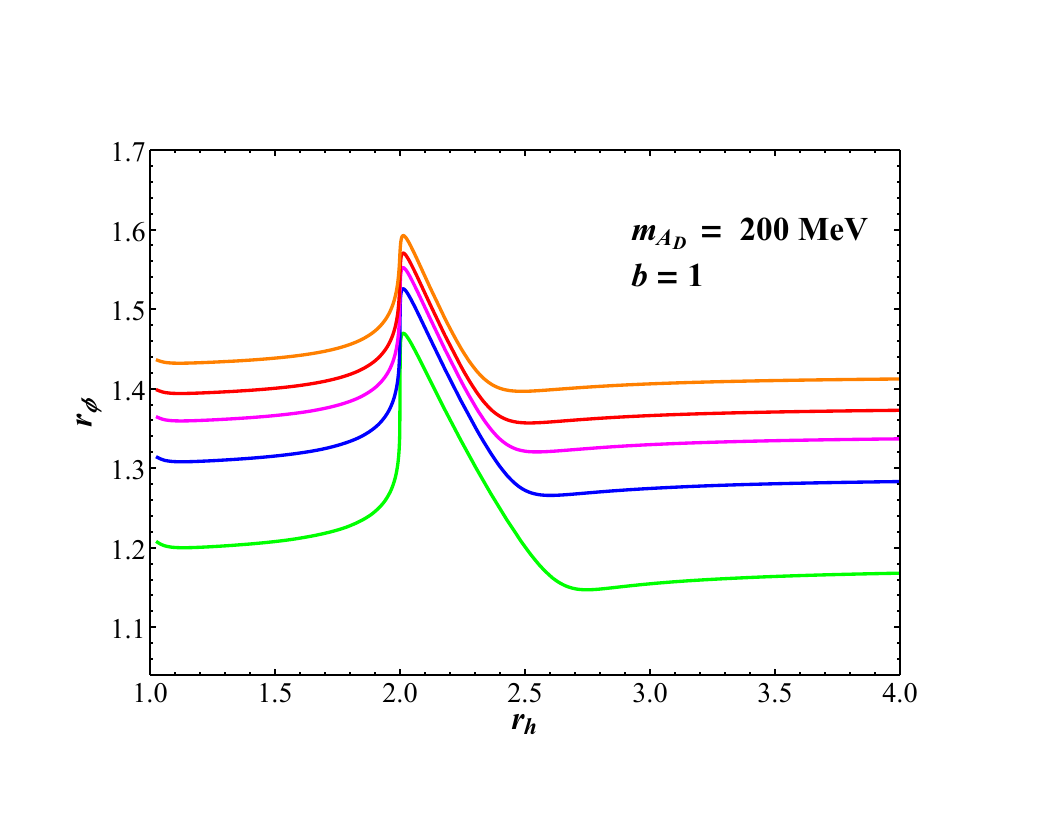}}
\vspace*{-0.75cm}
\centerline{\includegraphics[width=3.5in]{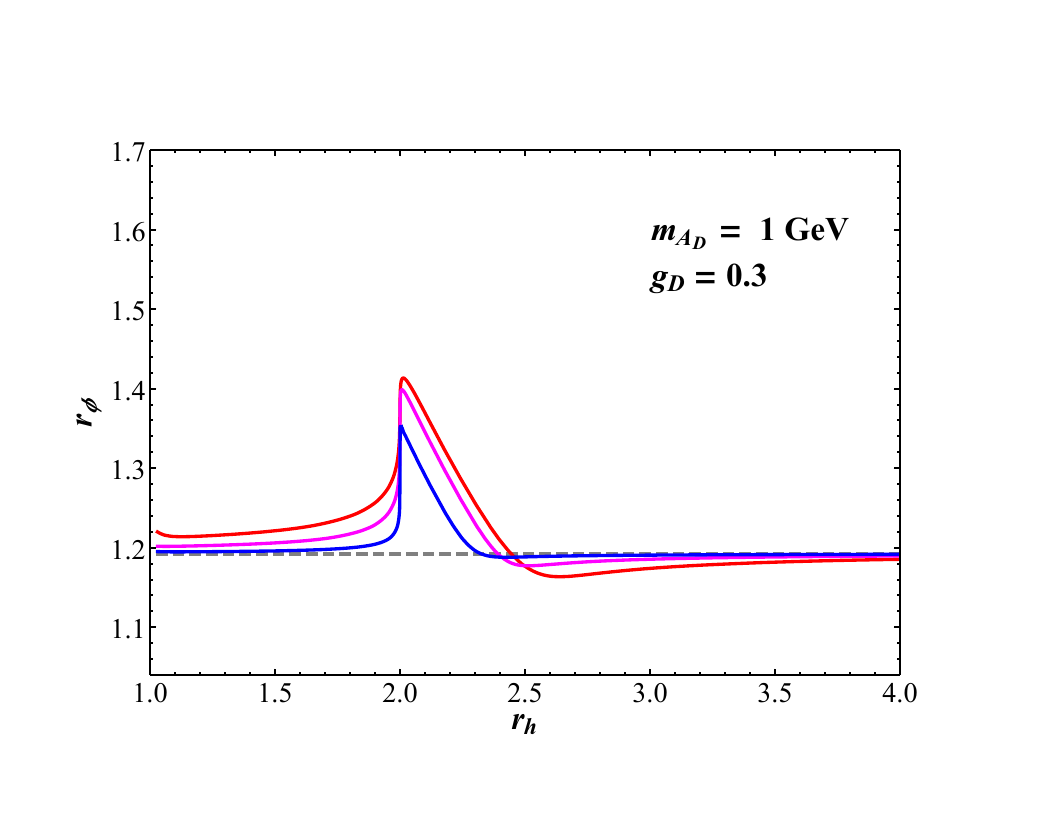}
\hspace{-0.75cm}
\includegraphics[width=3.5in]{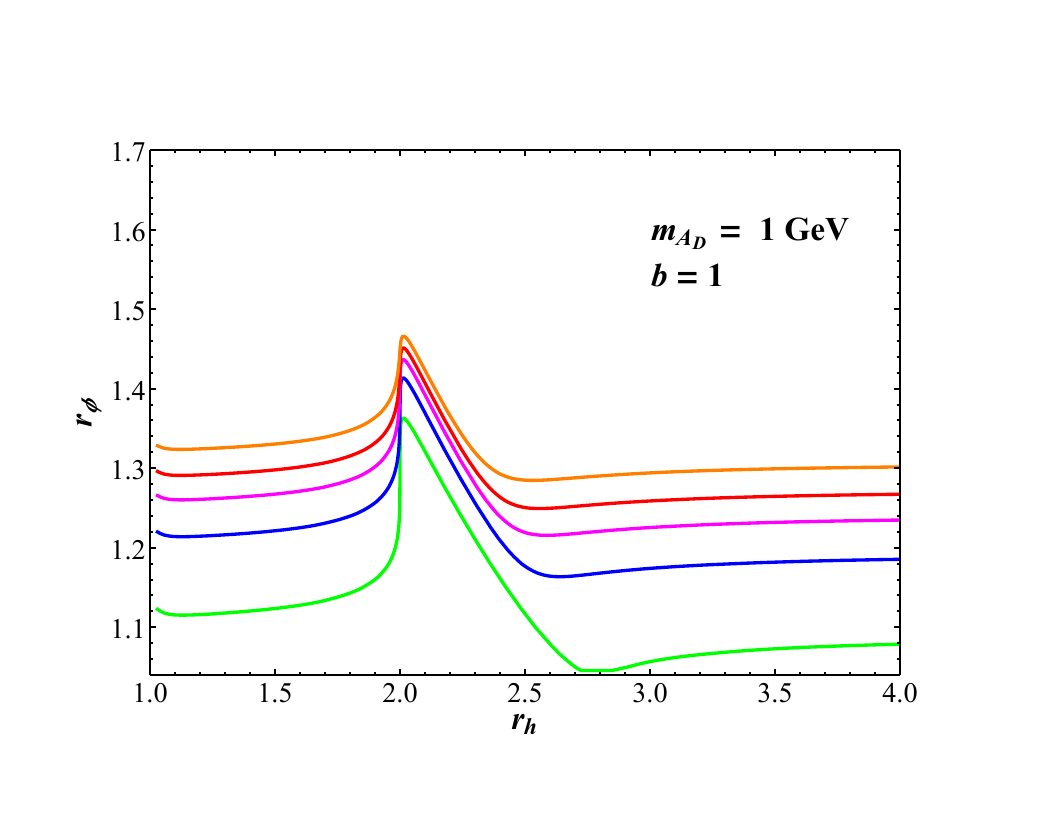}}
\caption{(Left) Contours in the $r_h$-$r_\phi$ plane that recreate the relic density assuming $g_D=0.3$ and various values of $m_{A_D}$. The contours are depicted for $b=1$ (Red), $b=0.4$ (Magenta), $b=0.1$ (Blue), and $b=0$ (Gray, dashed). (Right) Contours in the $r_h$-$r_\phi$ plane that recreate the relic density assuming $b=1$ and various values of $m_{A_D}$. The contours are depicted for $g_D=1$ (Orange), $g_D=0.7$ (Red), $g_D=0.5$ (Magenta), $g_D=0.3$ (Blue), and $g_D=0.1$ (Green). The discontinuity in the $g_D=1$ line in the $m_{A_D}= 100 \; \textrm{MeV}$ chart emerges because, for this dark photon mass and dark coupling, the observed DM relic abundance cannot be recreated with any $1<r_\phi<2$ when $1.9 \lesssim r_h \lesssim 2.2$.}
\label{fig8}
\end{figure}

In Figure \ref{fig8}, we depict contours in the $r_h$-$r_\phi$ plane that recreate the observed relic abundance, for differing values of $m_{A_D}$, $g_D$, and $b$. In these figures, we can see a number of the same characteristics already observed in Figures \ref{fig6} and \ref{fig7}: In particular, the resonance peak from the $s$-channel exchange of a dark Higgs in the cross section $\langle \sigma v \rangle_{A_D A_D}$ is readily apparent in all of the contours here: Near the $r_h=2$ resonance, the contours in the $r_h$-$r_\phi$ plane sharply move upward, indicating that a higher $r_\phi$ (and hence a more severe exponential Boltzmann suppression of the forbidden annihilation cross sections) is required in order to recreate the same relic abundance. By freely adjusting $r_h$, then, a much wider range of parameter space, in particular $r_\phi$ values, will recreate the observed DM relic abundance.

\begin{figure}[htbp]
\centerline{\includegraphics[width=2.5in]{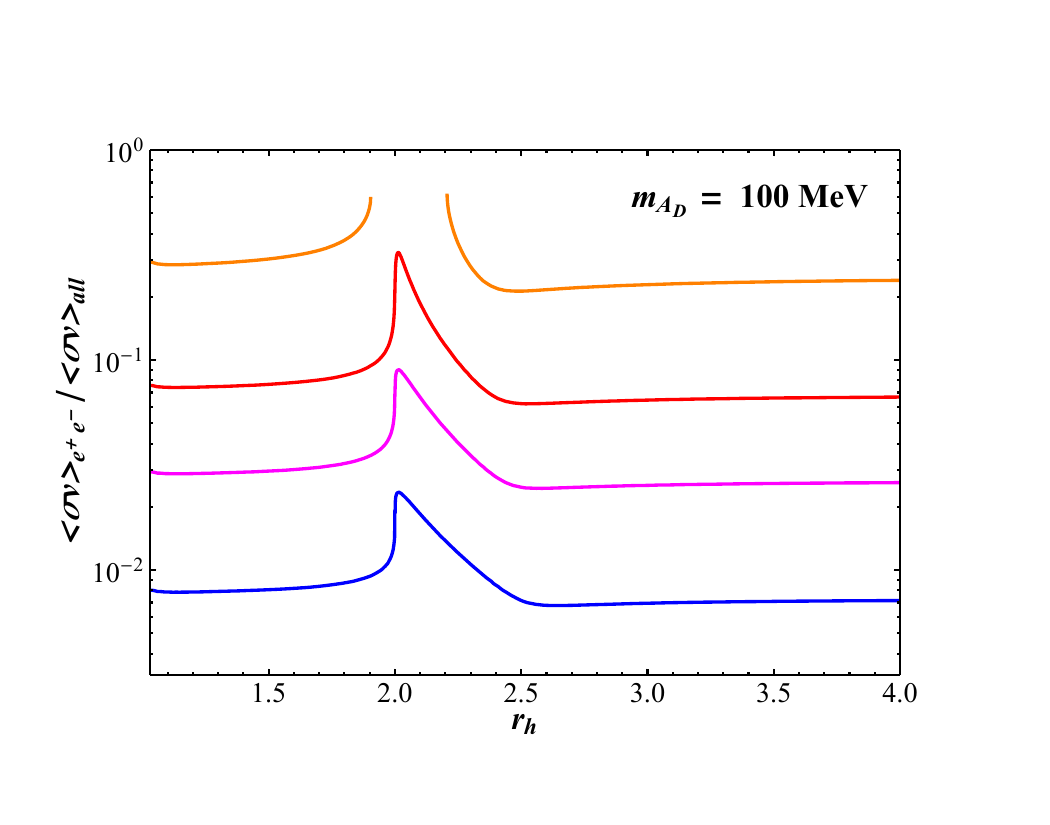}
\hspace{-0.75cm}
\includegraphics[width=2.5in]{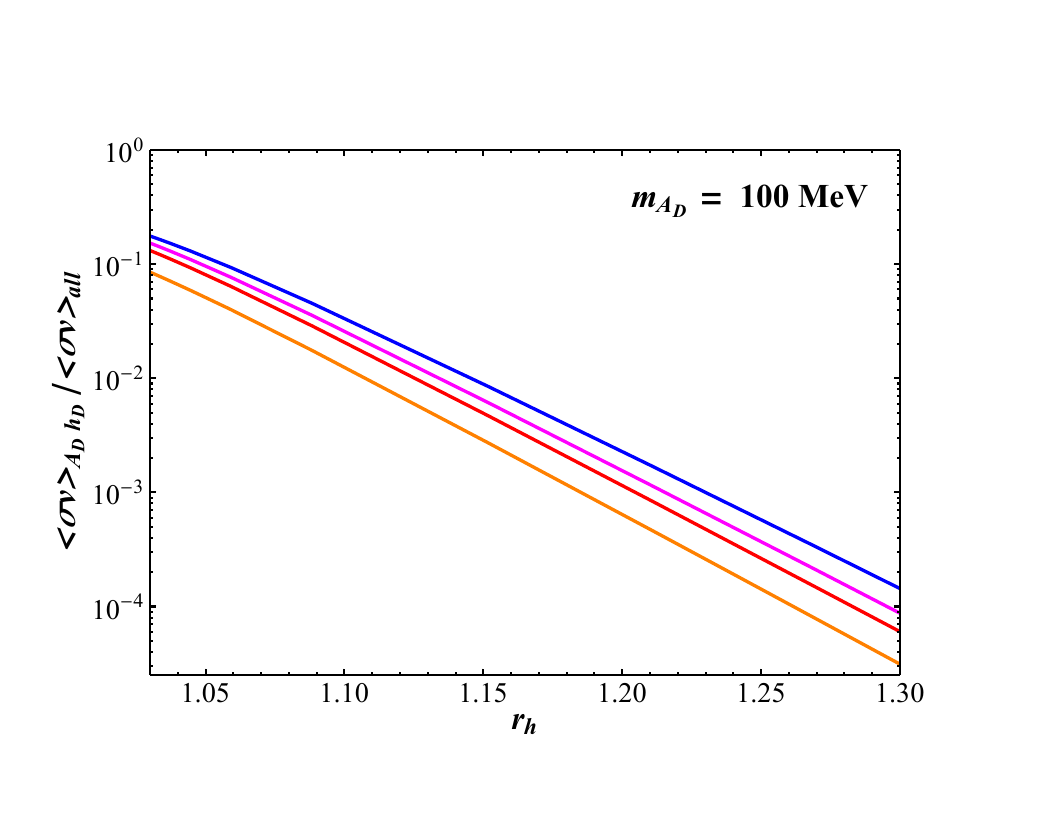}
\hspace{-0.75cm}
\includegraphics[width=2.5in]{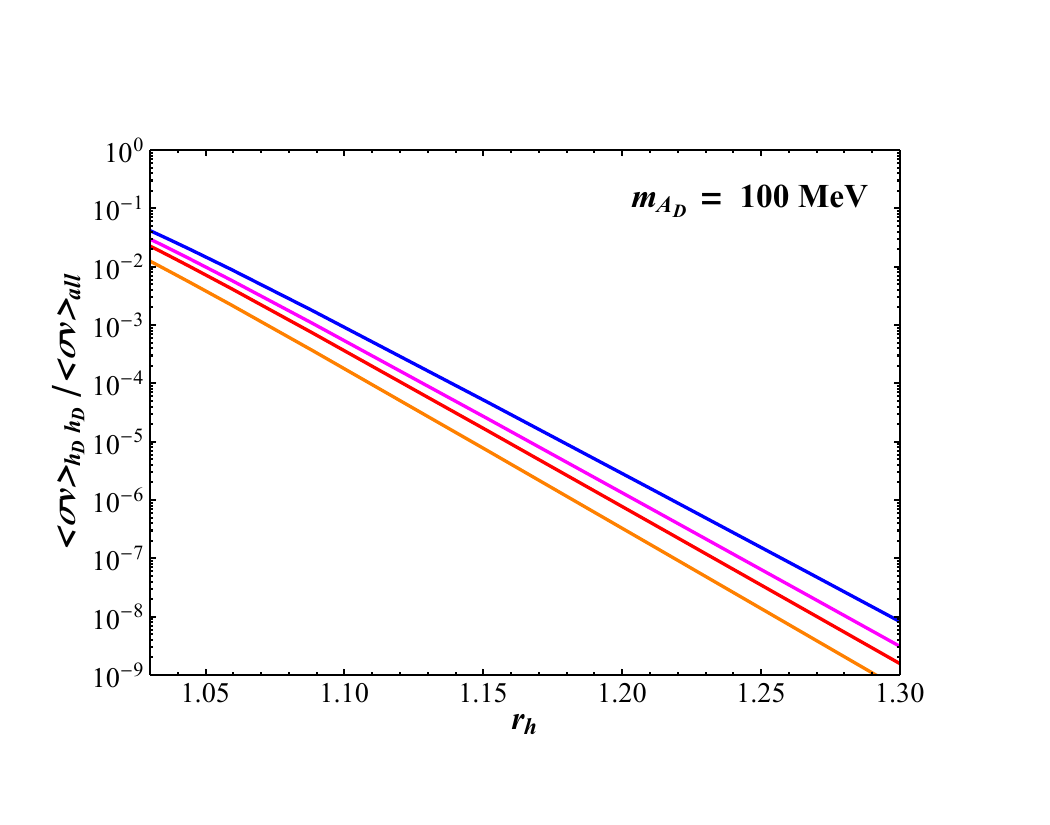}}
\vspace{-0.25cm}
\centerline{\includegraphics[width=2.5in]{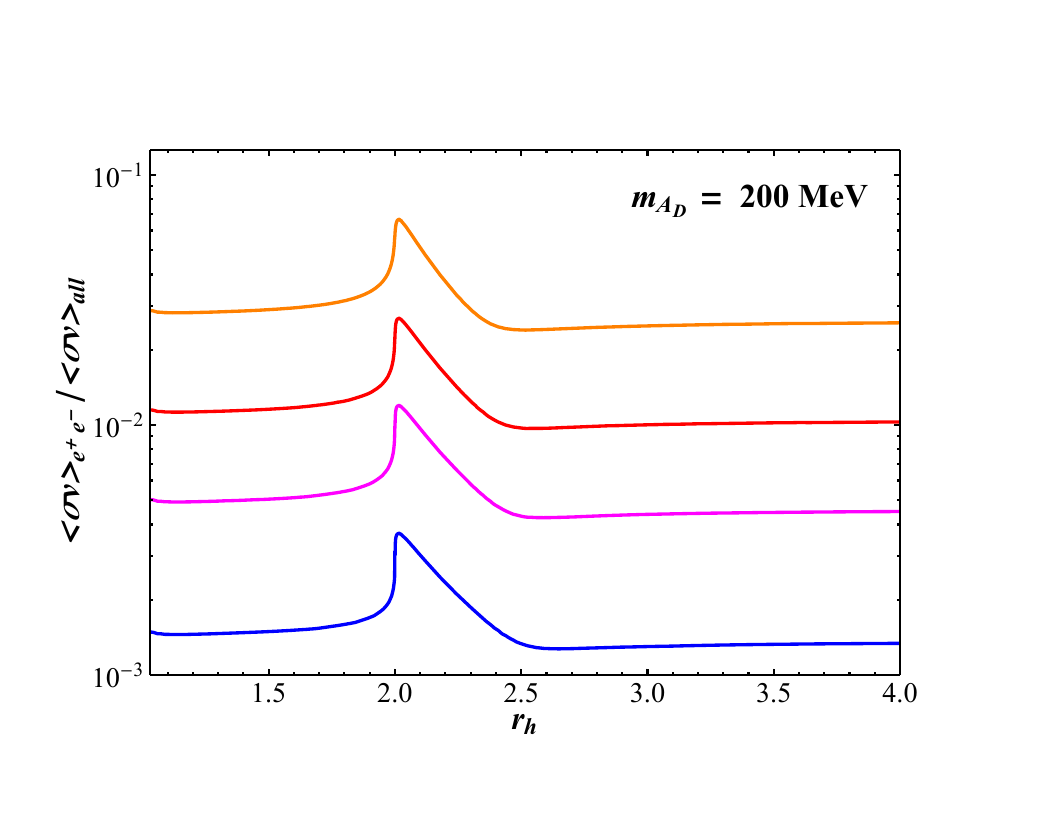}
\hspace{-0.75cm}
\includegraphics[width=2.5in]{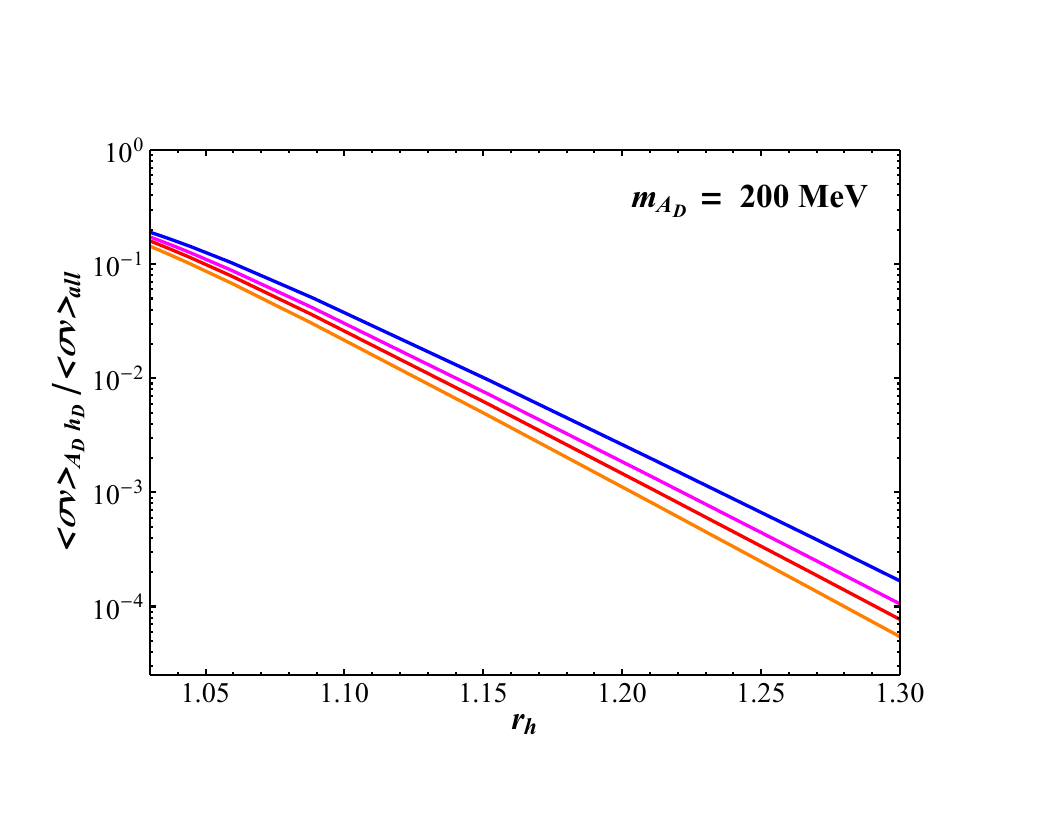}
\hspace{-0.75cm}
\includegraphics[width=2.5in]{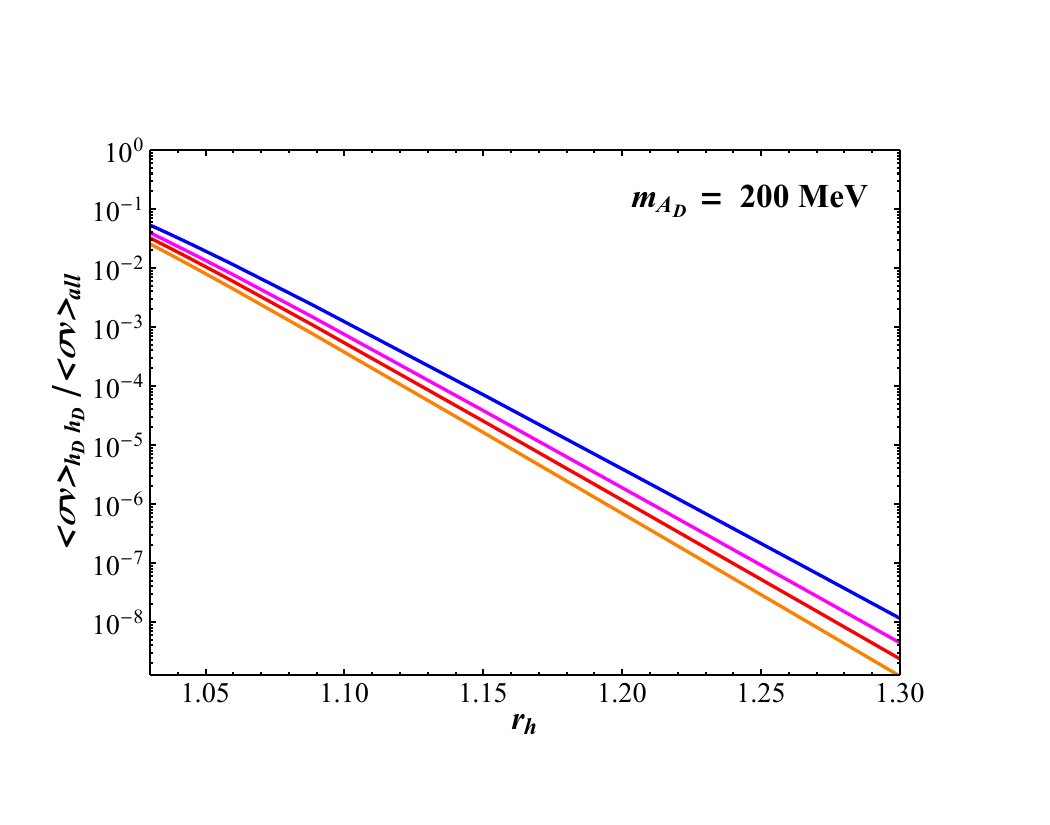}}
\vspace{-0.25cm}
\centerline{\includegraphics[width=2.5in]{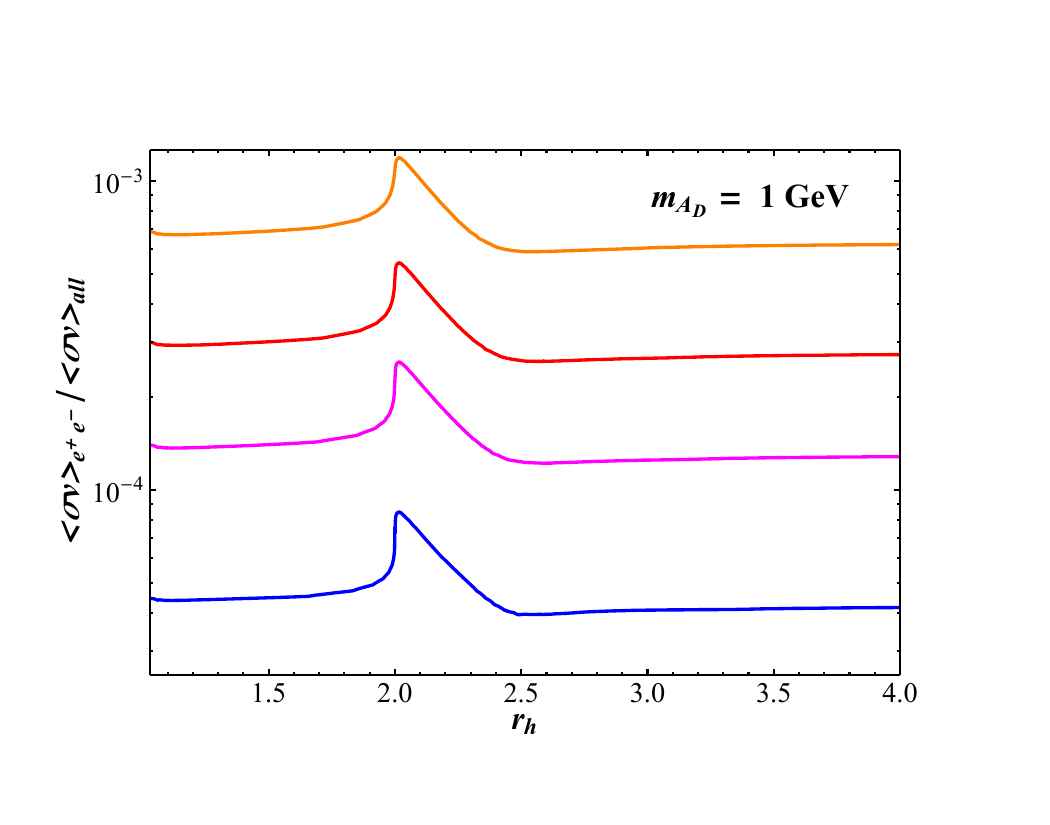}
\hspace{-0.75cm}
\includegraphics[width=2.5in]{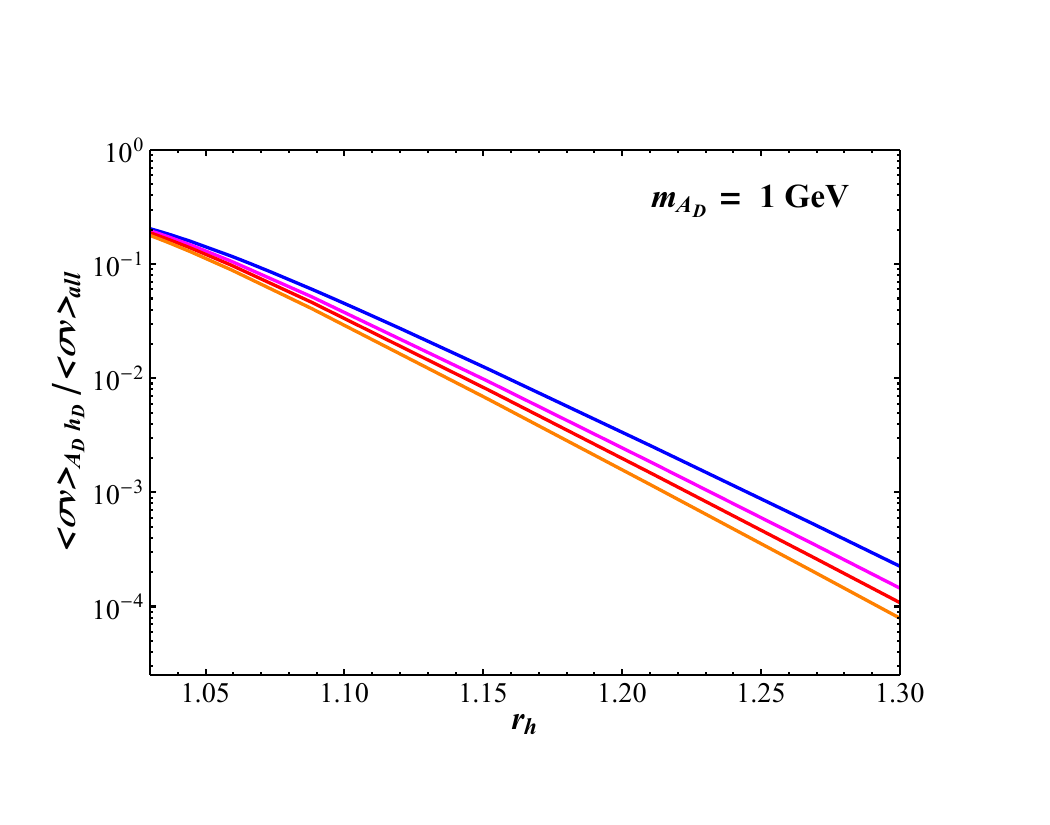}
\hspace{-0.75cm}
\includegraphics[width=2.5in]{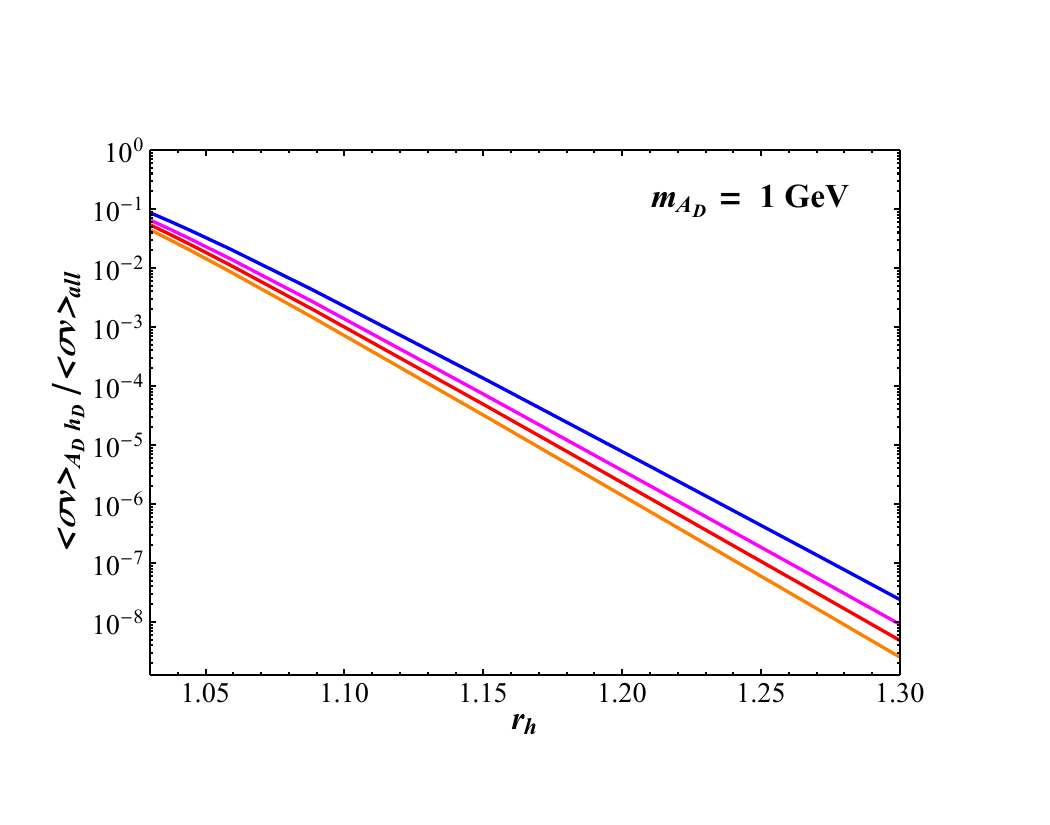}}
\caption{The values of the thermally averaged annihilation cross sections $\langle \sigma v \rangle_{e^+ e^-}$ (Left), $\langle \sigma v \rangle_{A_D h_D}$ (Center), and $\langle \sigma v \rangle_{h_D h_D}$ (Right) as a function of $r_h$, with $r_\phi$ adjusted to reproduce the observed DM relic density. Each chart assumes $b=1$ and $g_D=0.3$ (Blue), $0.5$ (Magenta), $g_D=0.7$ (Red), and $g_D=1$ (Orange). As before, the discontinuity in the $g_D=1$ line in the $m_{A_D}= 100 \; \textrm{MeV}$ chart emerges because, for this dark photon mass and dark coupling, the observed DM relic abundance cannot be recreated with any $1<r_\phi<2$ when $1.9 \lesssim r_h \lesssim 2.2$.}
\label{fig9}
\end{figure}

It is enlightening here to also depict the relative cross sections of the various annihilation processes that enter our calculation, namely the WIMP-like annihilation cross section $\phi^* \phi \rightarrow \overline{f} f$ and the three kinematically forbidden processes $\phi^* \phi \rightarrow A_D A_D,\, A_D h_D,\, h_D h_D$ at freeze-out explicitly for points in the parameter space at which the relic abundance is recreated. In Figures \ref{fig8} and \ref{fig9}, we depict the freeze-out thermal averages $\langle \sigma v \rangle$ for the processes $\phi^* \phi \rightarrow \overline{f} f, \, A_D h_D, \, h_D h_D$ as fractions of the total DM annihilation cross section at freeze-out, as a function of $r_h$ with $r_\phi$ adjusted to yield the observed relic density.

From Figure \ref{fig9}, we can draw several conclusions about the relative influences of the cross sections depicted on freeze-out. First, it is readily apparent that the processes $\phi^* \phi \rightarrow A_D h_D$ and $\phi^* \phi \rightarrow h_D h_D$ play very little role in determining the relic abundance unless $r_h$ is within a few percent of unity. Even then, we see that the ratio of these cross sections relative to the total annihilation cross section only approaches $\sim O(10\%)$: If we consider the instantaneous freeze-out approximation again, we might estimate that these processes could result in an $\sim O(10\%)$ decrease in the relic abundance near $r_h \sim 1$, which is consistent with the decrease of that magnitude we have previously observed in, \eg, Figure \ref{fig6}. The WIMP-like annihilation cross section $\langle \sigma v \rangle_{e^+ e^-}$, however, can potentially play a larger role. Curiously, the relative magnitude of $\langle \sigma v \rangle_{e^+ e^-}$ at freeze-out behaves counterintuitively with variations in the dark coupling $g_D$ and the mass ratio $r_h$: As $g_D$ is increased and/or $r_h$ is made to approach the $r_h=2$ resonance in $\langle \sigma v \rangle_{A_D A_D}$, both of which naively would result in an increased cross section $\langle \sigma v \rangle_{A_D A_D}$, and hence a decreased relative contribution of the cross section $\langle \sigma v \rangle_{e^+ e^-}$, the ratio $\langle \sigma v \rangle_{e^+ e^-}/\langle \sigma v \rangle_{all}$ actually \emph{increases}. This behavior appears to be a manifestation of the same phenomenon that mutes the effects of the variation of the relic density with changing $b$, namely, the exponential dependence of the forbidden cross section $\langle \sigma v \rangle_{A_D A_D}$ on the freeze-out temperature and the mass ratio $r_\phi = m_{A_D}/m_\phi$. In particular, as any factor which increases $\langle \sigma v \rangle_{A_D A_D}$ will correspondingly require a larger $r_\phi$, which exponentially reduces $\langle \sigma v \rangle_{A_D A_D}$ in order to recreate the observed relic abundance. This increase in $r_\phi$ not only results in a significant Boltzmann suppression of $\langle \sigma v \rangle_{A_D A_D}$ relative to the WIMP-like cross section $\langle \sigma v \rangle_{e^+ e^-}$, but also adds some polynomial enhancement of the WIMP-like process as $r_\phi$ gets closer to the $s$-channel resonance $r_\phi = 2$. The effect is particularly noticeable for lighter dark photon masses with higher dark couplings: For $g_D=1$ and $m_{A_D}=100 \; \textrm{MeV}$, we even see that the WIMP-like cross section $\langle \sigma v \rangle_{e^+ e^-}$ can even account for the majority of the total annihilation cross section at freeze-out, for certain values of $r_h$. However, we note that as $m_{A_D}$ becomes heavier, the maximum relevance of the WIMP-like process becomes much smaller. This validates our choice to omit other kinematically allowed WIMP-like channels, such as $\phi^* \phi \rightarrow \mu^+ \mu^-$, for our choice $m_{A_D}=1 \; \textrm{GeV}$: At most, the processes $\langle \sigma v \rangle_{e^+ e^-}$ accounts for $O(0.1\%)$ of the annihilation cross section at freeze-out, so the other WIMP-like processes should be similarly insignificant.

\section{Constraints: Direct Detection and CMB}\label{Section:Constraints}

Compared to, for example, the fermionic forbidden DM models in, \eg, \cite{DAgnolo:2015ujb,Cline:2017tka,Fitzpatrick:2020vba}, there are relatively few existing experimental constraints on the parameter space that we consider in the present model. In particular, constraints from the cosmic microwave background (CMB) are significantly relaxed because, unlike the Dirac fermion DM considered in those works, the cross sections for DM annihilation into SM particles are all either $p$-wave or suppressed by Boltzmann factors, implying that very little energy will be injected into the visible sector during the epoch of recombination (there is one exception, which we shall discuss below). Furthermore, detection prospects for the dark Higgs in, \eg, beam-dump or collider experiments are rather limited, since the dark Higgs has (other than a possible $\lesssim O(10^{-6})$ mixing term with the SM Higgs) no direct coupling with the SM.\footnote{A similar dark Higgs setup to the one we considered here is analyzed in \cite{Darme:2017glc}, where the dark Higgs can in fact have meaningful constraints arising from beam-dump experiments. There, however, the dark Higgs is lighter than the dark photon, so the dominant dark Higgs decay process is the $\epsilon^2$-suppressed $h_D \rightarrow e^+ e^-$. In that case, the dark Higgs is both long lived and has entirely visible decay products. This is not true in the mass range we consider in the present work, where the Higgs will promptly decay into either DM or on-shell/virtual dark photons.}

\begin{figure}[htbp]
\centerline{\includegraphics[width=5.0in]{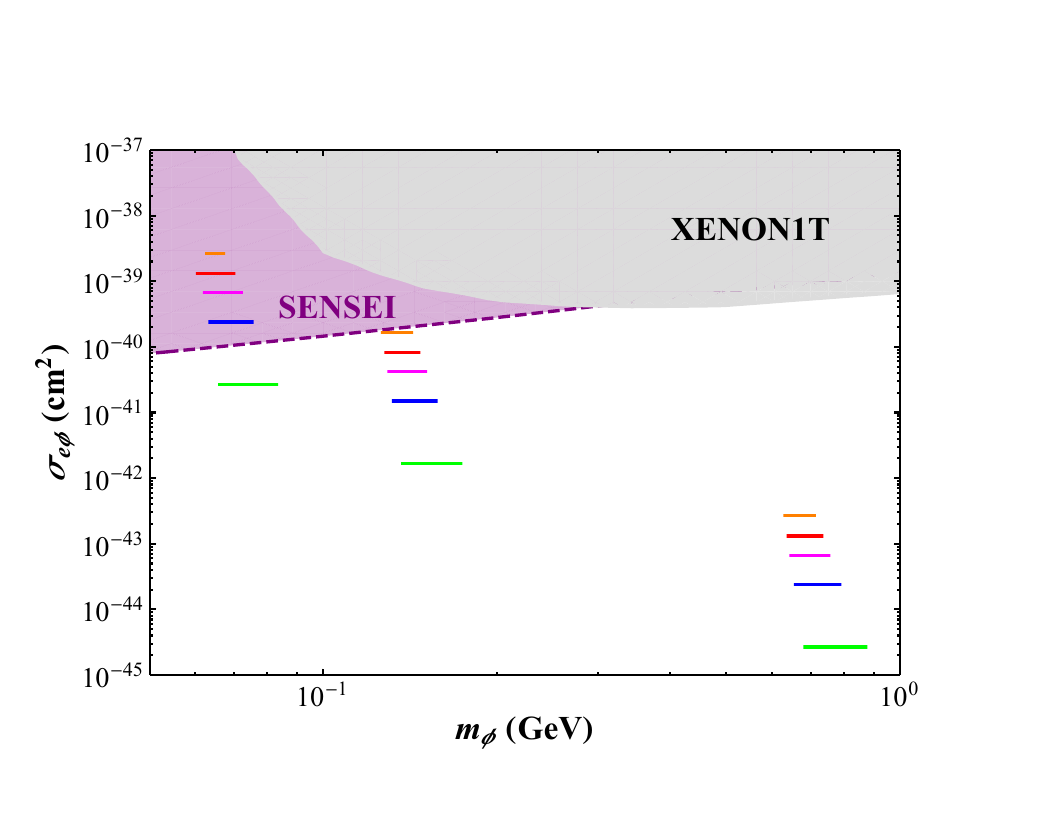}}
\caption{The direct detection cross sections $\sigma_{e \phi}$ (given in Eq.(\ref{eq:DirectDetection}), assuming $b=1$ for various selections of $m_{A_D}$ and $g_D$. Because this cross section is independent of the DM mass $m_\phi$, each benchmark point takes the form of a horizontal line in $m_\phi-\sigma_{e \phi}$ space, with the length determined by the range of $m_\phi$ for which the observed relic density can be recreated for some value of $r_h \equiv m_{h_D}/m_{A_D}$. The lines form 3 groups, which assume $m_{A_D}=100 \; \textrm{MeV}$ (Left), $m_{A_D}=200 \; \textrm{MeV}$ (Center), and $m_{A_D}=1 \; \textrm{GeV}$ (Right). Each line in a group assumes $g_D=0.1$ (Green), $g_D=0.3$ (Blue), $g_D=0.5$ (Magenta), $g_D=0.7$ (Red), and $g_D=1$ (Orange). The current $90\%$ CL constraint on this cross section from XENON1T ionization data \cite{Aprile:2019xxb} is depicted as a gray shaded region, notably, it does not exclude any of our benchmarks. The projected sensitivity of a 100$\textrm{g} \cdot \textrm{yrs}$ SENSEI null result \cite{Battaglieri:2017aum} is depicted as a purple line.}
\label{fig10}
\end{figure}

The first significant constraint on the parameter space of this setup comes from near-future direct detection experiments, in particular those stemming from dark-matter-electron scattering, such as SENSEI \cite{Crisler:2018gci}, SuperCDMS \cite{SuperCDMS:2018mne}, or DAMIC-M \cite{Castello-Mor:2020jhd}. When computing the scattering cross section of the DM with a free electron, we note that the small mixing of the dark Higgs with the SM Higgs scalar, combined with suppressed Yukawa couplings to light fermions, renders any contribution of the dark Higgs to this quantity utterly irrelevant. So, given that $m^2_\phi \gg m^2_e$ for the entire parameter space that we probe, the relevant scattering cross section for direct detection is simply the well-known result,
\begin{align}\label{eq:DirectDetection}
    \sigma_{e \phi} = \frac{4 \alpha_{\textrm{em}} m_e^2 g_D^2 \epsilon^2}{m_{A}^4} \approx \bigg( \frac{g_D}{0.3}\bigg)^2 \bigg( \frac{\epsilon}{3\times 10^{-4}}\bigg)^2 \bigg( \frac{\textrm{100 MeV}}{m_{A_D}}\bigg)^4 (2.4 \times 10^{-40} \; \textrm{cm}^2).
\end{align}
The cross section in Eq.(\ref{eq:DirectDetection}) is generally beyond current direct detection constraints. However, we can anticipate that relatively near-term experiments will have the capability to exclude significant portions of the parameter space. In Figure \ref{fig10}, we depict the cross section $\sigma_{e \phi}$ for our various benchmark points in parameter space, and compare these results to both existing constraints from XENON1T \cite{Aprile:2019xxb} and projected future constraints from the SENSEI experiment \cite{Battaglieri:2017aum}. Significantly, the XENON1T data does not exclude any regions of our parameter space. It should be noted however that the study in \cite{Aprile:2019xxb} gives two constraints from XENON1T data: A more conservative constraint assuming that electron recoil events with $\leq 12$ produced electrons are undetectable, and a less conservative one with no such cutoff. The authors impose the cutoff here because the liquid Xenon charge yield hasn't been measured below the approximate electron recoil energy required for $\sim 12$ electrons, and they only include the more aggressive bound because prior studies \cite{Essig:2012yx,DarkSide:2018ppu} do not impose a similar cutoff on electron number. In the interest of caution, we have presented the more conservative bound here-- when the electron number cutoff is removed, the scenarios with $m_{A_D}=100 \; \textrm{MeV}$ and $g_D \geq 0.5$ are excluded by the more aggressive bound, but the results are otherwise similar to what we have presented in Figure \ref{fig10}. Regardless of the interpretation of XENON1T data, it is clear that near-term null results from SENSEI over the course of the next several years may exclude most of the $m_{A_D} = 100 \; \textrm{MeV}$ results. Slightly longer-term upcoming experiments, such as SuperCDMS and DAMIC-1K may offer even stronger constraints on the model parameter space, excluding any benchmarks with $\sigma_{e \phi} \gsim 10^{-43} \; \textrm{cm}^2$ \cite{Battaglieri:2017aum}.

\begin{figure}[htbp]
\centerline{\includegraphics[width=5.0in]{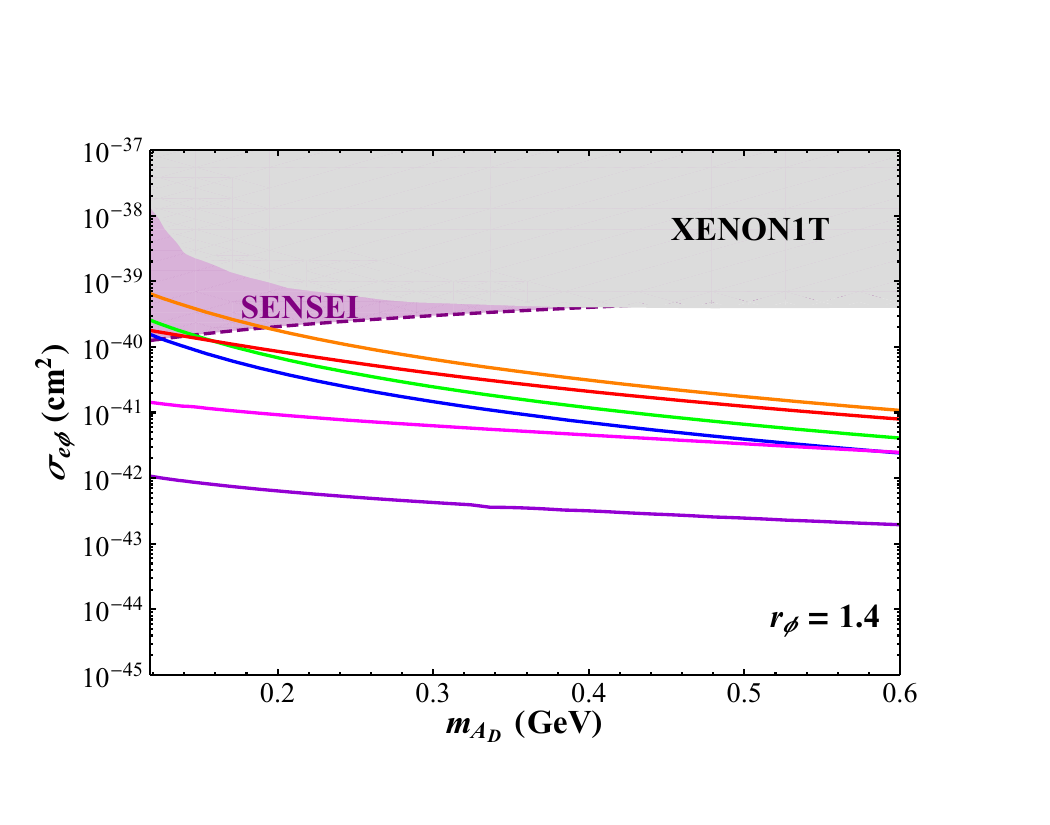}}
\caption{The direct detection cross sections $\sigma_{e \phi}$ (given in Eq.(\ref{eq:DirectDetection}), assuming $b=1$, $\epsilon=3\times 10^{-4}$, and $r_\phi=1.4$, for various selections of $r_h$ with $g_D$ fixed to recreate the observed dark matter relic abundance. Each contour assumes $r_h=1.7$ (Green), $r_h=1.9$ (Blue), $r_h=2.1$ (Violet), $r_h=2.2$ (Magenta), $r_h=2.3$ (Red), and $r_h=2.5$ (Orange). The XENON1T \cite{Aprile:2019xxb} and projected SENSEI \cite{Battaglieri:2017aum} sensitivities are included in the same way as in Figure \ref{fig10}.}
\label{fig10Extra}
\end{figure}

In an effort to further clarify the significance of direct detection constraints on our construction here, and in particular the role that the dark Higgs plays in broadening our parameter space, we can move beyond our benchmark points and briefly explore how these constraints look along a different plane. In Figure \ref{fig10Extra}, we depict the direct detection cross section $\sigma_{e \phi}$ as a function of the dark photon mass $m_{A_D}$ with $g_D$ selected to recreate the observed dark matter relic abundance, for $r_\phi=1.4$ (selected because this value permits the relic density to be recreated for a wide range of dark photon masses with values of $g_D \sim O(1)$) and various selections of $r_h$. We can again observe that while present experiments don't limit the parameter space at all, the projected SENSEI sensitivities may be capable of ruling out low dark photon masses $\lsim O(150-200 \; \textrm{MeV})$ in this scenario for most $r_h$ selections. When $r_h$ is such that resonant dark Higgs exchange dominates freeze-out (so, $r_h \gsim 2$), we see that the $g_D$ necessary to recreate the appropriate relic abundance is dramatically reduced: In turn, the direct detection cross section is reduced by as much as 2 to 3 orders of magnitude compared to the case when $r_h$ is far from resonance. In this plane, then, we can more clearly see the substantial role that the dark Higgs may play in broadening the allowable parameter space for this model.

\begin{figure}[htbp]
\centerline{\includegraphics[width=3.5in]{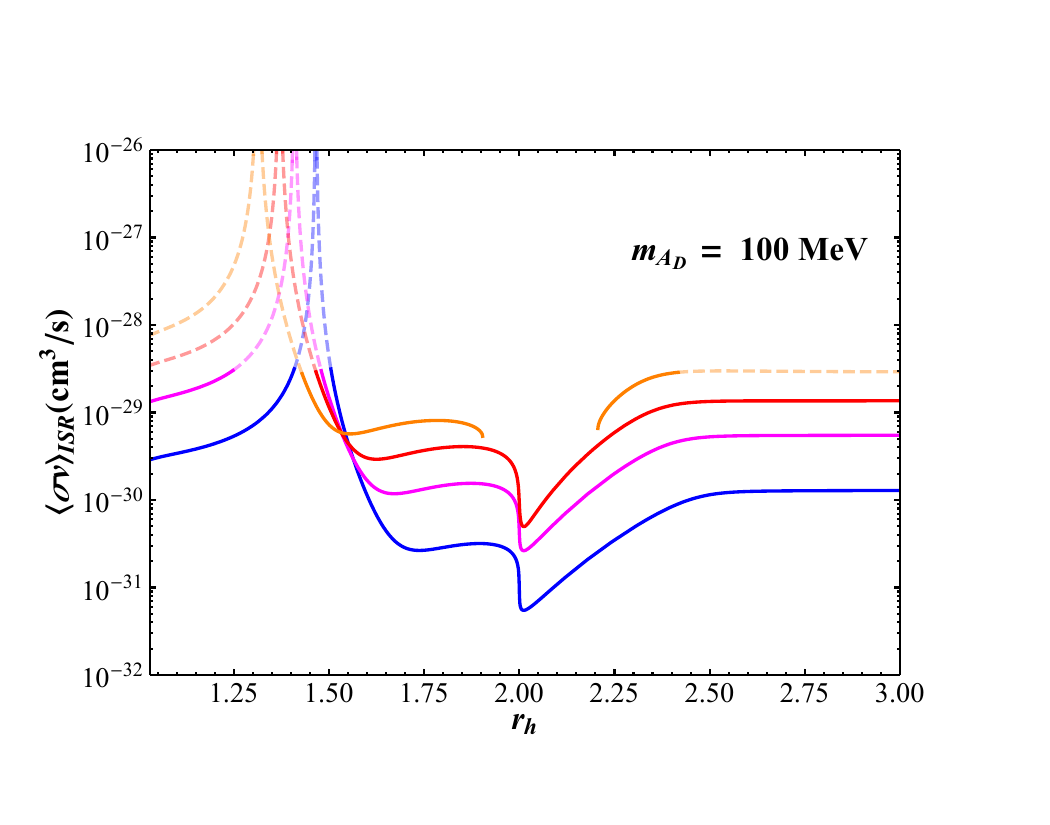}
\hspace{-0.75cm}
\includegraphics[width=3.5in]{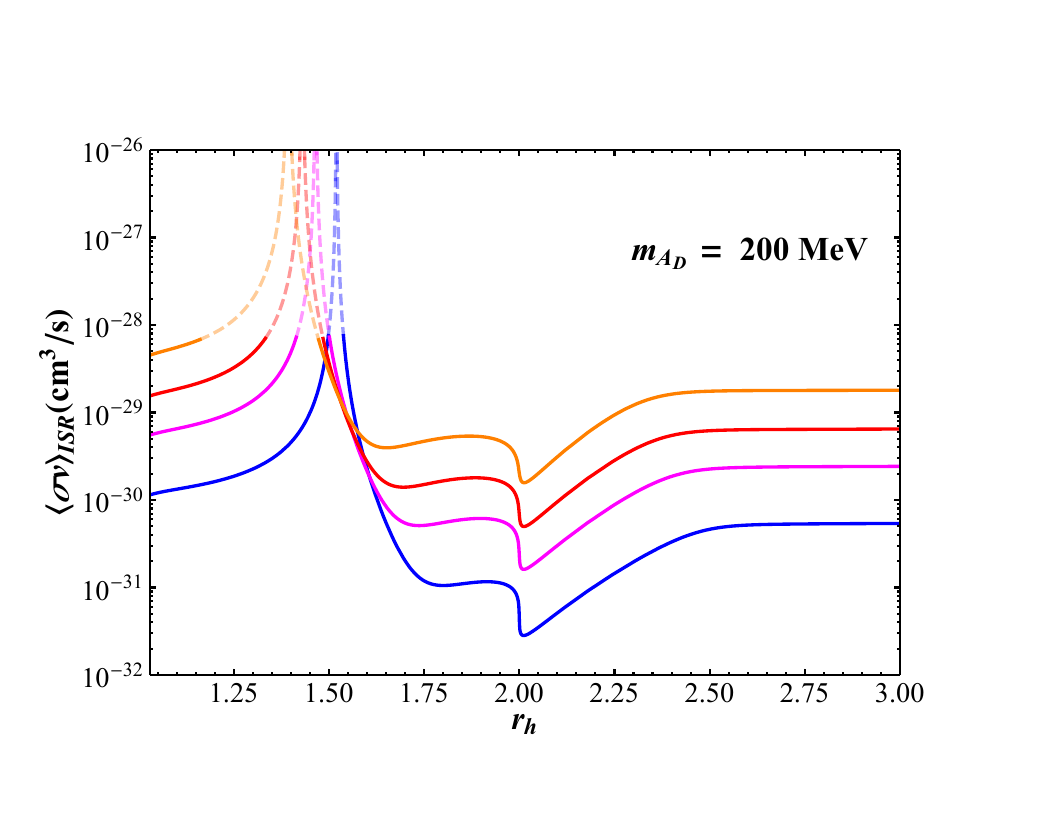}}
\caption{The cross section $\langle \sigma v \rangle_{\textrm{ISR}}$, as discussed in the text, for various benchmark points assuming $b=1$ and $m_{A_D} = 100 \; \textrm{MeV}$ (left) and $m_{A_D}= 200 \; \textrm{MeV}$ (right), as a function of $r_h$ with $r_\phi$ adjusted to recreate the observed relic abundance. The lines assume that $g_D=0.3$ (Blue), $g_D=0.5$ (Magenta), $g_D=0.7$ (Red), and $g_D=1$ (Orange). As before, the discontinuity in the $g_D=1$ curve for $m_{A_D}=100 \; \textrm{MeV}$ arises because near the $r_h=2$ resonance, there is no value of $r_\phi$ capable of reproducing the observed DM abundance for this benchmark. Solid curves indicate that this cross section satisfies the constraints from \emph{Planck} 2018 \cite{Planck:2018vyg,Planck:2019nip} and baryon acoustic oscillation \cite{BOSS:2016wmc,Beutler:BAO,Ross:2014qpa} data on this cross section extracted from \cite{Cang:2020exa}, while dashed curves indicate that this constraint is violated. Benchmark points for $g_D=0.1$ or $m_{A_D}=1 \; \textrm{GeV}$ are not pictured, because in these cases the constraints on $\langle \sigma v \rangle_{\textrm{ISR}}$ are trivially satisfied as long as $m_{h_D}/m_\phi -2 \gtrsim O(10^{-2})$.}
\label{fig11}
\end{figure}

Beyond the direct detection constraints, a mild constraint arises from the CMB, even with the velocity suppression of the annihilation processes that contribute to freeze-out. Specifically, it was found in \cite{Rizzo:2020jsm} that the $s$-wave process $\phi^* \phi \rightarrow A_D A_D^* \rightarrow A_D \overline{f} f$, in spite of suppression due to the three-body phase space and an additional coupling factor of $\alpha_D=g_D^2/(4 \pi)$ relative to the WIMP-like cross section, can have significant effects on the CMB by injecting energy into the visible sector at the epoch of recombination. Essentially, this is just the WIMP-like process $\phi^* \phi \rightarrow \overline{f} f$, with the addition of some initial-state dark radiation (dark ISR).\footnote{The astute reader may notice that the explicit inclusion of dark Higgs fields in our present construction has the potential to introduce another similar ``dark ISR'' process, $\phi^* \phi \rightarrow h_D A_D^* \rightarrow h_D \overline{f} f$. However, this process is not $s$-wave, and is therefore negligibly small during the epoch of recombination.} For simplicity, we will restrict our analysis of this process to the case in which the SM fermions $f$ are electrons, and work in the limit in which $m_\phi$, the DM mass, is much greater than the mass of the electron. Analytically, the cross section for this process is then given in the non-relativistic limit as
\begin{align}\label{eq:CMBxSection}
    (\sigma v)_{\textrm{ISR}} &\simeq \frac{g_D^4 \epsilon^2 \alpha_{\textrm{em}}}{96 \pi^2 m_\phi^4}\int_{0}^{(2m_\phi-m_{A_D})^2} d m_{ee}^2 \frac{m_{ee}^2\sqrt{(m_{ee}^2-4m_\phi^2+m_{A_D}^2)^2-4 m_{A_D}^2 m_{ee}^2}}{(m_{ee}^2-m_{A_D}^2)^2} \nonumber\\
    & \bigg\{ \bigg( 2+\frac{4 m_{A_D}^2 m_{ee}^2}{(m_{ee}^2-4m_\phi^2+m_{A_D}^2)}\bigg)+\frac{6 b m_\phi^2 Q_S^2 (4 m_\phi^2-m_{h_D}^2)}{(4 m_\phi^2(1+v^2/4)-m_{h_D}^2)^2+4 m_\phi^2(1+v^2/4) \Gamma_h(v)^2}\\
    &+ \frac{ b^2 m_\phi^4 Q_S^4  \bigg( 2+\frac{(m_{ee}^2-4 m_\phi^2+m_{A_D}^2)^2}{4 m_{A_D}^2 m_{ee}^2}\bigg)}{(4 m_\phi^2(1+v^2/4)-m_{h_D}^2)^2+4 m_\phi^2(1+v^2/4) \Gamma_h(v)^2} \bigg\}, \nonumber
\end{align}
where $m_{ee}^2$ is the squared invariant mass of the sum of the two electrons' four-momenta, $v$ is the relative velocity of the two DM particles in the center-of-mass frame, $\Gamma_h(v)$ is the decay width of the dark Higgs (the $v$ dependence emerges because in the non-relativistic limit of this process, the width from the decay $h_D \rightarrow \phi^* \phi$ will contribute a term proportional to $v$ to the decay width term in the Breit-Wigner form of the propagator), and we have dropped terms of $O(v^2)$ or higher, except where they appear in the propagator for an $s$-channel exchange of a dark Higgs, which for certain values of $m_\phi$ and $m_{h_D}$ may be near a resonance peak. To compare this expression to constraints from CMB data, we then merely need to find the thermal average of this cross section at the temperature of the DM during the epoch of recombination. In Figure \ref{fig11}, we depict the results for these cross sections assuming $b=1$ for various selections of $g_D$ and $m_{A_D}$, as a function of $r_h$ with $r_\phi$ adjusted to give the correct relic abundance. Cross sections which exceed the constraints given in \cite{Cang:2020exa} are plotted as dashed lines, while cross sections which are below these bounds are solid. Notably, we actually find that a rigorous thermal averaging of the cross section of Eq.(\ref{eq:CMBxSection}) is unnecessary here: Estimating the temperature of the DM during this era as $T_{DM } \sim T_{CMB}^2/(10^{-3} m_{\phi})$, where $10^{-3} m_{\phi}$ is roughly the temperature of DM kinetic decoupling from the SM \cite{Bringmann:2006mu}, while $T_{CMB} \sim 3000 \; \textrm{K}$ is the temperature of the SM thermal bath at the epoch of recombination, we find that taking the thermal average of the cross section at this approximate temperature gives a result that is visually indistinguishable (on our plots) to the results of simply identifying the thermal average as the cross section in the limit $v \rightarrow 0$. As a result, in Figure \ref{fig11} we only depict the cross section of Eq.(\ref{eq:CMBxSection}) in the limit of $v \rightarrow 0$, without more rigorously estimating of the DM temperature or performing any thermal averaging.

In Figure \ref{fig11}, we see that the constraint from the dark ISR process can actually significantly limit our parameter space -- in fact, this constraint represents the most stringent current limit on the benchmark points we have so far considered. Most notably, this occurs when the cross section enjoys a resonant enhancement of the exchange of a dark Higgs near $m_{h_D} = 2 m_\phi$. For light dark photons ($m_{A_D}= 100 \; \textrm{MeV}$), we observe that the resonance region excludes $1.4 \lesssim r_h \lesssim 1.55$ for $g_D=0.3$, $1.25 \lesssim 1.5$ for $g_D=0.5$, and even the entire regions $r_h \lesssim 1.45$ and $r_h \lesssim 1.4$ when $g_D=0.7$ and $g_D=1$, respectively.\footnote{This resonant enhancement also appears in the DM self-interaction process $\phi^* \phi \rightarrow \phi^* \phi$, however, because the experimental constraint on this cross section is very weak ($\sigma/m_\phi \lesssim 1 \; \textrm{cm}^2/\textrm{g}$, where $\sigma$ here is the self-interaction cross section \cite{Bondarenko:2020mpf}), the parameter space in which the self-interaction constraints are violated is always a small subset of the parameter space in which the CMB constraints are violated.} Even more interestingly, for $m_{A_D} = 100 \; \textrm{MeV}$ and $g_D=1$, the region of large $r_h$, where we find the dark Higgs is too heavy to play a significant role in the DM relic abundance calculation, the CMB constraint actually excludes the model. In other words, a dark Higgs that is light enough to significantly affect freeze-out is actually \emph{necessary} to render this benchmark point phenomenologically viable. The constraints arising from the CMB become far less significant for higher-mass dark photons (and hence higher-mass DM): For benchmark points with $m_{A_D} = 200 \; \textrm{MeV}$, the $g_D=1$ and $g_D=0.7$ points are only excluded for $1.15 \lesssim r_h \lesssim 1.45$ and $1.35 \lesssim r_h \lesssim 1.45$, respectively, while the constraints for smaller dark couplings are weaker still. For the benchmark points where $m_{A_D}=1 \; \textrm{GeV}$, the excluded regions are so narrow that we do not depict these cross sections in Figure \ref{fig11}.

We also note that all of the cross sections in Figure \ref{fig11} assume that $b=1$ (that is, the entirety of the DM's mass emerges from the vev of the dark Higgs) and that the kinetic mixing parameter $\epsilon$ is equal to $3 \times 10^{-4}$. With the exception of the benchmark $m_{A_D}=100 \; \textrm{MeV}$, $g_D=1$, we note that all benchmark points which fail the CMB constraint do so due to the $m_{h_D}/m_{\phi} \approx 2$ resonance peak, which scales as $b^2$, while the entire cross section $\langle \sigma v \rangle_{ISR}$ scales as $\epsilon^2$. Therefore, it is clear that these constraints, especially those requiring the mass parameters be very close to resonance, can be significantly weakened by simply reducing the value of $b$, which as we have mentioned before has a muted effect on the phenomenology of DM freeze-out, or significantly weakened \emph{or} strengthened by decreasing or increasing the kinetic mixing parameter $\epsilon$, which has a negligible effect on the relic abundance provided it remains large enough to keep the dark photon and dark Higgs in thermal equilibrium with the Standard Model.

Finally, it is significant to point out that when we are very near the $m_{h_D}/m_{\phi}$ resonance, our computation of the relic abundance itself may not be entirely reliable, since this annihilation cross section and even $3 \rightarrow 2$ processes such as $\phi^* \phi \phi \rightarrow A_D \phi$ enjoy enormous resonant enhancements which can render them comparable to the annihilation processes that we explicitly consider in our computation. We find, however, that any near-resonance points for which these processes have cross sections comparable to the $O(10^{-26}) \; \textrm{cm}^3/\textrm{s}$ cross sections that the combined WIMP-like and kinematically forbidden processes have at freeze-out require $m_{h_D}/m_{\phi}$ to be much closer to resonance than is required to exclude them with these CMB measurements. This can be readily seen for the cross section $\langle \sigma v \rangle_{\textrm{ISR}}$ in Figure \ref{fig11}, where we can note that the $s$-wave cross section must be several orders of magnitude below $O(10^{-26}) \; \textrm{cm}^3/\textrm{s}$ in order to satisfy these CMB constraints. For the dominant $3 \rightarrow 2$ process, $\phi^* \phi \phi \rightarrow A_D \phi$, the case is slightly more complicated, but no more consequential: We find that this cross section's contributions to the collision term in the Boltzmann equation only competes with those of the $2\rightarrow 2$ WIMP-like and forbidden processes at freeze-out for the benchmark points where $m_{A_D}= 100 \; \textrm{MeV}$ and $g_D=1$ (where the $r_\phi$ value required to recreate the relic abundance is highest among our benchmark points, and therefore the $3 \rightarrow 2$ process doesn't suffer an exponential Boltzmann suppression relative to the forbidden cross section, as discussed in Section \ref{Section:RelicDensity}), and then only for the narrow region $1.3 \lesssim r_h \lesssim 1.35$. Outside of this region the $3 \rightarrow 2$ cross section at freeze-out's contribution to the collision term is at least an order of magnitude below the combined $2 \rightarrow 2$ cross sections' contribution. We can clearly see in Figure \ref{fig11} that this range of $r_h$ values is well within the region already excluded by CMB measurements, so our omission of the $3 \rightarrow 2$ processes from the freeze-out calculation remains justified for any phenomenologically viable points in our parameter space.

\begin{figure}[htbp]
\centerline{\includegraphics[width=3.5in]{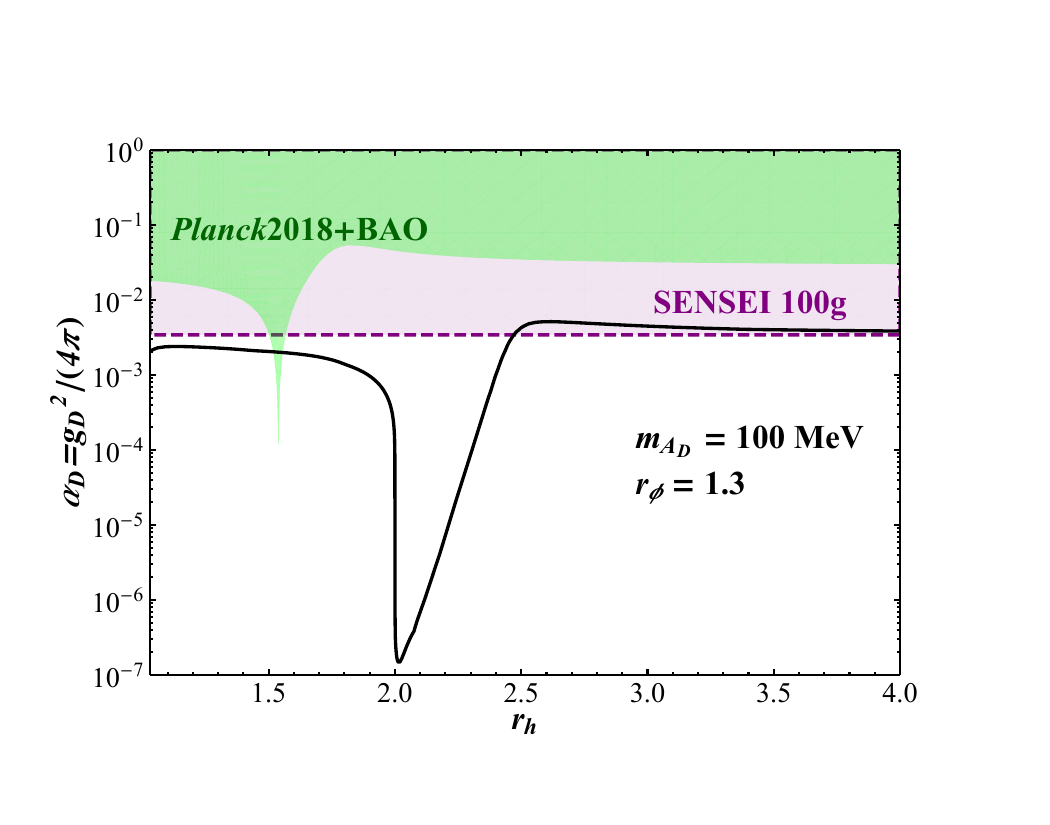}
\hspace{-0.75cm}
\includegraphics[width=3.5in]{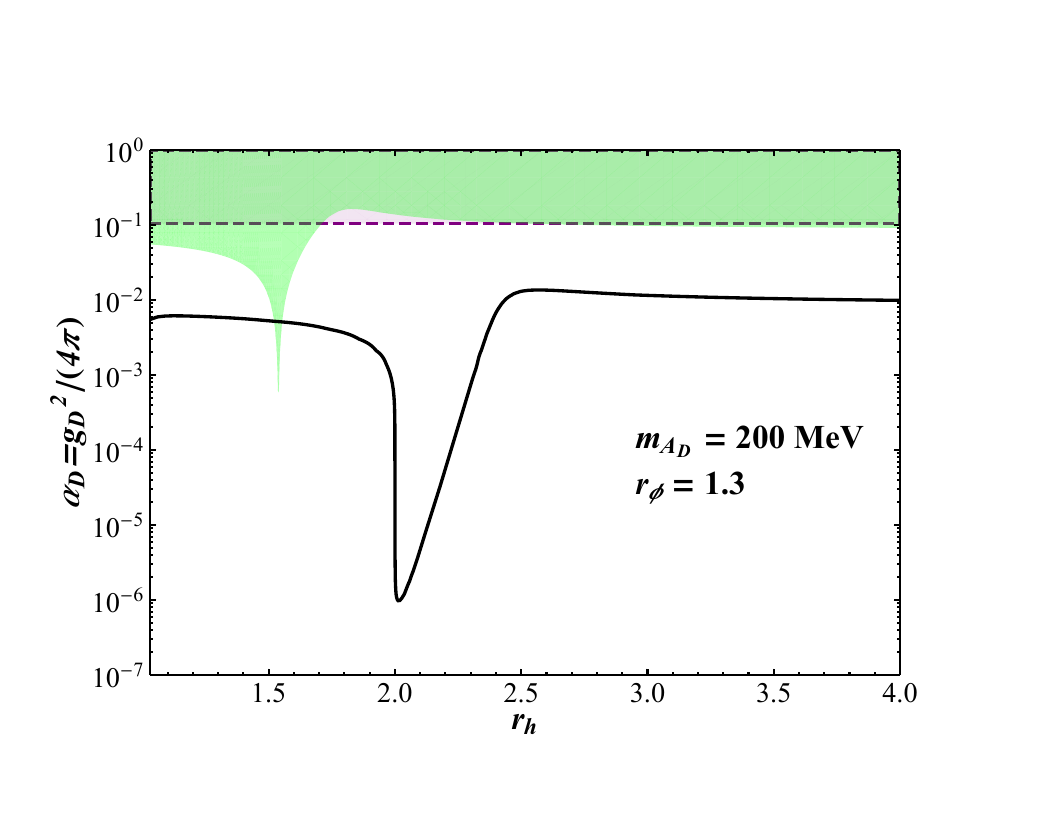}}
\vspace*{-0.75cm}
\centerline{\includegraphics[width=3.5in]{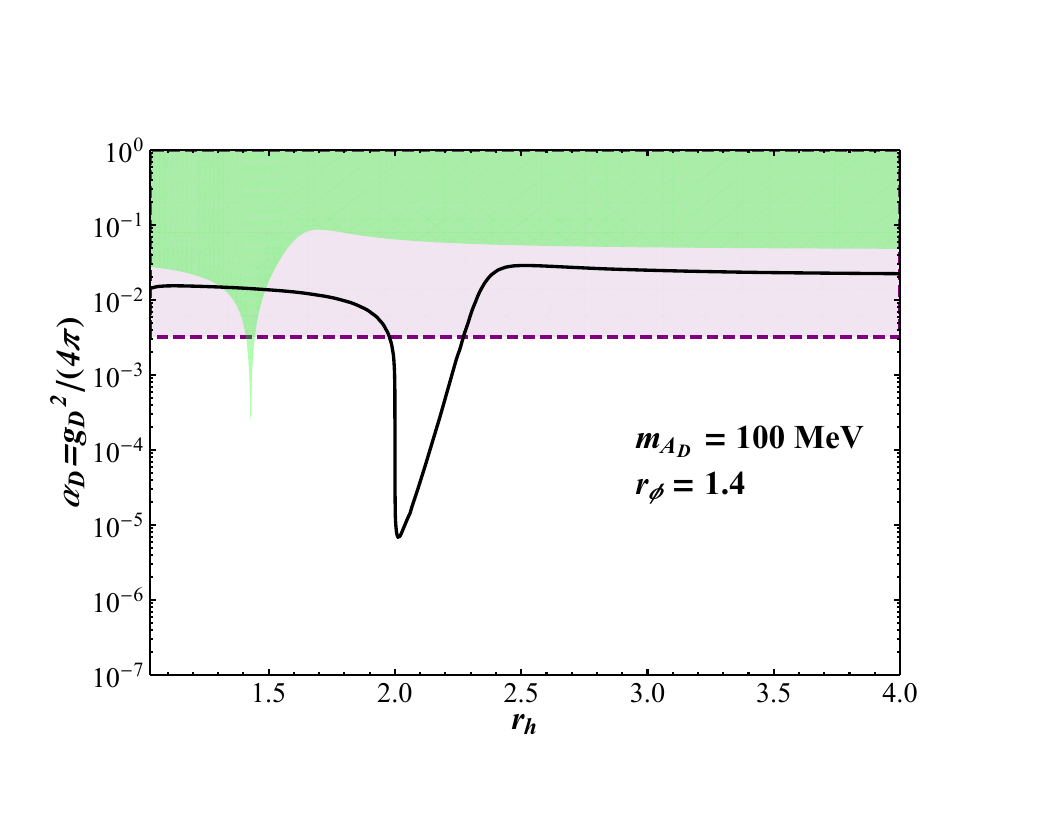}
\hspace{-0.75cm}
\includegraphics[width=3.5in]{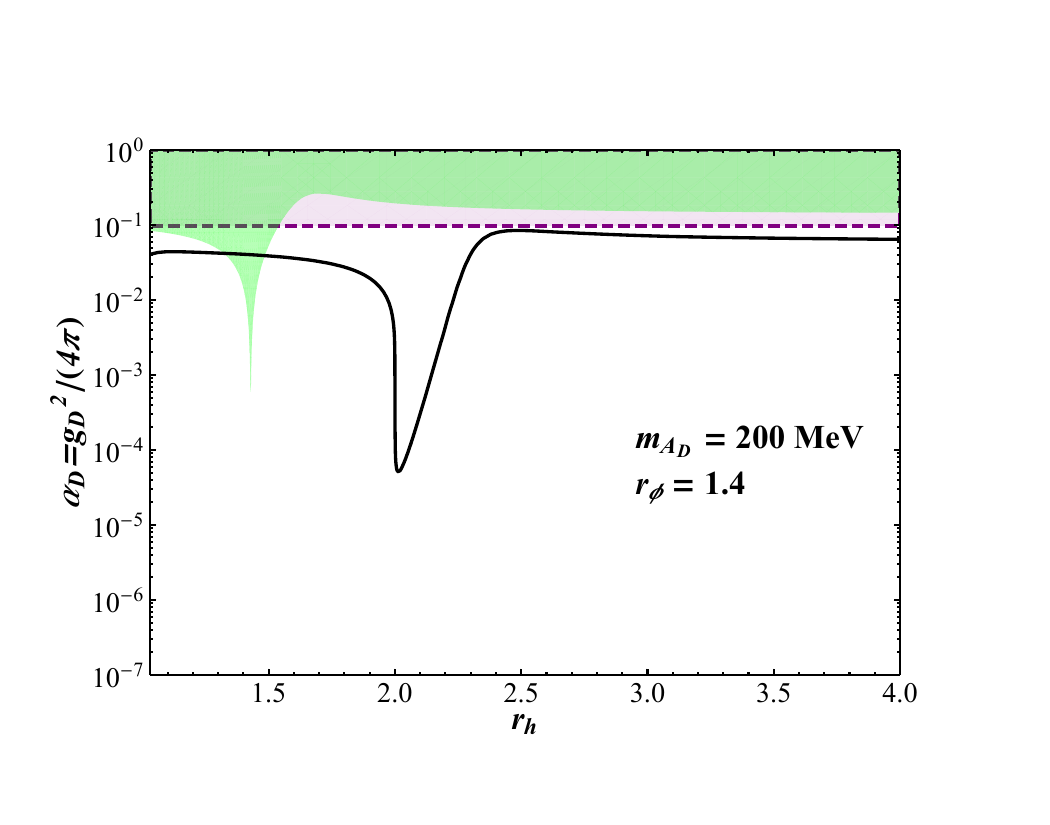}}
\vspace*{-0.75cm}
\centerline{\includegraphics[width=3.5in]{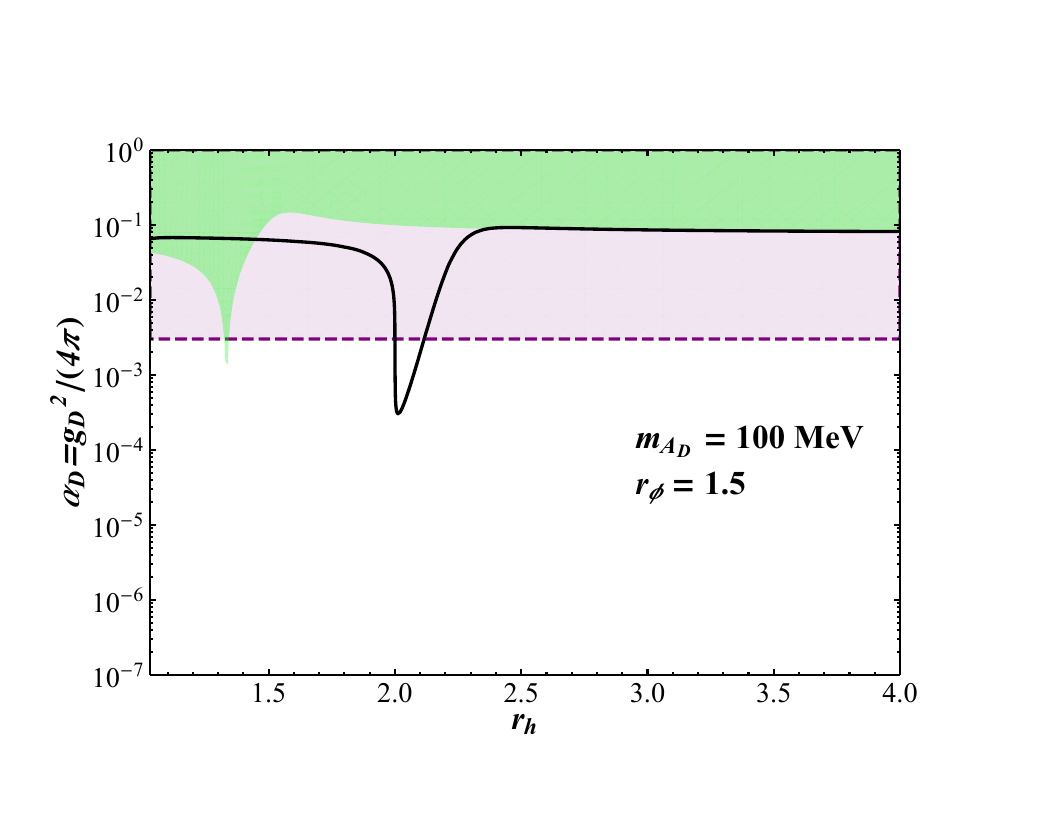}
\hspace{-0.75cm}
\includegraphics[width=3.5in]{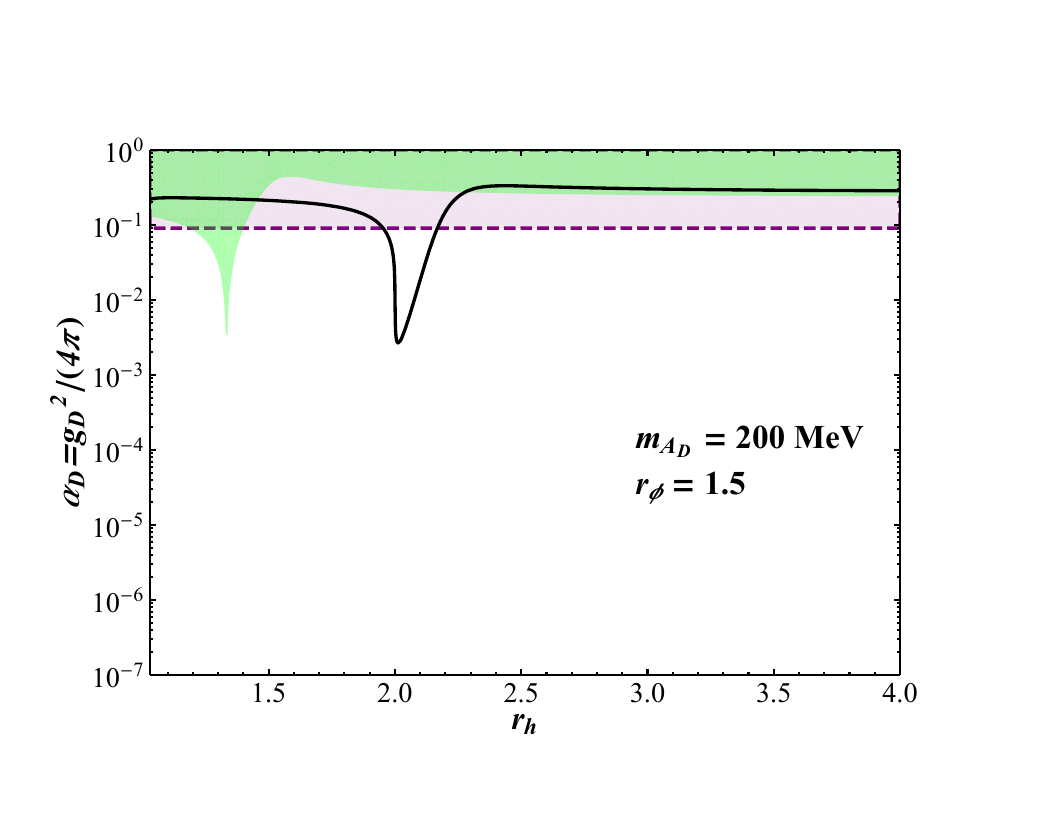}}
\caption{Contours in the $\alpha_D-r_h$ plane, where $\alpha_D=g_D^2/(4 \pi)$ that recreate the observed dark matter relic abundance assuming $b=1$ and $\epsilon = 3\times 10^{-4}$, for various choices of $r_\phi$ and $m_{A_D}$. Regions excluded by CMB limits on $(\sigma v)_{\textrm{ISR}}$ \cite{Cang:2020exa} are depicted as a green shaded region, while projected constraints from a null result of SENSEI after 100 $\textrm{g} \cdot \textrm{years}$ of exposure \cite{Battaglieri:2017aum} are depicted as a purple shaded region with a dashed boundary.}
\label{fig12}
\end{figure}

Before concluding this Section, it is helpful to depict the combined constraints from both direct detection and the CMB. To that end, in Figure \ref{fig12} we depict points that recreate the observed dark matter relic abundance as contours in the $\alpha_D-r_h$ plane, where $\alpha_D \equiv g_D^2/(4 \pi)$ is the dark coupling fine structure constant, for various selections of $r_\phi$ and $m_{A_D}$.\footnote{It should be noted that even the smallest values of $\alpha_D$ depicted in Figure \ref{fig12} are many orders of magnitude in excess of the approximate condition given in \cite{Cline:2017tka} to ensure that $\phi e^- \rightarrow \phi e^-$ occurs sufficiently quickly to allow the dark matter to achieve thermal equilibrium with the SM in the early universe. Furthermore, we find it unlikely that the contours in this Figure would be substantially affected by considering the possibility that a low $\alpha_D$ allows the dark Higgs $h_D$ to freeze out before the dark matter: Since the small couplings only appear for the region of the contours very near $r_h = 2$, we can expect that any dark matter annihilation processes which depend on an out-of-equilibrium dark Higgs number density will be secondary to the resonantly-enhanced annihilation process $\phi^* \phi \rightarrow A_D A_D$.} We have then overlayed the dominant present-day constraint on these models (from the effects of $(\sigma v)_{\textrm{ISR}}$ on the CMB), as well as the near-future constraint from a null result from SENSEI. Here, we can see that with the exception of the region near the $(\sigma v)_{\textrm{ISR}}$ resonance peak, the projected limits from SENSEI will generally always be more stringent than those which emerge from the CMB. Furthermore, since both $(\sigma v)_{\textrm{ISR}}$ and $\sigma_{\phi e}$ possess the same dependence on the kinetic mixing parameter $\epsilon$, this dominance will hold even as these constraints are altered by adjustments to $\epsilon$. In Figure \ref{fig12} we can also again clearly see the role that the dark Higgs can play in broadening the allowed parameter space of this model: Even modestly near the $r_h = 2$ resonance of the $\phi^* \phi \rightarrow A_D A_D$ process, regions of parameter space that are otherwise excluded by the CMB and/or a SENSEI null result can easily evade these constraints.

\section{Summary and Conclusions}\label{Section:Conclusion}

In this work, we have explored the effect of including a dark Higgs, $h_D$, on a simple realization of the vector portal/kinetic mixing DM framework with a complex scalar DM candidate $\phi$ and a dark photon $A_D$, specifically focusing on the so-called ``forbidden DM'' (FDM) regime in which the dominant process driving freeze-out is $\phi^* \phi \rightarrow A_D A_D$ (DM annihilation into a pair of dark photons), rather than the more conventional WIMP-like annihilation processes such as $\phi^* \phi \rightarrow e^+ e^-$. This setup, in particular the selection of a complex scalar for the DM candidate, represents the simplest construction within the FDM paradigm featuring non-trivial coupling between the dark Higgs and the DM.

We then outlined the mechanism by which the DM relic abundance is produced in this model, following a simple freeze-out process. Even without the inclusion of dark Higgs effects, we noted that there exist some significant phenomenological differences between the freeze-out process of our complex scalar DM construction here and that of the more well-studied FDM scenario with Dirac fermion DM \cite{DAgnolo:2015ujb,Cline:2017tka,Fitzpatrick:2020vba}. In particular, the complex scalar scenario is not subject to the same CMB constraints that favor extremely small kinetic mixing parameters $\epsilon < 10^{-6}$ for the Dirac fermion construction. Instead, we can allow $\epsilon \sim 10^{-(3-4)}$, which is large enough that some phenomenological complications which arise in the Dirac fermion case, such as the possibility of non-equilibrium dark photon number densities or the potential significance of $3\rightarrow 2$ processes such as $\phi^* \phi \phi \rightarrow A_D \phi$, play no significant role here. Motivated by both qualitative and numerical arguments, we ultimately argued that four processes may contribute significantly to the relic abundance in our calculations, specifically the WIMP-like process $\phi^* \phi \rightarrow e^+ e^-$ (and, if kinematically accessible, annihilation to other SM fermion pairs) and the three kinematically forbidden processes $\phi^* \phi \rightarrow A_D A_D$, $\phi^* \phi \rightarrow A_D h_D$, and $\phi^* \phi \rightarrow h_D h_D$.

After giving expressions for the thermally averaged cross sections for these processes in the presence of the dark Higgs, we began a numerical study of the DM relic abundance realized in this model and its dependence on various parameters, assuming a dark gauge coupling constant $0.1 \leq g_D \leq 1$, and dark photon masses in the range between $100 \; \textrm{MeV}$ and $1 \; \textrm{GeV}$. In the regime in which the kinematically forbidden transitions dominate, we found that the most significant effect of the dark Higgs $h_D$ on the DM relic abundance lay in its contribution to the cross section of $\phi^* \phi \rightarrow A_D A_D$ via an $s$-channel dark Higgs exchange: With only modest $O(10\%)$ tuning of the dark Higgs mass near the resonance peak $r_h \equiv m_{h_D}/m_{A_D} \approx 2$, this effect can reduce the relic abundance by 1-3 orders of magnitude compared to the identical system with the dark Higgs omitted. We also explored the sensitivity of this resonance effect to the coupling strength between the dark Higgs and the scalar DM. We described this coupling via the dimensionless parameter, $b$, where $b$ intuitively represents the fraction of $\phi$'s mass squared that emerges from coupling with the dark Higgs. Assuming no fine tuning in the scalar potential of the dark Higgs and the DM, $b$ should be a significant fraction of unity. Using both analytical and numerical arguments, we found that the resonant dark Higgs exchange's contribution to freeze-out demonstrates remarkable robustness to variation of $b$: This parameter can be as small as $b=0.05$ and the dark Higgs resonance can still reduce the DM yield by an order of magnitude compared to a construction with no dark Higgs. Other effects of the dark Higgs on the relic abundance were explored and generally produced far more limited effects: In particular, the effect of the processes $\phi^* \phi \rightarrow h_D A_D$ and $\phi^* \phi \rightarrow h_D h_D$ were negligible for the mass range we considered.

In order to explore the effect of the dark Higgs in broadening the viable parameter space of this FDM construction, we next depicted model parameter selections that recreated the observed DM relic abundance in the $r_h-r_\phi$ plane for various selections of $g_D$, $m_{A_D}$, and $b$. As can be expected by the magnitude of the resonance effect we observed previously, the favored $r_\phi$ value for a given set of model parameters is highly sensitive to $r_h$ even moderately close to resonance and, in contrast to constructions without a dark Higgs, a significant range of $r_\phi$ values (and therefore DM masses $m_\phi$) are feasible even for fixed choices of $m_{A_D}$ and the dark coupling $g_D$. We then completed our discussion with a brief survey of other experimental constraints on the points in parameter space for which the observed DM relic abundance is recreated. These constraints are quite mild: In particular, we argued that in the parameter space we consider, the lack of direct tree-level couplings between the dark Higgs and the SM severely limits detection prospects of $h_D$ itself, \eg through beam-dump or collider experiments. Instead, the most significant constraints on our construction emerge from CMB measurements and direct detection searches for the DM particle $\phi$. We then explored both of these effects quantitatively.
The model's CMB constraints stem from the $\epsilon^2$-suppressed $s$-wave process $\phi^* \phi \rightarrow A_D A_D^* \rightarrow A_D e^+ e^-$. For the lightest dark photon mass we consider ($m_{A_D}= 100 \; \textrm{MeV}$), these restrictions can be significant if $g_D$ is also chosen to be large ($g_D \geq 0.7$), in which case the entire region $r_h \lesssim 1.5$ is in general excluded. For smaller $g_D$ or larger $m_{A_D}$, these constraints only exclude a narrow region around the resonance peak $m_{h_D}/m_\phi = r_h r_\phi \approx 2$. We further argued that the only other significant experimental constraint on this model's parameter space stems from direct detection searches, which may be capable of probing/excluding most realizations of this model with $m_{A_D} = \textrm{100} \; \textrm{MeV}$ in the immediate future with upcoming SENSEI measurements and constraining benchmark points we consider with heavier dark photons with only slightly longer-term experiments such as SuperCDMS or DAMIC-1K.

Overall, we have found that this simple scalar DM realization of the FDM paradigm is subject to significant effects from the existence of the dark Higgs, in particular due to the $s$-channel exchange of a dark Higgs in the process $\phi^* \phi \rightarrow A_D A_D$. With only modest resonant enhancement near $m_{h_D} \approx 2 m_{A_D}$, this exchange becomes the dominant contribution to the $\phi^* \phi \rightarrow A_D A_D$ cross section for virtually any natural value of the coupling constant between $\phi$ and $h_D$, resulting in relic abundances that can differ by as much as 3 orders of magnitude from a corresponding scenario without dark Higgs effects included.
Furthermore, while we note that in this work we have considered the simplest possible construction in which this effect is present, a wider array of more complicated frameworks exist which should yield qualitatively similar results regarding the importance of this resonance (for example, a construction with pseudo-Dirac fermion DM in which the dark Higgs field imparts a Majorana mass to the DM, or one in which the kinetic mixing parameter $\epsilon$ is allowed to be much lower, potentially resulting in non-equilibrium number densities of dark photons and dark Higgses during freeze-out). Given the scale of the potential effect of these dark Higgs exchanges and the broad range of parameter space in our construction over which this effect is applicable, our results suggest that for a wide range of realizations of the forbidden DM paradigm, the effect of the dark Higgs, so often neglected, can easily be enormous.

\section*{Acknowledgements}
The authors would like to thank D. Rueter for discussions. This work was supported by the Department of Energy, Contract DE-AC02-76SF00515.



\end{document}